\documentclass[a4paper,fleqn,usenatbib]{mnras}

\usepackage{newtxtext,newtxmath}

\usepackage{soul}

\usepackage{tikz}
\usetikzlibrary{shapes,arrows}

\tikzstyle{line} = [draw, -latex']
\tikzstyle{cloud} = [draw, ellipse, 
 node distance=2cm, minimum height=2em, text centered, text width=2 cm]
\usepackage[T1]{fontenc}
\usepackage{ae,aecompl}

\usepackage{graphicx}	
\usepackage{amsmath}	
\usepackage{amssymb}	
\usepackage{hyperref}      

\newcommand{\mtrx}[1]{\mathsf{\mathbf{#1}}}
\newcommand{\vctr}[1]{\mathbf{#1}}

\defcitealias{klein19}{K19}
\defcitealias{bocquet19}{Bo19}
\defcitealias{bulbul19}{Bu19}
\defcitealias{saro15}{S15}

\title[MARD-Y3 and SPT-SZ Validation]{Validation of Selection Function, Sample Contamination and Mass Calibration in Galaxy Cluster Samples}

\author[S. Grandis et al.]{S.~Grandis,$^{1,2}$\thanks{E-mail: s.grandis@physik.lmu.de}
M.~Klein,$^{1,3}$
J.~J.~Mohr,$^{1,2,3}$
S.~Bocquet,$^{1,2}$
M.~Paulus,$^{1,3}$ 
\newauthor
T.~M.~C.~Abbott,$^{4}$
M.~Aguena,$^{5,6}$
S.~Allam,$^{7}$
J.~Annis.$^{7}$
B. A. Benson,$^{7, 8, 9}$
E.~Bertin,$^{10,11}$
\newauthor
S.~Bhargava,$^{12}$
D.~Brooks,$^{13}$
D.~L.~Burke,$^{14,15}$
A.~Carnero~Rosell,$^{16}$
M.~Carrasco~Kind,$^{17,18}$
\newauthor
J.~Carretero,$^{19}$
R. Capasso,$^{20}$
M. Costanzi,$^{21,22}$
L.~N.~da Costa,$^{23,24}$
J.~De~Vicente,$^{16}$
\newauthor
S.~Desai,$^{25}$
J.~P.~Dietrich,$^{1}$
P.~Doel,$^{13}$
T.~F.~Eifler,$^{26,27}$
A.~E.~Evrard,$^{28,29}$
B.~Flaugher,$^{7}$
\newauthor
P.~Fosalba,$^{30,31}$
J.~Frieman,$^{7,32}$
J.~Garc\'{i}a-Bellido,$^{33}$
E.~Gaztanaga,$^{30,31}$
D.~W.~Gerdes,$^{28,29}$
\newauthor
D.~Gruen,$^{34,14,15}$
R.~A.~Gruendl,$^{17,18}$
J.~Gschwend,$^{6,24}$
G.~Gutierrez,$^{7}$
W.~G.~Hartley,$^{13,35}$
\newauthor
S.~R.~Hinton,$^{36}$
D.~L.~Hollowood,$^{37}$
K.~Honscheid,$^{38,39}$
D.~J.~James,$^{40}$
T.~Jeltema,$^{37}$
\newauthor
K.~Kuehn,$^{41,42}$
N.~Kuropatkin,$^{7}$
M.~Lima,$^{6,7}$
M.~A.~G.~Maia,$^{7,24}$
J.~L.~Marshall,$^{43}$\newauthor
P.~Melchior,$^{44}$
F.~Menanteau,$^{17,18}$
R.~Miquel,$^{45,19}$
R.~L.~C.~Ogando,$^{23,24}$
A.~Palmese,$^{7,8}$\newauthor
F.~Paz-Chinch\'{o}n,$^{17,18}$
A.~A.~Plazas,$^{44}$
A.~K.~Romer,$^{12}$
A.~Roodman,$^{14,15}$
E.~Sanchez,$^{16}$\newauthor
A.~Saro,$^{21,22,46}$
V.~Scarpine,$^{7}$
M.~Schubnell,$^{29}$
S.~Serrano,$^{30,31}$
E.~Sheldon,$^{47}$
M.~Smith,$^{48}$\newauthor
A.~A.~Stark,$^{40}$
E.~Suchyta,$^{49}$
M.~E.~C.~Swanson,$^{18}$
G.~Tarle,$^{29}$
D.~Thomas,$^{50}$\newauthor
D.~L.~Tucker,$^{7}$
T.~N.~Varga,$^{3,51}$
J.~Weller,$^{3,51}$
R.~Wilkinson$^{12}$
}

\date{Accepted XXX. Received YYY; in original form ZZZ}

\pubyear{2018}

\hypersetup{draft}  
\begin{document}
\maketitle

\begin{abstract}
We construct and validate the selection function of the MARD-Y3 sample. 
This sample was selected through optical follow-up of the 2nd ROSAT faint source catalog (2RXS) 
 with Dark Energy Survey year 3 (DES-Y3) data. The selection function is modeled  by combining an empirically constructed X-ray selection function with an incompleteness model for the optical follow-up. We validate the joint selection function by testing the consistency of the constraints on the X-ray flux--mass and richness--mass scaling relation parameters derived from different sources of mass information:  (1) cross-calibration using SPT-SZ clusters, (2) calibration using number counts in X-ray, in optical and in both X-ray and optical while marginalizing over cosmological parameters, and (3) other published analyses.
We find that the constraints on the scaling relation from the number counts and SPT-SZ cross-calibration agree, indicating that our modeling of the selection function is adequate.  Furthermore, we apply a largely cosmology independent method to validate selection functions via the computation of the probability of finding each cluster in the SPT-SZ sample in the MARD-Y3 sample and vice-versa.  This test reveals no clear evidence for MARD-Y3 contamination, SPT-SZ incompleteness or outlier fraction.
 Finally, we discuss the prospects of the techniques presented here to limit systematic selection effects in future cluster cosmological studies.
\end{abstract}




\vskip20in

\section{Introduction}

The number of galaxy clusters as a function of mass and redshift is generally accepted as being one of the major sources of 
information on the composition and evolution of the Universe \citep[see, for instance,][and references therein]{haiman01, albert06, allen11}. Cluster numbers can be predicted by multiplying the number density of halos, the "halo mass function" (HMF), with the volume sampled. The HMF is highly sensitive to the matter density and the amplitude of 
matter fluctuations and can be accurately calibrated by simulations \citep[e.g.,][]{jenkins01,warren06,tinker08, bocquet16, mcclintock19b}. The cosmological volume is dependent on the expansion history.  Together with the redshift evolution of the amplitude of fluctuations, this makes the
 redshift evolution of the number of clusters very sensitive to the yet unexplained late time accelerated expansion of the 
Universe.

Clusters can be selected in large numbers through their observational signatures at different wavelengths. In X-rays, the intra cluster 
medium (ICM), heated by having fallen into the cluster's gravitational potential emits a strong and diffuse thermal emission in X-rays
\citep[see][for selections based on this signature]{vikhlinin98,boehringer01,romer01,clerc14,klein19}. At 
optical wavelengths, clusters can be identified as over-densities of red galaxies \citep[for recent applications to wide photometric surveys, see e.g.][]{Koester07, rykoff16}. In the 
millimeter 
regime inverse Compton scattering 
of the cosmic microwave background (CMB) photons with the ICM makes clusters detectable as extended shadows in CMB maps. This phenomenon is called the Sunyaev-Zel'dovich 
effect \citep[SZE, ][]{sunyeavzeldovich72}. Scanning CMB-surveys for such shadows enables the detection of near-complete, approximately mass-limited cluster samples
\citep{Hasselfield13, bleem15, planck16_sze, hilton18, huang19} .

Inference of the mass distribution, and thereby cosmological constraints, is limited by the ability to characterise incompleteness, 
contamination, and the observable--mass mapping in each of these selection techniques. This leads to the problem of determining the 
selection function of any cluster sample. The problem is split into to parts: determining how the selection function depends on the X-ray, optical, or SZE observables, and calibrating the scaling relation between that observable and cluster mass. The latter is called 
\textit{mass calibration} \citep[for a review, see][]{pratt19}. It is tackled by measuring the cluster gravitational potential either 
through the coherent distortion of 
background galaxies due to weak gravitational lensing \citep[e.g.,][]{bardeau07, okabe10, hoekstra12, 
applegateetal14, israel14, Melchior15, okabesmith16, melchior17, schrabback18a, dietrich19, stern19, mcclintock19}, 
or the analysis of the projected phase 
space distribution of member galaxies 
whose velocities are measured by spectroscopic observations \citep{sifon13, bocquet15, zhang17, capasso19a, capasso19b}. 
Such techniques are direct probes of the clusters' gravitational potential.

Both X-ray and SZE selections are known to provide cluster candidate lists with less contamination than those carried out
in optical surveys which suffer from projection effects \citep[see for instance][and reference therein]{costanzi19}.
In X-ray studies at sufficiently high detection significance, extent 
information can be used to 
control contamination \citep{vikhlinin98, pacaud06}. Nevertheless, optical confirmation is still required to estimate the redshift of 
the candidates and to reduce the 
contamination. Traditionally, targeted imaging of individual cluster candidates was performed to this end.  

Such campaigns of pointed follow-up have recently been superseded by automated optical 
confirmation, as for instance by the Multi-wavelength Matched Filter 
tool \citep[MCMF,][]{klein18}. It scans photometric data along the line 
of sight toward X-ray or SZE cluster candidates with a spatial and 
color filter to identify cluster galaxies and determine the clusters redshift. Such 
tools have the advantage of exploiting the ever larger coverage 
of deep and wide photometric surveys such as the 
Dark Energy Survey\footnote{\url{https://www.darkenergysurvey.org}} 
\citep[DES,][]{DES16} or the upcoming
Euclid Mission\footnote{\url{http://sci.esa.int/euclid/42266-summary/}} \citep{laureijs11} and 
the Rubin Observatory Legacy Survey of Space and Time\footnote{\url{https://www.lsst.org/}} \citep[LSST,][]{Ivezic08}) to follow up candidate lists 
whose size would make pointed observations impractical.

In this work we seek to construct and validate the selection function of an X-ray selected and optically cleaned sample. We focus on 
the MARD-Y3 sample 
\citep[hereafter \citetalias{klein19}]{klein19} which is constructed by following up the highly contaminated 2nd ROSAT faint source catalog 
\citep[2RXS, ][]{boller16} with the DES data of the first 3 years of observations \citep[between August 2013 and February 2016, DES-Y3, ][]{desy3}. Strict optical cuts lead to a final contamination of $2.5\%$ at the cost of optical 
incompleteness, which we model alongside the X-ray selection. The optical follow up also provides measurements of the optical richness 
of the selected clusters.

First, we confirm the selection function model by constraining the scalings between X-ray flux and mass, and 
richness and mass in different ways.  We perform a cross-calibration of our observables with the indirect mass information 
contained in the SZE-signature of SPT selected clusters by cross-matching the two samples to derive the flux and 
richness scaling relation parameters. Then we constrain the same scaling relations by fitting for the number counts of 
MARD-Y3 clusters while marginalizing over cosmology. The former method 
is largely independent of the selection function for the MARD-Y3 sample, while the latter is strongly dependent on it. 
Consequently, consistent scaling relation parameter constraints from the two methods validate our selection function model.

We also test the selection functions by further developing the formalism of cross-matching and detection 
probabilities introduced by  \citet[hereafter \citetalias{saro15}]{saro15}. 
We thereby constrain the probability of MARD-Y3 contamination, the
SPT-SZ incompleteness and the probability of outliers from 
the scaling relations. This also allows us to identify a population of clusters that exhibit either a surprisingly high X-ray flux
or surprisingly low SZE-signal.

The paper is organised as follows. In Section~\ref{sec:framework} presents the conceptual framework within which we model galaxy cluster samples;  Section~\ref{sec:method} presents the specific validation methods used in this work; and Section~\ref{sec:dataNpriors} contains a description of the MARD-Y3 and the SPT-SZ cluster samples as well as the priors adopted for the analysis.  Our main results are presented in Section~\ref{sec:results}, 
comprising the different cross-checks on scaling relation parameters that validate our selection 
function modeling and that enable further checks of the selection functions. 
The results are then discussed in Section~\ref{sec:discussion}, 
leading to our conclusions in Section~\ref{sec:conclusions}. 
The
appendices contain more extensive descriptions of the construction
and validation of the X-ray observational error model  (Appendix~\ref{app:X_meas_uncert}) and a gallery of multi-wavelength images used for visual inspection
(Appendix~\ref{app:gallery}).

\section{Conceptual Framework for Cluster Cosmology Analyses}
\label{sec:framework}

In the following section we will present in mathematical detail the model of the cluster population used in this work to describe the 
properties of the cluster samples. This discussion follows the Bayesian hierarchical framework established by \citet{bocquet15}.
The cluster population is modeled by a forward modeling approach that transforms the differential number of clusters as a function 
of halo mass $M_\text{500c}$\footnote{$M_{500,\text{c}}$ is the mass enclosed in a spherical over density with average density 500 times the critical density of the Universe. For sake of brevity we will refer to $M_{500,\text{c}}$ as $M$ for the rest of this work.} and redshift $z$ to the space of observed cluster properties, 
such as the measured X-ray flux $\hat f_\text{X}$, the measured richness $\hat \lambda$ and the measured SZE signal-to-noise $\xi$. 
This transformation is performed in two steps. 
First,  scaling relations having intrinsic scatter are utilized to estimate the cluster numbers as a function of intrinsic flux, richness, and SZE signal-to-noise.
These relations have several free parameters such as amplitude, mass and redshift trends,
intrinsic scatter around the mean relation, and correlation coefficients among the intrinsic scatter on different observables.
Constraining these free parameters is the aim of this work, as these constraints characterise the systematic uncertainty in the 
observable--mass relations.
Second, we apply models of the measurement uncertainty
to construct the cluster number density as a function of their properties.
We also present the modeling of the selection function and of the likelihood used to infer the parameters governing the scaling 
relations.
  
\subsection{Modeling the cluster population}\label{sec:cluster_pop_model}

The starting point of our modeling of the cluster population is their differential number as function of 
halo mass $M_\text{500c}$ and redshift $z$, given by 
\begin{equation}
\frac{\text{d}N}{\text{d}M} \Big|_{M, z} = \frac{\text{d}n}{\text{d}M}\Big|_{M, z} \frac{\text{d}^2 V}{\text{d}z\text{d}
\Omega} \text{d}z \text{d}\Omega, 
\end{equation}
where $\frac{\text{d}n}{\text{d}M}\big|_{M, z}$ is the halo mass function describing the differential number density of halos at 
mass $M$ and redshift $z$, as presented by \citet{tinker08}; $\frac{\text{d}^2 V}{\text{d}z\text{d}\Omega}\text{d}z \text{d}\Omega$ is the cosmological volume 
subtended by the redshift bin $\text{d}z$ and the survey angular footprint $\text{d}\Omega$.

The mapping from halo mass to intrinsic cluster properties is modeled by scaling relations,
which are characterised by a mean relation with free parameters and a correlated scatter.
The mean intrinsic relations we use read as
\begin{equation}\label{eq:xraysacling}
\langle f_\text{X} \rangle = \frac{L_0\,  A_\text{X} }{4 \pi d_L^2(z)}\Big(\frac{M\,h}{M_{0,\text{X}}}\Big)^{B_\text{X}} 
\Big(\frac{E(z)}{E(z_{0,\text{X}})}\Big)^2 \Big(\frac{1+z}{1+z_{0,\text{X}}}\Big)^{C_\text{X}}
\end{equation}
for the X-ray flux\footnote{The flux in this form makes explicit the cosmological dependencies due to distances and to self-similar evolution while allowing for departures from that self-similar evolution.},

\begin{equation}
\langle \lambda \rangle = A_\lambda \Big(\frac{M\,h}{M_{0,\lambda}}\Big)^{B_\lambda} \Big(\frac{E(z)}
{E(z_{0,\lambda})}\Big)^{C_\lambda}
\label{eq:richnessscaling}
\end{equation}
for the richness, and 
\begin{equation}\label{eq:SZscaling}
\langle \zeta \rangle = A_\text{SZ} \Big(\frac{M\,h}{M_{0,\text{SZ}}}\Big)^{B_\text{SZ}} \Big(\frac{E(z)}
{E(z_{0,\text{SZ}})}\Big)^{C_\text{SZ}}
\end{equation}
for the SZE signal-to-noise in a reference field. $h$ is the present day expansion rate in units of $100$~km~s$^{-1}$~Mpc$^{-1}$, and $E(z)$ the ratio between the expansion rate at redshift $z$ and the current day expansion rate. The form of the redshift evolution adopted in equations~(\ref{eq:richnessscaling}) and (\ref{eq:SZscaling}) has explicit cosmological dependence in the redshift evolution that is not well motivated  \citep[see discussion in][hereafter \citetalias{bulbul19}]{bulbul19}, 
but we nevertheless adopt these 
forms for consistency with previous studies (e.g. \citetalias{saro15}).
The pivot points in mass $M_{0,\text{X}}=6.35\times10^{14} h \, M_\odot h^{-1}$, $M_{0,\lambda}=3.\times 10^{14} M_\odot h^{-1}= M_{0,\text{SZ}}$, in luminosity  $L_0=10^{44}$ erg s$^{-1}$,
and in redshift $z_{0,\text{X}}=0.45$, $z_{0,\lambda}=0.6=z_{0,\text{SZ}}$ are constants in our analysis. 
In contrast, the parameters $A_\aleph$, $B_\aleph$ and $C_\aleph$ for $\aleph \in (\text{X}, \lambda, \text{SZ})$ are free parameters of the likelihoods described in section~\ref{sec:method}. 
These parameters encode the systematic uncertainty in the mass derived from each observable.

The inherent stochasticity in the cluster populations is modeled by assuming that the intrinsic observable scatters log-normally 
around the mean intrinsic relation\footnote{The notation utilised here is imprecise. The scaling relation 
describes the mean of the natural logarithm of the intrinsic observable, not the natural logarithm of the mean, as suggested by the 
notation. Not interpreting $\langle\cdot\rangle$ as an average ensures a fully consistent notation.}.
Consequently, given mass $M$ and redshift $z$, the probability for the intrinsic cluster observables ($f_\text{X}$, $\lambda$, 
$\zeta$) is given by 
\begin{equation}\label{eq:P_introbs_m}
P(f_\text{X}, \lambda, \zeta|M, z) = \frac{1}{\sqrt{(2\pi)^3 \det \mtrx{C} } } \frac{1}{f_\text{X} \lambda \zeta} \exp \Big\{ -\frac{1}{2} \vctr{\Delta x}^T 
\mtrx{C}^{-1}  \vctr{\Delta x} \Big\},
\end{equation}
with 
\begin{equation}
\vctr{\Delta x}^T = (\ln f_\text{X} - \ln \langle f_\text{X} \rangle, \ln \lambda - \ln \langle \lambda \rangle,  \ln \zeta - \ln 
\langle \zeta \rangle)
\end{equation}
and 
\begin{equation}
\mtrx{C} = \begin{bmatrix}
    \sigma_\text{X}^2                                           & \sigma_\text{X} \sigma_\lambda \rho_{\text{X}, \lambda}    & 
\sigma_\text{X} \sigma_\text{SZ} \rho_{\text{X}, \text{SZ}}  \\
    \sigma_\text{X} \sigma_\lambda \rho_{\text{X}, \lambda}     & \sigma_\lambda^2                                           &  
\sigma_\lambda \sigma_\text{SZ} \rho_{\lambda, \text{SZ}}   \\
    \sigma_\text{X} \sigma_\text{SZ} \rho_{\text{X}, \text{SZ}} &  \sigma_\lambda \sigma_\text{SZ} \rho_{\lambda, 
\text{SZ}} & \sigma_\text{SZ}^2
\end{bmatrix},
\end{equation}
where $\sigma_\aleph$ for  $\aleph \in (\text{X}, \lambda, \text{SZ})$  encodes the magnitude of the intrinsic 
log-normal scatter in the respective observable, while the correlation coefficients $\rho_{\aleph, \aleph^\prime}$ encode 
the degree of correlation between the intrinsic scatters on the respective observables. The scatter parameters and the 
correlation coefficients are free parameters of our analysis. 

The assumption of log-normality is motivated theoretically by two facts: the intrinsic observables are strictly larger than zero, and a log-normal scatter is the simplest model. Operationally, it has the added benefit of allowing one to introduce correlated scatter in a well defined way. Observationally, deviations from log-normality have not been detected  (e.g. \citet{mantz16} did not measure any significant skewness in several different observable--mass relations). In this work, we introduce a new framework to test log-normality (c.f. Section~\ref{sec:consistencymethod}).

The differential number of objects as a function of intrinsic observables can be computed by applying the stochastic mapping between mass and intrinsic observables to the differential number of clusters as a function of mass, i.e
\begin{equation}\label{eq:d3Ndfxdlamdzeta}
\frac{\text{d}^3 N}{\text{d}f_\text{X}\text{d}\lambda\text{d}\zeta}\Big|_{f_\text{X}, \lambda, \zeta, z} = \int \text{d}M 
P(f_\text{X}, \lambda, \zeta|M, z) \frac{\text{d}N}{\text{d}M} \Big|_{M, z}.
\end{equation}

In some parts of our subsequent analysis, we do not require the distribution in SZE signal-to-noise. The differential 
number of objects as a function of intrinsic X-ray flux and richness can be obtained either by marginalising 
equation~(\ref{eq:d3Ndfxdlamdzeta}) over the intrinsic SZE signal $\zeta$, or by defining $P(f_\text{X}, \lambda|M, z)$ just for 
the X-ray and optical observable by omitting the SZE part,
\begin{equation}\label{eq:d2Ndfxdlam}
\begin{split}
 \frac{\text{d}^2 N}{\text{d}f_\text{X}\text{d}\lambda}\Big|_{f_\text{X}, \lambda, z} &= \int \text{d}\zeta \frac{\text{d}^3 N}
{\text{d}f_\text{X}\text{d}\lambda\text{d}\zeta}\Big|_{f_\text{X}, \lambda, \zeta, z} \\
  &= \int \text{d}M P(f_\text{X}, \lambda|M, z) \frac{\text{d}N}{\text{d}M} \Big|_{M, z}.
\end{split}
\end{equation}

\subsection{Modeling measurement uncertainties}\label{sec:treat_meas_err}

The intrinsic cluster observables are not directly accessible as only their measured values are known. We thus need to 
characterise the mapping between intrinsic and measured observables.

For the X-ray flux, we assume that the relative error on the flux $\hat \sigma_\text{X}$ is the same as the relative error 
in the count rate. For each object $(i)$ in our catalog we can determine 
\begin{equation}\label{eq:P_X_meas_i}
P(\hat f_\text{X}^{(i)}|f_\text{X}) = \frac{1}{\sqrt{2 \pi (\hat \sigma_\text{X}^{(i)})^2}} \frac{1}{\hat f_\text{X}^{(i)}} \exp \Big\{ -\frac{1}{2}\frac{(\ln \hat 
f_\text{X}^{(i)}-\ln f_\text{X})^2}{(\hat \sigma_\text{X}^{(i)})^2} \Big\}.
\end{equation}
For an application described below, it is necessary to know the measurement uncertainty on the X-ray flux for 
arbitrary $\hat f_\text{X}$, also those fluxes for which there is no corresponding entry in the catalog.  As described in more 
detail in Appendix~\ref{app:X_meas_uncert}, we extrapolate the relative measurement uncertainty rescaled to the 
median exposure time in our footprint, creating a function $\hat \sigma_\text{X}^2 (\hat f_\text{X}, z, t_\text{exp})$, 
which in turn allows us to compute
\begin{equation} \label{eq:P_meas_X_arb}
\begin{split}
P(\hat f_\text{X}|f_\text{X}, z, t_\text{exp}) = & \frac{1}{\sqrt{2 \pi \hat \sigma_\text{X}^2 (\hat f_\text{X}, z, t_\text{exp})}}  \frac{1}{\hat f_\text{X}}\\
& \exp \Big\{ -\frac{1}{2}\frac{(\ln \hat f_\text{X}-\ln f_\text{X})^2}{\hat \sigma_\text{X}^2 (\hat f_\text{X}, z, t_\text{exp})} 
\Big\}.
\end{split}
\end{equation}

Following \citetalias{saro15}, the measurement uncertainty on the optical richness is modeled as Poisson noise in the 
Gaussian limit, that is 
\begin{equation}\label{eq:Pmeasopt}
P(\hat \lambda | \lambda) = \frac{1}{\sqrt{2 \pi \lambda}} \exp\Big\{ -\frac{1}{2}\frac{(\hat \lambda - \lambda)^2}
{\lambda}\Big\}.
\end{equation}

The measurement uncertainty on the SZE signal-to-noise follows the prescription of \citet{vanderlinde10}, who have determined the 
relation between measured SZE signal-to-noise $\xi$ and the intrinsic signal-to-noise, as a function of the effective field depth 
$\gamma_\text{f}$\footnote{In \citet{dehaan16} these factors are presented as renormalizations of the amplitude of the 
SZE-signal--mass relation. Our notation here is equivalent, but highlights that they describe a property of the mapping between 
intrinsic SZE-signal and measured signal, and not between intrinsic signal and mass.}, namely
\begin{equation}\label{eq:P_SZ_meas}
P(\xi|\zeta, \gamma_\text{f}) = \frac{1}{\sqrt{2\pi}}\exp\Big\{-\frac{1}{2}\Big( \xi - \sqrt{\gamma^2_\text{f} \zeta^2 + 3} \Big)^2 \Big\}.
\end{equation}

\subsection{Modeling selection functions}\label{sec:selection_function1}

The selection functions in optical and SZE observables are easy to model as the mapping between measured and intrinsic observables is known and the selection criterion is a sharp cut in the measured observable.  For the optical case,
the removal of random superpositions by imposing $f_\text{cont}<0.05$ in the optical follow-up 
leads to a redshift dependent minimal measured richness 
$\lambda_\text{min}(z)$, as discussed in \citetalias{klein19}. This leads to an optical selection function which is a 
step function in measured richness 
\begin{equation}
P(\text{DES}|\hat \lambda, z) = \Theta(\hat \lambda - \lambda_\text{min}(z)),
\end{equation}
where $\Theta(x)$ is the Heavyside step function with value 0 at $x<0$, and 1 at $x\ge0$. 
Using the measurement uncertainty 
on richness (equation~\ref{eq:Pmeasopt}), we construct the optical selection function in terms of intrinsic richness $\lambda$ as
\begin{equation}\label{eq:Pdet_opt}
\begin{split}
P(\text{DES}|\lambda, z) &= P(\hat \lambda>\lambda_\text{min}(z) | \lambda) = \int_{\lambda_\text{min}(z)}^\infty \text{d}\hat \lambda P(\hat \lambda | \lambda)
\end{split}
\end{equation}

The SPT catalog we use in this work is selected by a lower limit to the measured signal-to-noise $\xi>4.5$, 
which, analogously to the optical case, is a step function in $\xi$ and leads to an SZE selection function on $\zeta$ 
given by 
\begin{equation}
\begin{split}
P(\text{SPT}|\zeta, \gamma_\text{f}) &= P(\xi>4.5|\zeta, \gamma_\text{f}) = \int_{4.5}^\infty \text{d} \xi P(\xi|\zeta, \gamma_\text{f}).
\end{split}
\end{equation}

\begin{figure}
	\includegraphics[width=\columnwidth]{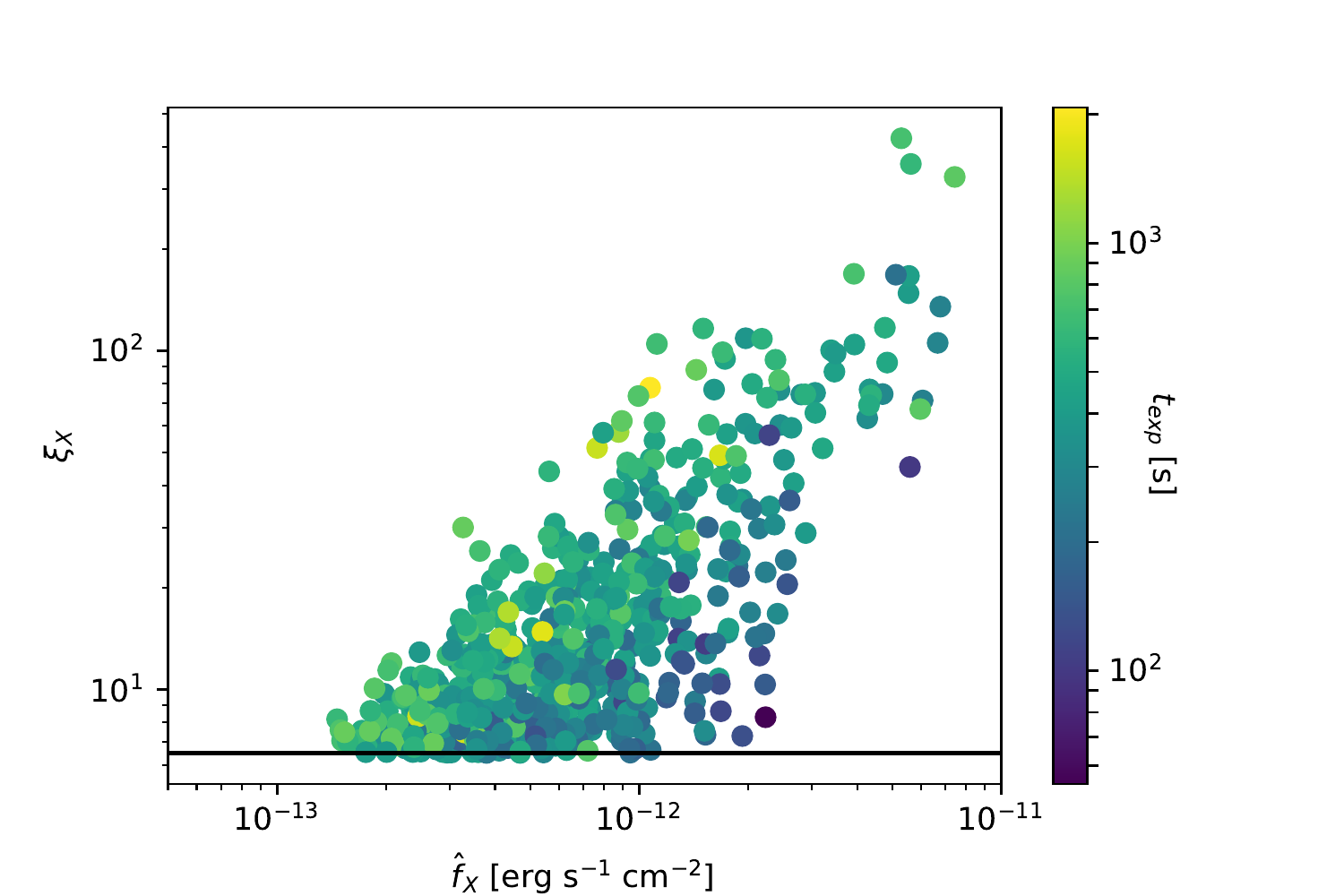}
	\vskip-0.10in
    \caption{The measured X-ray flux $\hat f_\text{X}$ and the X-ray detection significance $
\xi_\text{X}$, color coded by the exposure time. The black horizontal line indicates the X-ray selection criterion $\xi_\text{X}>6.5$. While X-ray flux 
and significance clearly display scaling, the scatter around this scaling correlates with exposure time. This relation and the scatter around it can be used to estimate the X-ray selection function (c.f. section~\ref{sec:X_selection}).}
    \label{fig:Xcompl_data}
\end{figure}

\subsubsection{Constraining the X-ray selection function}
\label{sec:X_selection}

The selection criterion used to create the 2RXS catalog is given by the cut $\xi_\text{X}>6.5$, where $\xi_\text{X}$ is the significance of 
existence of a source, computed by maximizing the likelihood that a given source is not a background fluctuation \citep{boller16}. In the space of 
this observable, the selection function is a simple step function.  In X-ray studies however, the selection function in the space of 
intrinsic X-ray flux is traditionally determined by image simulations \citep{vikhlinin98, pacaud06, clerc18}. In such an analysis, 
the emission from simulated clusters is used to create simulated X-ray images or event files, which are then 
analyzed with the same source extraction tools that are employed on the actual data.  As a function of intrinsic flux, the 
fraction of recovered clusters is then used to estimate the selection function $P(\text{X-det}|f_\text{X}, ...)$.  This process captures, to
the degree that the adopted X-ray surface brightness model is consistent with that of the observed population, the impact of
morphological variation on the selection.

In this work, we take a novel approach, inspired by the treatment of optical and SZE selection functions outlined above. This approach is based on the concept that the traditional selection function can be 
described as a combination of two distinct statistical processes: the mapping between measured detection significance $\xi_\text{X}$ and measured flux $\hat f_\text{X}$, and the mapping between measured flux $\hat f_\text{X}$ and intrinsic flux $f_\text{X}$, i.e.
\begin{equation}
\begin{split}
P(\text{X-det}|f_\text{X}, ...) = \int \text{d}\hat f_\text{X} P(\text{X-det}|\hat f_\text{X},..) P(\hat f_\text{X}|f_\text{X}, ...),
\end{split}
\end{equation}
where the second part of the integrand is the description of the measurement uncertainty of the X-ray flux. This mapping is needed to perform the number counts and any mass calibration, so it needs to be determined anyway. Its construction is described in Appendix~\ref{app:X_meas_uncert}. The first term can be easily computed from the 
mapping between measured flux $\hat f_\text{X}$ and X-ray significance $\xi_\text{X}$, $P(\xi_\text{X}|\hat 
f_\text{X}, ..)$. Indeed, it is just the cumulative distribution of that mapping for $\xi_\text{X}>6.5$.

The mapping between measured flux $\hat f_\text{X}$ and X-ray significance $\xi_\text{X}$ can be seen in 
Fig.~\ref{fig:Xcompl_data} for the MARD-Y3 clusters, where we plot the detection significance against the measured fluxes. The 
relation displays significant scatter, which is partially due to the different exposure times (color-coded). Also clearly 
visible is the selection at $\xi_\text{X}>6.5$ (black line). As an empirical model for this relation we make the ansatz
\begin{equation}
\langle \xi_\text{X} \rangle = \xi_0(z) e^{\alpha_0} \Big(\frac{\hat f_\text{X} }{ f_0(z) }\Big)^{\alpha_1} 
\Big(\frac{ t_\text{exp} }{400 \text{s}}\Big)^{\alpha_2},
\end{equation}
where $\xi_0(z)$ and $f_0(z)$ are the median significance and measured flux in redshift bins. To reduce measurement noise, we 
smooth them in redshift. We then assume that the significance of each cluster scatters around the mean significance 
with a log-normal scatter $\sigma_\alpha$. This provides the distribution $P(\xi_\text{X}|\hat f_\text{X}, z, 
t_\text{exp})$.

To fit the free parameters of this relation, namely ($\alpha_0$, $\alpha_1$, $\alpha_2$, $\sigma_\alpha$), we 
determine the likelihood of each cluster $i$ as 
\begin{equation}\label{eq:anc_like}
L_{\alpha, i} = \frac{P(\xi_\text{X}^{(i)}|\hat f_\text{X}^{(i)}, z^{(i)}, t_\text{exp}^{(i)})}{P(\xi_\text{X}>6.5|\hat f_\text{X}
^{(i)}, z^{(i)}, t_\text{exp}^{(i)})},
\end{equation}
where the numerator is given by evaluating $P(\xi_\text{X}|\hat f_\text{X}, z, t_\text{exp})$ for each cluster, while the 
denominator ensures proper normalization for the actually observable data, i.e. clusters with $\xi_\text{X}>6.5$. In 
properly normalising we account for the Malmquist bias introduced by the X-ray selection. Note also that we do not 
require the distribution of objects as a function of $\hat f_\text{X}$ to perform this fit, as it would multiply both the 
numerator and the denominator and hence cancel out.

The total log-likelihood of the parameters ($\alpha_0$, $\alpha_1$, $\alpha_2$, $\sigma_\alpha$) is given by the sum 
of the log-likelihoods $\ln L_\alpha = \sum_i \ln L_{\alpha, i}$. This likelihood provides stringent constraints on the 
parameters  ($\alpha_0$, $\alpha_1$, $\alpha_2$, $\sigma_\alpha$).
We find the best fitting values $\alpha_0=-0.113\pm0.020$, $\alpha_1=1.275\pm0.031$, $\alpha_2=0.799\pm0.038$ and $
\sigma_\alpha=0.328\pm0.012$. Noticeably, the constraints are very tight, indicating that the sample itself provides 
precise information about this relation. 

Given this relation, the X-ray selection function 
can be computed as 
\begin{equation}
\begin{split}
P(\text{RASS}|\hat f_\text{X}, z, t_\text{exp}) &= P(\xi_\text{X}>6.5|\hat f_\text{X}, t_\text{exp}, z)  \\ 
 &= \int_{6.5}^\infty \text{d}\xi_\text{X}P(\xi_\text{X}|\hat f_\text{X}, t_\text{exp}, z).
\end{split}
\end{equation}
Whenever the X-ray selection function is required, we sample the extra nuisance parameters with the ancillary likelihood (Eq.~\ref{eq:anc_like}), marginalizing over
the systematic uncertainties in this element of the X-ray selection function.
Further discussion of the parameter posteriors 
and their use to test for systematics in the selection function can be found in section~\ref{sec:X_selection_disc}.

\begin{figure}
	\includegraphics[width=\columnwidth]{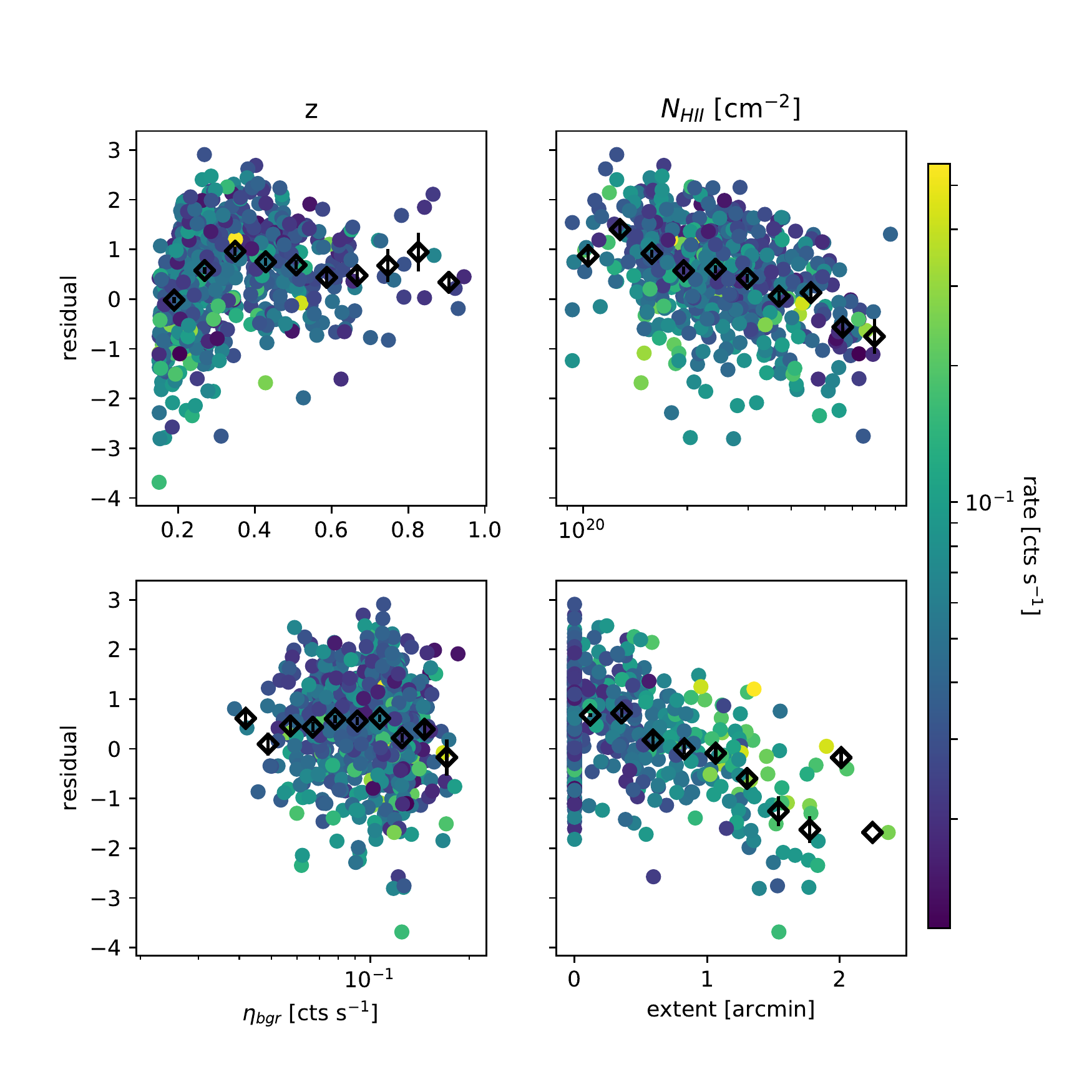}
	\vskip-0.10in
    \caption{Residuals of the fitted significance--flux relation against redshift (upper left panel), Galactic foreground neutral hydrogen column density (upper right), background counts rate in the aperture (lower left) and measured extent. Color coded is the count rate of the objects. The residuals are normalized by the best fit value of the intrinsic scatter around the mean relation. As black point we show the means of these points in bins along the x-axis. This plot indicates that the addition of a redshift trend or and extent trend would be natural extension of our model. The current level of systematic and statistical uncertainties however does not require these extensions.}
    \label{fig:Xcompl_fit_trends}
\end{figure}

\subsubsection{Testing for additional dependencies}\label{sec:residual_trends}

Empirically calibrating the relation governing the X-ray selection function has three benefits. (1) We take full account 
of the marginal uncertainty in the X-ray selection function. (2) Compared to image simulation, we do not rely on the 
realism of the clusters put into the simulation. Indeed, we use the data themselves to infer the relation. Together with the 
aforementioned marginalisation this ensures that we do not artificially bias our selection function. (3) We can 
empirically explore any further trends of the residuals of the significance--flux relation with respect to other quantities. 

Trends of the residuals are shown in Fig.~\ref{fig:Xcompl_fit_trends}, where the residual $\sigma_\alpha^{-1} \ln(\xi_\text{X}^{(i)}/\langle  
\xi_\text{X}\rangle^{(i)})$ is plotted against redshift (upper left panel), Galactic hydrogen column density (upper right), 
background count rate in an aperture of 5' radius (lower left) and measured extent (lower right). As black dots we show the means of the populations in bins along the x axis. We find a weak trend 
with hydrogen column density.  For simplicity we let this trend contribute to the overall scatter $
\sigma_\alpha$. We find no correlation with the background brightness. There is a clear trend with measured extent, as can be expected for extended sources like clusters. We do not, however, follow up on this trend, as 442 of the 708 cluster that we consider have a measured extent of 0 (due to the large PSF of RASS).

Disturbingly, we find a trend with redshift which is not captured by our model, as can be seen in the upper left 
panel of Fig.~\ref{fig:Xcompl_fit_trends}. At the lowest redshifts, we tend to over predict the significance given flux and 
exposure time, while at intermediate and high redshifts we tend to underestimate it. This residual systematic 
manifests itself at different stages in our analysis, and we discuss this as it arises and again in section~\ref{sec:X_selection_disc}.

\section{Validation methods}
\label{sec:method}

As described above, the selection model for the clusters is specified by the form of the mass--observable relation and the 
intrinsic and observational scatter of the cluster population around the mean relation.  The choice of the form of this relation
should be driven primarily by what the data themselves demand, with guidance from the principle of preferred simplicity (Occam's razor) 
and informed by predictions from structure formation simulations.  This scaling relation should be empirically calibrated using methods 
such as weak lensing and dynamical masses, whose systematics can be calibrated and corrected using by comparison with structure formation simulations.  Finally, a key step in cosmological analyses of cluster samples is to check consistency of the cluster sample 
with the best fit model of cosmology and mass--observable relation \citep[e.g., goodness of fit; see][]{bocquet15}.

The mass-observable relation can be calibrated using multiple sources of mass information, including direct mass information and 
from the cluster counts themselves (which is the distribution in observable and redshift of the cluster sample). thus, 
there is ample opportunity for validation of the scaling relation and the selection
function.  In future cosmology analyses, blinding of the cosmological and nuissance parameters will be the norm, and cluster cosmology 
is no exception.  The validation of a cluster sample through the requirement that all different reservoirs of information about the 
scaling relation lead to consistent results can be carried out in a blinded manner and should lead to improved stability and robustness
in the final, unblinded cosmological results.  We note that given the sensitivity of mass measurements to the distance-redshift relation and
the sensitivity of the counts to both distance-redshift and growth of structure, these blinded tests should in general be carried out within each 
family of cosmological models considered (e.g., flat or curved $\nu$-$\Lambda$CDM, flat or curved $\nu$-wCDM, etc).

In this work we seek to perform the following tests to validate the selection function modeling of the MARD-Y3 sample:
(1) we investigate whether the X-ray flux--mass and richness--mass relation obtained by cross-calibration using SPT-SZ mass information
is consistent with the relation derived from the number counts of the MARD-Y3 sample;
(2) we compare the scaling relation constraints from different flavours of number counts with each other (e.g., number counts in X-ray flux and redshift, in optical richness and redshift, 
and in both X-ray flux, optical richness and redshift); 
(3) finally, we constrain the probability of incompleteness in the SPT-SZ sample or contamination in the MARD-Y3 sample by comparing the clusters with and without counterparts in the other survey to the probabilities of having 
or not having counterparts as estimated using the selection functions.
We take advantage in these validation tests of the fact
that these scaling relations have been previously studied, and so we can compare our results not only internally but also externally to
the literature.  Finally, a key validation test could be carried out with the 
weak lensing information from DES, but we delay that to a future
analysis where we hope also to present unblinded cosmological results.

Given the stochastic description of the cluster population outlined above, we set-up different likelihood functions for each of
these tests. These likelihoods are functions of the parameters determining the mapping between intrinsic observables and mass, 
the scatter around these relations and the correlation coefficients among the 
different components of scatter. 
Consequently, sampling these likelihoods with the data constrains the parameters.
In the following sections we present the likelihoods used for each of the three 
validation methods listed above.

\subsection{SPT-SZ cross-calibration}\label{sec:xcalmethod}

For each object in the matched MARD-Y3 -- SPT-SZ sample (see section~\ref{sec:matched_sample}), we seek to predict the likelihood of observing the measured 
SZE signal-to-noise $\xi^{(i)}$ given the measured X-ray flux $\hat f_\text{X}^{(i)}$, measured richness $\hat 
\lambda^{(i)}$ and the scaling relation parameters. This likelihood is constructed by first making a prediction of the intrinsic SZE-signal-to-noise $\zeta$ that is 
consistent with the measured X-ray flux $\hat f_\text{X}^{(i)}$ and measured richness $\hat \lambda^{(i)}$, depending 
on the scaling relation parameters. To this end, the joint distribution of intrinsic properties is evaluated at the intrinsic 
fluxes and richnesses consistent with the measurements
\begin{equation}\label{eq:P_zeta_xopt}
\begin{split}
P(\zeta | \hat f_\text{X}^{(i)}, \hat \lambda^{(i)}, z^{(i)})  \propto \int \text{d}\lambda & P(\hat \lambda^{(i)} | \lambda)  
 \int  \text{d} f_\text{X} P(\hat f_\text{X}^{(i)}|f_\text{X}) \\
&  \frac{\text{d}^3 N}{\text{d}f_\text{X}\text{d}\lambda\text{d}\zeta}\Big|_{f_\text{X}, \lambda, \zeta, z^{(i)}} .
\end{split}
\end{equation}
This expression of expected intrinsic SZE-signal takes account of the Eddington bias induced by the observational 
and intrinsic scatter in the X-ray and optical observable acting in combination with the fractionally larger number of objects at low mass, 
encoded in the last term of the expression. 

To evaluate the likelihood of the measured SZE signal-to-noise $\xi^{(i)}$ given the measured X-ray flux $\hat 
f_\text{X}^{(i)}$ and measured richness $\hat \lambda^{(i)}$, we need to compare the predicted distribution 
$P(\zeta | \hat f_\text{X}^{(i)}, \hat \lambda^{(i)}, z^{(i)})$ with the likely values of intrinsic SZE-signal derived from the 
measurement $\xi^{(i)}$ and the measurement uncertainty.  This is written
\begin{equation}\label{eq:indiv_like_sptcc}
\begin{split}
P(\xi^{(i)}| \hat f_\text{X}^{(i)}, \hat \lambda^{(i)}, &\ z^{(i)}) = \\
& \frac{ \int \text{d}\zeta\, P(\xi^{(i)}|\zeta, \gamma_\text{f}^{(i)}) P(\zeta 
| \hat f_\text{X}^{(i)}, \hat \lambda^{(i)}, z^{(i)}) }{ \int \text{d}\zeta\, P(\text{SPT}|\zeta, \gamma_\text{f}^{(i)}) P(\zeta | \hat 
f_\text{X}^{(i)}, \hat \lambda^{(i)}, z^{(i)})  }.
\end{split}
\end{equation}
Notably, the denominator ensures the proper normalisation and also takes into account the Malmquist bias\footnote{In cluster population studies redshift information is usually available. Thus, the term "Malmquist bias" does not refer to the larger survey volume at which high flux objects can be detected, when compared to low flux objects. It refers to that fact that in the presence of scatter among observables, up-scattering objects are more likely to pass any selection criterion than down-scattered objects. This biases observable—observable plots close to the selection threshold.} introduced 
by the SPT-SZ selection. Also note that the normalization cancels the dependence of this likelihood on the amplitude of the number of 
objects at the redshift $z^{(i)}$, measured flux $\hat f_\text{X}^{(i)}$ and measured richness $\hat \lambda^{(i)}$. This strongly weakens its cosmological dependence and makes it independent of the X-ray and the optical selection 
function \citep[see also][]{liu15}. For sake of brevity we omitted that this likelihood depends on the scaling relation parameters 
and the cosmological parameters, all needed to compute the distribution of intrinsic properties. 

The total 
log-likelihood of SPT-SZ cross-calibration over the matched sample is given by the sum of the individual log likelihoods
\begin{equation}\label{eq:lnLsptcc}
\ln L_\text{SPTcc} = \sum_{i\in \text{matched}} \ln P(\xi^{(i)}| \hat f_\text{X}^{(i)}, \hat \lambda^{(i)}, z^{(i)}),
\end{equation}
which is a function of the scaling relation parameters and cosmology. Sampling it with priors on the SZE scaling 
relation parameters that come from an external calibration will then 
transfer that calibration to the X-ray flux and richness scaling relations.

\subsection{Calibration with number counts}\label{sec:numbercountsmethod}

The number of clusters as a function of measured observable and redshift is a powerful way to constrain the mapping 
between observable and mass, 
because the number of clusters as a function of mass is known for a given cosmology 
\citep[see self-calibration discussions in][]{majumdar03,hu03,majumdar04}. 

\subsubsection{X-ray number counts}\label{sec:XNC}

The 
likelihood of number counts is given by
\begin{equation}\label{eq:lnL_NC_X}
\ln L_{\text{nc X}} = \sum_i \ln N \big|_{\hat f_\text{X}^{(i)}, z^{(i)}} - N_\text{tot},
\end{equation}
where the expected number of objects as a function of measured flux $\hat f_\text{X}^{(i)}$ and redshift $z^{(i)}$ is 
\begin{equation}
\begin{split}
 N \big|_{\hat f_\text{X}^{(i)}, z^{(i)}} &= P(\text{RASS}|\hat f_\text{X}^{(i)}, z^{(i)}, t_\text{exp}^{(i)}) \int \text{d}f_\text{X} 
P(\hat f_\text{X}^{(i)}|f_\text{X})\\
 & \int \text{d}\lambda\,  P(\text{DES}|\lambda, z^{(i)}) \frac{\text{d}^2 N}{\text{d}f_\text{X}\text{d}\lambda}\Big|
_{f_\text{X}, \lambda, z^{(i)}} \text{d}\hat f_\text{X},
\end{split}
\end{equation}
where the first factor takes into account the X-ray selection, the second factor models the measurement uncertainty 
on the X-ray flux and the 
third factor models the optical incompleteness.

The total number of objects is computed as 
\begin{equation}\label{eq:ntot}
\begin{split}
N_\text{tot}= &\int\text{d}t_\text{exp} P(t_\text{exp}) \int \text{d}z \int \text{d}\hat f_\text{X} P(\text{RASS}|\hat  f_\text{X}, z, t_\text{exp}) \\
& \int \text{d}f_\text{X} P(\hat f_\text{X}|f_\text{X}, z, t_\text{exp})  \int \text{d}\lambda P(\text{DES}|\lambda, z) \\
&\frac{\text{d}^2 N}{\text{d}f_\text{X}\text{d}\lambda}\Big|_{f_\text{X}, \lambda, z}, 
\end{split}
\end{equation}
where $P(t_\text{exp})$ is the solid angle weighted exposure time distribution. We highlight here that, unlike previous work, we explicitly model not only the selection on the X-ray observable, but also fold in the 
incompleteness correction due to the MCMF optical cleaning via the term $P(\text{DES}|\lambda, z)$.

\subsubsection{Optical number counts}\label{sec:optNC}

While not customary for a predominantly X-ray selected sample, the number counts of clusters can also be represented as a function of optical richness.. In this case, the likelihood reads
\begin{equation}\label{eq:lnL_NC_opt}
\ln L_{\text{nc } \lambda} = \sum_i \ln N \big|_{\hat \lambda^{(i)}, z^{(i)}} - N_\text{tot},
\end{equation}
where $N_\text{tot}$ is given by equation(~\ref{eq:ntot}), whereas the expected number of clusters as a function of 
measured richness $\hat \lambda^{(i)}$ and redshift $z^{(i)}$ is computed as follows
\begin{equation}
\begin{split}
N \big|_{\hat \lambda^{(i)}, z^{(i)}} = & \int \text{d} t_\text{exp} P(t_\text{exp}) \int \text{d}\hat f_\text{X} P(\text{RASS}|
\hat f_\text{X}, z, t_\text{exp}) \\
& \int \text{d}f_\text{X} P(\hat f_\text{X}|f_\text{X}, z, t_\text{exp}) \\
&  \int \text{d}\lambda P(\hat \lambda^{(i)} |\lambda, z) 
\frac{\text{d}^2 N}{\text{d}f_\text{X}\text{d}\lambda}\Big|_{f_\text{X}, \lambda, z} \text{d}\hat \lambda, 
\end{split}
\end{equation}
where the first three integrals take account of the X-ray selection, 
while the last integral models the measurement uncertainty on the 
richness.

\subsubsection{Combined X-ray and optical number counts}

Besides determining the number counts in only one observable, one can also determine the number counts in more than 
one observable \citep[e.g.][]{mantz10}, in our case by fitting for the number of objects as a function of both measured flux $\hat f_\text{X}^{(i)}
$ and richness $\hat \lambda^{(i)}$. We call this flavour of number counts \text{2D number counts}, as opposed to the \text{1D number counts} in either X-ray flux (c.f. section~\ref{sec:XNC}) or richness (c.f. section~\ref{sec:optNC}).  The likelihood of 2D number counts reads
\begin{equation}
\ln L_{\text{nc X,}\lambda} = \sum_i \ln N \big|_{\hat f_\text{X}^{(i)}, \hat \lambda^{(i)}, z^{(i)}} - N_\text{tot},
\end{equation}
where the expected number of objects as a function of measured flux $\hat f_\text{X}^{(i)}$ and richness $\hat 
\lambda^{(i)}$ is
\begin{equation}\label{eq:lnL_NC_2d}
\begin{split}
 N \big|_{\hat f_\text{X}^{(i)}, \hat \lambda^{(i)}, z^{(i)}} = & P(\text{RASS}|\hat f_\text{X}^{(i)}, z^{(i)}, t_\text{exp}^{(i)}) \\
& \int \text{d}f_\text{X} P(\hat f_\text{X}^{(i)}|f_\text{X})\\
& \int \text{d}\lambda P(\hat \lambda^{(i)}|\lambda) \frac{\text{d}^2 N}{\text{d}f_\text{X}\text{d}\lambda}\Big|_{f_\text{X}, 
\lambda, z^{(i)}} \text{d}\hat f_\text{X} \text{d}\hat \lambda,
\end{split}
\end{equation}
computed by folding the intrinsic number density with the measurement uncertainties on flux and richness.

\subsection{Consistency check using two cluster samples}\label{sec:consistencymethod}

Given the selection functions for two cluster samples, 
the probability that any member of one sample is present in the
other can be calculated. Thus, two distinct tests can be set up: 1) for each object in the sample A, we can compute the probability of being detected by the sample B, and compare this probability to the actual occurrence of matches; 2) inversely, we can start from the sample B, compute the probabilities of detection by A, and compare that to the occurrence of matches. This provides a powerful consistency check of the two selections functions, and 
if anomalies are found, this approach can be used, for example, to probe for contamination or unexplained incompleteness in
the cluster samples.

\begin{figure}
    \resizebox {\columnwidth} {!} {
        \begin{tikzpicture}[scale=1, auto]
        
            \node [cloud, ] (c1) {SPT-SZ cluster $\xi^{(i)}$, $z^{(i)}$};
            \node [cloud, below of=c1, left of=c1] (c2) {should not be in MARD-Y3};
            \node [cloud, below of=c1, right of=c1] (c3) {should be in MARD-Y3};
            \node [cloud, below of=c2] (c4) {is not in MARD-Y3};
            \node [cloud, below of=c3] (c5) {is in MARD-Y3};

            \path [line] (c1) -- node [anchor=east, xshift=-0.3cm] {$1-p_\text{M|S}^{(i)}$} (c2);
            \path [line] (c1) -- node [anchor=west, xshift=0.1cm]{$p_\text{M|S}^{(i)}$} (c3);
            \path [line] (c2) -- node [anchor=east] {$1-\pi_\text{t}$}  (c4);
            \path [line] (c2) -- node [anchor=west, xshift=0.2cm] {$\pi_\text{t}$} (c5);
            \path [line] (c3) -- node [anchor=west] {$1$} (c5);
        
            \node [below of=c4, node distance=0.8cm] {$(1-p_\text{M|S}^{(i)})(1-\pi_\text{t})$};
            \node [below of=c5, node distance=0.8cm] {$(1-p_\text{M|S}^{(i)}) \pi_\text{t} + p_\text{M|S}^{(i)} $};
        \end{tikzpicture}
    }
    \vskip-0.1in
    \caption{Probability tree describing the probability of an SPT-SZ cluster being detected in MARD-Y3. 
    Besides the matching probability computed from the scaling between intrinsic observables and mass, the scatter around this relation, 
    the observational uncertainties on the observables and the selection functions $p_\text{M|S}^{(i)}$, 
    we also introduce the chance of either X-ray flux and richness boosting or SZE signal dimming $\pi_\text{t}$, 
    which would lead to the MARD-Y3 detection of SPT-SZ cluster that should otherwise not have been matched. 
    Summarized at the end of each branch are the probabilities of matching or of not matching.    \label{fig:mardydetspt}}
\end{figure}
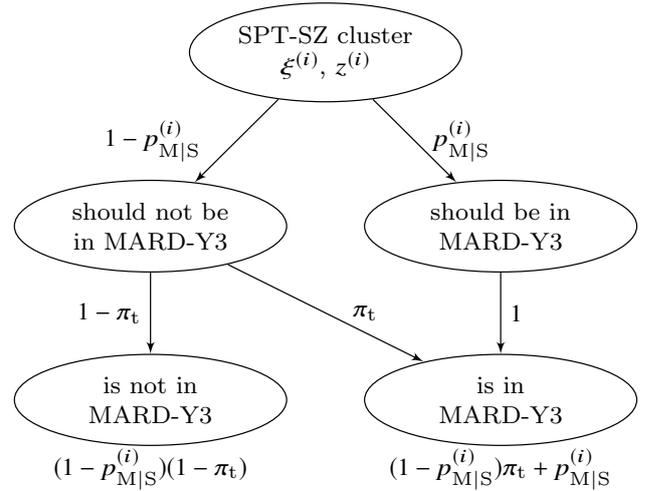

\subsubsection{MARD-Y3 detection probability for SPT-SZ clusters}\label{sec:mardy3detonspt}
For any SPT-SZ cluster with measured SZE signal-to-noise $\xi^{(i)}$ and redshift $z^{(i)}$ in the joint SPT--DES-Y3 
footprint we can compute the probability of being detected by MARD-Y3 as follows. We first predict the probability distribution of 
intrinsic fluxes and richnesses associated with the measured SZE-signal-to-noise as
\begin{equation}\label{eq:P_fx_lam_given_zeta}
\begin{split}
P(f_\text{X}, \lambda|\xi^{(i)}, z^{(i)}) \propto \int \text{d}\zeta\, & P(\xi^{(i)}|\zeta, \gamma_\text{f}^{(i)}) \\
& \frac{\text{d}^3 N}{\text{d}
f_\text{X}\text{d}\lambda\text{d}\zeta}\Big|_{f_\text{X}, \lambda, \zeta, z^{(i)}}.
\end{split}
\end{equation}
This expression needs to be properly normalised to be a distribution in intrinsic flux and richness. This is achieved by 
imposing $\int \text{d}f_\text{X}\int \text{d}\lambda \, P(f_\text{X}, \lambda|\xi^{(i)}, z^{(i)}) =1$, which sets the 
proportionality constant of the equation above. Note that this normalization cancels the dependence of this expression on 
the number of clusters observed.

The predicted distribution of intrinsic fluxes and richnesses needs to be folded with the selection functions to compute 
the detection probability. The optical selection function is simply given by equation~(\ref{eq:Pdet_opt}) evaluated at the 
cluster redshift $z^{(i)}$. On the other hand, when computing the X-ray selection function we take the RASS exposure 
time at the SPT-SZ position into account, while marginalising over all possible measured fluxes. The X-ray selection 
function thus reads
\begin{equation}
\begin{split}
P(\text{RASS}|f_\text{X}, t_\text{exp}^{(i)}, z^{(i)}) = \int \text{d}\hat f_\text{X} \,& P(\text{RASS}|\hat f_\text{X}, 
t_\text{exp}^{(i)}, z^{(i)}) \\
&P(\hat f_\text{X}|f_\text{X},  t_\text{exp}^{(i)}, z^{(i)})~,
\end{split}
\end{equation}
where the second factor is taken from equation~(\ref{eq:P_meas_X_arb}), the expression for the X-ray measurement error at 
arbitrary measured flux $\hat f_\text{X}$.

The probability of detecting in MARD-Y3 a SPT-SZ cluster with measured SZE-signal-to-noise $\xi^{(i)}$ and redshift 
$z^{(i)}$ can then be computed by folding the predicted distribution of fluxes and richnesses with the selection 
functions in flux and richness as follows
\begin{equation}\label{eq:p_mardy3_spt}
\begin{split}
p_\text{M|S}^{(i)} & := P(\text{RASS}, \text{DES}|\xi^{(i)}, z^{(i)}) \\ 
&=  \int \text{d}f_\text{X} P(\text{RASS}|f_\text{X}, t_\text{exp}^{(i)}, z^{(i)})  \\
& ~~~ \int \text{d}\lambda P(\text{DES}|\lambda, z^{(i)}) P(f_\text{X}, \lambda|\xi^{(i)}, z^{(i)})~,
\end{split}
\end{equation}
where we omit the dependence on the SPT-SZ field depth $\gamma_\text{f}^{(i)}$ 
at the position of the SPT-SZ selected cluster. 

Given these probabilities, we can define two interesting classes of objects: 
(1) \textit{unexpected MARD-Y3 confirmations of SPT-SZ detections}, i.e. SPT-SZ objects that should not have a MARD-Y3 match given their low probability but have been 
nonetheless matched , and (2) \textit{missed MARD-Y3 confirmations of SPT-SZ detections}, i.e SPT-SZ objects with a very high chance of being matched by MARD-Y3
that have nonetheless not been matched.  For the discussion in this paper we adopt a low probability threshold of
$p_\text{M|S}^{(i)}<0.025$ for the unexpected confirmations, and we adopt a high probability threshold of 
$p_\text{M|S}^{(i)}>0.975$ for the missed confirmations. 

Anticipating that we find a few unexpected MARD-Y3 confirmations and no missed confirmations, we introduce here the probability $\pi_\text{t}$ that an SPT-SZ cluster that should not be confirmed based on his $p_\text{M|S}^{(i)}$ is confirmed nonetheless. The likelihood of 
$\pi_\text{t}$ can be computed by following the probability tree shown in Fig~\ref{fig:mardydetspt}. 
The probability of being matched is $(1-p_\text{M|S}^{(i)}) \pi_\text{t} + p_\text{M|S}^{(i)}$, while the probability of not 
being matched is $(1-p_\text{M|S}^{(i)})(1-\pi_\text{t})$. Thus, the log-likelihood is given by
\begin{equation}\label{eq:lnL_ptail}
\begin{split}
\ln L (\pi_\text{t}) &= \sum_{i\in \text{match}} \ln \big((1-p_\text{M|S}^{(i)}) \pi_\text{t} + p_\text{M|S}^{(i)}\big) + \\
 & + \sum_{i \in \text{!match}}\ln \big( (1-p_\text{M|S}^{(i)})(1-\pi_\text{t}) \big)
\end{split}
\end{equation}
This likelihood also depends on the scaling relation parameters through the detection probabilities $p_\text{M|S}^{(i)}$. Marginalizing 
over these scaling relation parameters accounts for the systematic uncertainty on the observable--mass relations.

\begin{figure}
    \resizebox {\columnwidth} {!} {  
        \begin{tikzpicture}[scale=1, node distance = 1cm, auto]
        
            \node [cloud, ] (c1) {MARD-Y3 object $\hat f_\text{X}^{(i)}$, $\hat \lambda^{(i)}$, $z^{(i)}$};
            \node [cloud, below of=c1, left of=c1] (c1a) {is contaminant};
            \node [cloud, below of=c1, right of=c1] (c1b) {is cluster};
            \node [cloud, below of=c1a, left of=c1b] (c2) {should not be in SPT-SZ};
            \node [cloud, below of=c1a, right of=c1b] (c3) {should be in SPT-SZ};
            \node [cloud, below of=c2] (c4) {is not in SPT-SZ};
            \node [cloud, below of=c3] (c5) {is in SPT-SZ};
            
            \path [line] (c1) -- node [anchor=east, xshift=-0.3cm] {$\pi_\text{c}$} (c1a);
            \path [line] (c1) -- node [anchor=west, xshift=0.1cm] {$1-\pi_\text{c}$} (c1b);
            \path [line] (c1b) -- node [anchor=east, xshift=-0.3cm] {$1-p_\text{S|M}^{(i)}$} (c2);
            \path [line] (c1b) -- node [anchor=west, xshift=0.1cm]{$p_\text{S|M}^{(i)}$} (c3);
            \path [line] (c2) -- node [anchor=east] {$1$}  (c4);
            \path [line] (c3) -- node [anchor=west, xshift=0.2cm] {$\pi_\text{i}$} (c4);
            \path [line] (c3) -- node [anchor=west] {$1-\pi_\text{i}$} (c5);
            \path [line] (c1a) |- node [anchor=west, yshift=1.8cm, xshift=-0.38cm] {1} (c4);

            \node [below of=c4, node distance=0.8cm, xshift=-0.5cm] {$\pi_\text{c} + (1-\pi_\text{c})(1-p_\text{S|M}^{(i)} + 
\pi_\text{i} p_\text{S|M}^{(i)})$};
            \node [below of=c5, node distance=0.8cm] {$(1-\pi_\text{c})p^{(i)}_\text{S|M}(1-\pi_\text{i}) $};
        \end{tikzpicture}
    }
    \vskip-0.10in
    \caption{Probability tree describing the probability of a MARD-Y3 cluster being detected by SPT. 
    Besides the matching probability computed from the scaling between intrinsic observables and mass, the scatter around this relation, 
    the observational uncertainties on the observables and the selection functions $p_\text{S|M}^{(i)}$, 
    we also introduce the chance that a MARD-Y3 cluster is a contaminant $\pi_\text{c}$, 
    and the chance that SPT-SZ misses a cluster that it should detect, 
    indicating incompleteness in SPT, $\pi_\text{i}$. 
    Summarized at the end of each branch are the probabilities of being matched or not matched.    
    \label{fig:sptdetmardy}}

\end{figure}
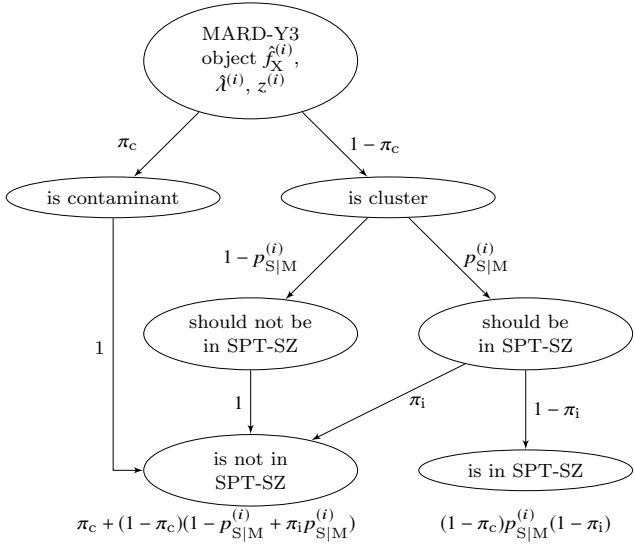

\subsubsection{SPT-SZ detection probability for MARD-Y3 clusters}\label{sec:sptdetonmardy3}

Similarly to the case in the previous section, for each MARD-Y3 cluster with measured X-ray flux $\hat f_\text{X}
^{(i)}$, measured richness $\hat \lambda^{(i)}$ and redshift $z^{(i)}$ in the joint SPT-SZ--DES-Y3 footprint, we can 
compute the probability of it being detected by SPT
\begin{equation}\label{eq:psptmardy3}
\begin{split}
p_\text{S|M}^{(i)} :&=  P(\text{SPT}|\hat f_\text{X}^{(i)}, \hat \lambda^{(i)}, z^{(i)}) \\
&= \int \text{d}\zeta P(\text{SPT}| \zeta, \gamma_\text{f}^{(i)}) P(\zeta | \hat f_\text{X}^{(i)}, \hat \lambda^{(i)}, z^{(i)}),
\end{split}
\end{equation}
where the first factor in the integrand is the SPT-SZ selection function evaluated for the field depth at the MARD-Y3 cluster 
position, while the second factor is the prediction for the intrinsic SZE-signal-to-noise consistent with the measured X-ray
 and optical properties. The latter is taken from equation~(\ref{eq:P_zeta_xopt}) while ensuring that it is properly normalized, 
$\int \text{d}\zeta P(\zeta | \hat f_\text{X}^{(i)}, \hat \lambda^{(i)})=1$.

We introduce the probability of each individual MARD-Y3 
cluster being a contaminant $\pi_\text{c}$, and the probability that a SPT-SZ cluster that should be detected has not 
been detected $\pi_\text{i}$. From the probability tree shown in Fig.~\ref{fig:sptdetmardy} we can 
determine the probability of a MARD-Y3 cluster being matched by SPT-SZ as $(1-\pi_\text{c})p^{(i)}_\text{S|M}(1-
\pi_\text{i})$, and the probability of a cluster not being matched as $\pi_\text{c} + (1-\pi_\text{c})(1-p_\text{S|M}^{(i)} + 
\pi_\text{i} p_\text{S|M}^{(i)})$. Thus, the log-likelihood is given by
\begin{equation}\label{eq:lnL_cont_inc}
\begin{split}
\ln L (\pi_\text{c}, \pi_\text{i}) = & \sum_{i\in \text{match}} \ln \left((1-\pi_\text{c})p^{(i)}_\text{S|M}(1-\pi_\text{i})\right) + \\
 & \sum_{i \in \text{!match}}\ln (\pi_\text{c} + (1-\pi_\text{c})  (1-p_\text{S|M}^{(i)} + \pi_\text{i} p_\text{S|M}^{(i)})).
\end{split}
\end{equation}
This likelihood also depends on the scaling relation parameter through the detection probabilities $p_\text{M|S}^{(i)}$. Marginalizing 
over the scaling relation parameters accounts for the the systematic uncertainties on the observable--mass relations.
Finally, note that the probability of MARD-Y3 contamination $\pi_\text{c}$ and of SPT incompleteness $\pi_\text{i}$ are perfectly degenerate in this context. We find that the likelihood of SPT confirmation of MARD-Y3 clusters (Eq.~\ref{eq:lnL_cont_inc}) effectively only constrains the difference between the two probabilities. That is $\pi_\text{c}=0.1$ and $\pi_\text{i}=0.0$ is approximately as likely as $\pi_\text{c}=0.0$ and $\pi_\text{i}=0.1$. 

\subsubsection{Physical Interpretation}

Several physical effects might bias cluster observables in an unusually significant level compared to the exception from the scatter in observables. In the case of the X-ray flux these effects are, for instance, AGN contamination and cluster core phenomena. Line of sight projections might bias the richness of an object, while extreme astrophysical contamination from correlated radio or dusty emission might bias the SZE signal. The object classes defined above (section~\ref{sec:mardy3detonspt}~and~\ref{sec:sptdetonmardy3}) allow one to select likely candidates for these effects from the comparison of two surveys. This is especially useful in the low signal-to-noise regime where the mass incompleteness of cluster samples in large. In this regime, physical effects within clusters are not resolved. Selecting target lists for high signal-to-noise follow-up might thus further our understanding on the mass incompleteness.

For instance, the classification as an unexpected MARD-Y3 confirmations of an SPT-SZ object can be due to an underestimated detection probability caused by an unexpectedly low SZE signal, or to the X-ray flux and richness being biased high, leading to an actual detection despite the low detection probability. It would thus be indicative of interesting physical properties such as extremely cool cluster cores, strong astrophysical contamination of the SZE signal or strong projection effects in the optical. The presence and impact of these effects would have to be studied with high resolution X-ray or (sub-)millimeter imaging, or spectroscopic follow-up of the cluster members, respectively. Also note that this class of objects in unlikely to be a MARD-Y3 contaminant, as we find an SPT-SZ object at the same position. Given that the SPT-SZ objects and the putative MARD-Y3 contaminants are both rare on the plane of the sky, the chance of randomly superposing two objects from these classes is small.

As another example, missed MARD-Y3 confirmations of SPT-SZ objects can be due to high SZE signals biasing the detection probability high, or to the X-ray flux and richness being biased low, leading to an nondetection despite the high detection probability. This circumstance is less likely to occur, as astrophysical SZE contaminants usually bias the SZE fluxes low, projection effects bias the richness high, and AGN contamination and cluster core emission bias the X-ray fluxes high. Nevertheless, such an object would be an interesting candidate for an SPT-SZ contaminant, or a case of excess incompleteness in the MARD-Y3 sample. Following the same logic, a missed SPT-SZ confirmation of MARD-Y3 object would indicate either the presence of physical effects that bias the X-ray flux and the richness high, astrophysical contamination that biases the SZE signal low, MARD-Y3 contamination or SPT-SZ excess incompleteness.

\section{Dataset and Priors}\label{sec:dataNpriors}

We present here the cluster samples and then the priors used in obtaining the results presented in the following section.

\subsection{Cluster samples}\label{sec:data}

Here we summarize not only the main properties of the prime focus of our validation, the MARD-Y3 cluster sample, but also the
SPT-SZ sample that we use for validation and for  cross-matching with MARD-Y3.

\begin{figure}
	\includegraphics[width=\columnwidth]{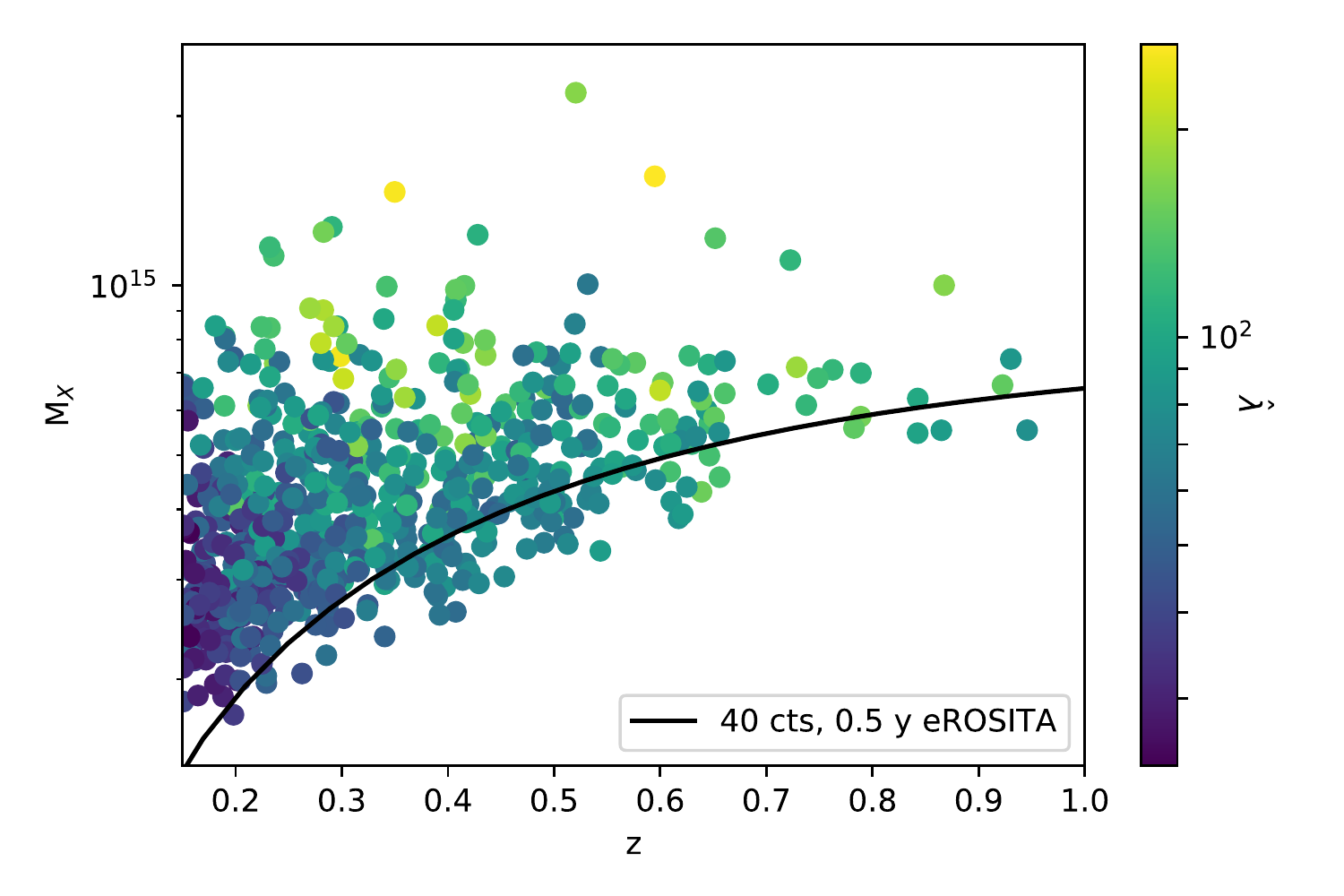}
	\vskip-0.10in
    \caption{MARD-Y3 sample of 708 clusters constructed by cleaning the 2nd RASS faint source catalog with DES data. 
    While not used in the rest of the analysis, the X-ray inferred mass $M_\text{X}$ is used here to highlight the mass range of our sample.  The color encodes the measured richness of the counterpart in the DES data. 
    The black line indicates the forecast of the estimated mass corresponding to 40 photon counts in the first eROSITA full sky survey after half a year of observing time.}
    \label{fig:sample}
\end{figure}

\subsubsection{MARD-Y3 X-ray selected clusters}\label{sec:mardysample}

In this work we seek to validate the mass information and the selection function modeling of the MARD-Y3 cluster sample, presented in \citetalias{klein19}. 
In that work, optical follow-up with the MCMF algorithm \citep[][]{klein18} of the RASS 2nd faint source 
catalog \citep[hereafter 2RXS,][]{boller16} is performed
by scanning the DES photometric data with a spatial filter centered on the X-ray candidate position and inferred mass, and a 
color filter based on the 
red-sequence model at a putative redshift.
This process provides a cluster richness estimate $\hat\lambda$ and photometric redshift $z$.
Comparison to the richness distribution in lines of sight without X-ray candidates allows one to 
estimate the probability $f_\text{cont}$ that the X-ray source and optical system identified by MCMF 
are a random superposition (contamination).
In cases of multiple richness peaks along a line of sight toward a 2RXS candidate, 
the redshift with lowest $f_\text{cont}$ is identified as the optical counterpart. 
The redshifts display sub-percent level scatter with respect to spectroscopic redshifts, and the richnesses $\hat\lambda$ 
can be adopted as an additional cluster mass proxy.

In this work we focus on the $z>0.15$ sample with $f_\text{cont}<0.05$ plus an additional rejection of luminosity--richness 
outliers with an infra-red signature compatible with an active galactic nucleus (c.f. 
\citetalias{klein19}, Section 3.11).  Our MARD-Y3
sample is then 708 clusters in a footprint of 5204 deg$^2$ with an expected 
contamination of 2.6\% \citepalias{klein19}.

For these clusters, several other X-ray properties, such as the detection significance 
$\xi_\text{X}$ and the RASS exposure time $t_\text{exp}$ 
are available from 2RXS. \citet{boller16} originally report luminosities in the [0.1,2.4] keV band extracted from a fixed aperture with radius of 5$'$. \citetalias{klein19} rescaled these luminosities  to luminosities in the rest frame [0.5,2] keV band and  within $R_{500c}$, the radius enclosing 
an over-density of 500 w.r.t. to the critical density. The rescaling is derived from the cross-matching with the MCXC catalog by \citet{piffaretti11}.
This correction is only reliable at $z>0.15$. Using the luminosity--mass scaling relation by \citetalias{bulbul19} and correcting for Eddington bias, \citetalias{klein19} gave an point estimate of the X-ray inferred mass $M_\text{X}$, that we use for plotting purposes.
The X-ray flux $\hat f_\text{X}$ we employ is computed 
as $\hat f_\text{X} = L_\text{X}/(4 \pi d_\text{L}^2(z))$, 
where $ L_\text{X}$ is the X-ray luminosity within $R_\text{500c}$, 
and $ d_\text{L}(z)$ is the luminosity distance evaluated at the reference cosmology.
This leads to the fact that technically our X-ray flux corresponds to the rest frame [0.5,2] keV. 
The transformation from the observed [0.1,2.4] keV band to this band is discussed in \citetalias{klein19}. 
It is also noteworthy that MCMF allows one to detect the presence of more than one significant optical structure along the line of 
sight toward an X-ray candidate.

In Figure~\ref{fig:sample} we show the redshift--X-ray inferred mass distribution of this 
sample, color coded to reflect the cluster richnesses.
We also show as a black line the mass corresponding to 40 photon counts in the first 
eROSITA full sky survey (eRASS1), computed using the 
eROSITA count rate--mass relation forecast by \citet{grandis19}.  This indicates that the MARD-Y3 sample we study here
is comparable to the one we expect to study in the eRASS1 survey.

\subsubsection{SPT-SZ SZE selected clusters}

We adopt the catalog of clusters selected via their SZE signatures in the SPT-SZ 2500~deg$^2$ survey \citet{bleem15}.
Utilising this sample to an SZE signal-to-noise of 4.5, we confirm the clusters in the DES-Y3 footprint using MCMF \citep{kleinip}.
The low contamination level of the parent sample allows one to achieve a low level of contamination 
by imposing the weak cut of $f_\text{cont}<0.2$.
Above a redshift of $z>0.2$ this provides us with a sample of 436 clusters.
The X-ray properties, as well as the optical properties of these objects have been extensively studied 
\citep[see for instance][and references therein]{mcdonald14, saro15, hennig17, chiu18, bulbul19, capasso19a}.
Furthermore, successful cosmological studies have been performed with the $\xi>5$ subsample \citep{bocquet15, dehaan16, bocquet19}, 
indicating that the survey selection function is well understood and that the mass information derived from the SZE is reliable.
This motivates us to employ this sample as a reference for our validation of the observable mass relations and the selection function of 
the MARD-Y3 sample.

\subsubsection{Cross-matched sample}\label{sec:matched_sample}

To identify clusters selected both by SPT-SZ and by MARD-Y3, we perform a positional matching within 
the angular scale of 2 Mpc at the MARD-Y3 cluster redshift.  We match 120 clusters in the redshift range $z\in(0.2, 1.1)$.
We identify 3 clusters where the redshift determined by the MCMF run on RASS, $z_\text{RASS}$, is significantly different from the 
redshift MCMF assigns for the SPT-SZ candidate, $z_\text{SPT}$. Specifically, for these objects $ |z_\text{RASS} -z_\text{SPT}| > 0.02 (z_\text{RASS} +z_\text{SPT})/2 $, which is equivalent to more then 3 sigma w.r.t. to the typical MCMF photometric redshift accuracy \citep{klein18, klein19}.
While for all three cases $z_\text{RASS}<z_\text{SPT}$, in all cases the MCMF run on the SPT-SZ candidate list 
identifies optical structures at $z_\text{RASS}$ as well.
Both their X-ray fluxes and SZE signals are likely biased w.r.t. to the nominal relation for individual clusters due to the presence of 
several structures along the line of sight.
Disentangling the respective contributions of the different structures along the line of sight is complicated by different scaling of 
X-ray flux and SZE signal with distance. We exclude these objects from the matched sample.

In only one case, two MARD-Y3 clusters are associated with the same SPT-SZ cluster: `SPT-CL J2358-6129', $z_\text{SPT}=0.403$. 
Visual inspection (c.f. Fig~\ref{fig:gal1}) reveals that one of the MARD-Y3 clusters, $z_\text{RASS}=0.398$, is well centered on the 
SZE signal, and also coincides with a peak in the galaxy density distribution. 
The second MARD-Y3 cluster in the north--northwest, $z_\text{RASS}=0.405$, is offset from the peaks in galaxy density, and does not 
correspond to any SZE signal. Given the lack of the SZ-counterpart, we do consider this MARD-Y3 cluster not being matched by SPT. 
We also identify a pair of SPT clusters (`SPT-CL J2331-5051', `SPT-CL J2332-5053') matched to the same MARD-Y3 cluster (2RXS-J233146.5-505227), shown in Fig.~\ref{fig:gal_doublespt}. Both SPT clusters are at redshift $\sim 0.57$, as is the MARD-Y3 clusters. The X-ray emission is blended into one source in the RASS image, but \textit{Chandra} follow-up by \citep{andersson11} clearly shows that $~95\%$ of the X-ray originates from `SPT-CL J2331-5051', which is also more significant in the SZe. We therefore take that to be the match.
Our final matched sample therefore contains 123 clusters.  

\begin{table}
	\centering
	\caption{Summary of the priors employed in this work. These priors are implemented as Gaussian probability distributions, where we present the mean $\mu$ and the standard deviation $\sigma$ as $\mu\pm\sigma$. 
}

	\label{tab:priors}
	\begin{tabular}{lcc} 
		\hline
		\multicolumn{3}{l}{Cosmological Parameters}\\
		\hline
		$H_\text{0}$ & 70.6$\pm$2.6  & \citet{rigault18}\\		
		$\Omega_\text{M}$ & 0.276$\pm$0.047 & SPT \citepalias{bocquet19}\\ 
		$S_8 = \sigma_8 \big( \Omega_\text{M}/0.3\big)^{0.2}$ &  $0.766\pm 0.025$ & SPT \citepalias{bocquet19}\\ 		\hline
		\multicolumn{3}{l}{SZE $\zeta$--mass Relation}\\
		\hline
 		$A_\text{SZ}$ & 5.24$\pm$0.85 & SPT \citepalias{bocquet19} \\
 		$B_\text{SZ}$ & 1.53$\pm$0.10 & \\
 		$C_\text{SZ}$ & 0.47$\pm$0.41 & \\
 		$\sigma_\text{SZ}$ & 0.16$\pm$0.08 & \\
 		\hline
		\multicolumn{3}{l}{X-ray $L_\mathrm{X}$--mass Relation}\\
		\hline
 		$A_\text{X}$ & 4.20$\pm$0.91 & SPT--XMM\citepalias{bulbul19} \\
 		$B_\text{X}$ & 1.89$\pm$0.18 & \\
 		$C_\text{X}$ & -0.20$\pm$0.50 & \\
 		$\sigma_\text{X}$ & 0.27$\pm$0.10 & \\
 		\hline
		\multicolumn{3}{l}{Optical $\lambda$--mass Relation}\\
		\hline
 		$A_\lambda$ & 71.9$\pm$6.1 & SPT-DES \citepalias{saro15} \\
 		$B_\lambda$ & 1.14$\pm$0.20 & \\
 		$C_\lambda$ & 0.73$\pm$0.76 & \\
 		$\sigma_\lambda$ & 0.15$\pm$0.08 & \\
 		\hline
	\end{tabular}
\end{table}

\subsection{Priors}\label{sec:priors}

In this section, we present the priors used in the likelihood analysis. We first discuss the cosmological priors assumed. Then we describe the priors on the SZE signal--mass relation, the X-ray luminosity--mass relation and the richness--mass relation. These priors are summarized in Table~\ref{tab:priors}. In the respective sub-sections, we describe in which analysis step the specific prior is used.

\subsubsection{Priors on cosmology}

Throughout this work, we marginalise over the following cosmological parameters to propagate our uncertainty on 
these parameters. The X-ray flux--mass relation has a distance dependence making it dependent on the present day expansion rate, also called the Hubble constant $H_\text{0}$. We therefore adopt the prior $H_\text{0}
= 70.6 \pm 2.6 $ km s$^{-1}$ Mpc$^{-1}$ from cepheid calibrated distance ladder measurements presented by 
\citet{rigault18}\footnote{Given the still unresolved controversy on the exact value of the Hubble constant, the value 
adopted here has the benefit of not being in significant tension with any other published result.}. 

For our number counts analysis in Section 5.2, we constrain scaling relation parameters by comparing the measured cluster number counts to a prediction based on our scaling relation model with assumed cosmological priors.  We assume priors
$\Omega_\text{M}=0.276\pm0.047$ and $S_8 = \sigma_8 \big( \Omega_\text{M}/0.3\big)^{0.2} = 0.766\pm 0.025$, derived by \citet[hereafter \citetalias{bocquet19}]{bocquet19} from the number counts analysis of 343 SZE selected 
galaxy clusters supplemented with gas mass measurements for 89 clusters and weak lensing shear profile 
measurement for 32 clusters. Note that the aforementioned $H_\text{0}$ prior is consistent with the constraints from \citetalias{bocquet19}.

\subsubsection{Priors on SZE $\zeta$-mass relation}

When performing the SPT-SZ cross-calibration (section~\ref{sec:SPTSZxCal}) we assume priors on the SZE scaling relation parameters to infer the X-ray 
flux--mass and richness--mass scaling relation parameters. These priors are derived from the X-ray and WL calibrated 
number counts of SPT-SZ selected clusters as described in \citetalias{bocquet19}. The adopted values are reported in 
Table~\ref{tab:priors}. These priors were derived simultaneously with the cosmological priors discussed above, and 
both rely on the assumption that the SPT-SZ selection function is well characterised and that the SZE-signal--mass 
relation is well described by equation~(\ref{eq:SZscaling}). These priors are also used when estimating the outlier fraction, the MARD-Y3 contamination and the  SPT incompleteness (section~\ref{sec:xmatch}). Note that \citetalias{bocquet19} only considered 
SPT-SZ clusters with SZE-signal-to-noise $\xi>5$ and $z>0.25$, while we adopt their results to characterize a sample 
with $\xi>4.5$ and $z>0.2$.  Considering that this is an extrapolation from typical masses of $\sim 3.6\, 10^{14}$ M$_\odot$ for $\xi=5$ to $\sim 3.3\, 10^{14}$ M$_\odot$ for $\xi=4.5$,  we view this as a minor change.

\subsubsection{Priors on X-ray $L_\mathrm{X}$-mass relation}

The X-ray luminosity--mass relation (c.f. Table~\ref{tab:priors}) used as comparison for the luminosity--mass relations we derive from the SPT-SZ cross-calibration (section~\ref{sec:SPTSZxCal}) and the number count fits (section~\ref{sec:numcounts}) has been determined 
by \citetalias{bulbul19}, who studied the X-ray luminosities of 
59 SPT-SZ selected clusters observed with XXM-Newton\footnote{We use the relation of type II for the core included luminosity within the [0.5,2] keV band.}. The authors 
then use priors on the SZE
-signal--mass relation to infer the luminosity--mass relation parameters. 
These measurements are also used as priors for the optical number counts (section~\ref{sec:XNC}) and when determining the systematic uncertainty on the outlier probability $
\pi_\text{t}$, the MARD-Y3 contamination $\pi_\text{c}$ and the SPT-SZ incompleteness $\pi_\text{i}$ (section~\ref{sec:xmatch}). 

\subsubsection{Priors on optical $\lambda$-mass scaling relation}

The richness--mass relation used as comparison for the richness--mass relations we derive from the SPT-SZ cross-calibration (section~\ref{sec:SPTSZxCal}) and the number count fits (section~\ref{sec:numcounts}) was
derived by \citetalias{saro15} from a sample of 25 SPT-SZ selected cluster, matched with DES redmapper 
selected clusters. In that work the SZE-signal--mass relation parameters were determined by fitting the SPT-SZ selected 
cluster number counts at fixed cosmology. The resulting constraints on the richness--mass relation are reported in 
Table~\ref{tab:priors}. These measurements are also used as prior for the X-ray number counts (section~\ref{sec:opt_nc}) and when determining the systematic uncertainty on the outlier probability $
\pi_\text{t}$, the MARD-Y3 contamination $\pi_\text{c}$ and the SPT-SZ incompleteness $\pi_\text{i}$ (section~\ref{sec:xmatch}).

\begin{figure*}
	\includegraphics[width=2\columnwidth]{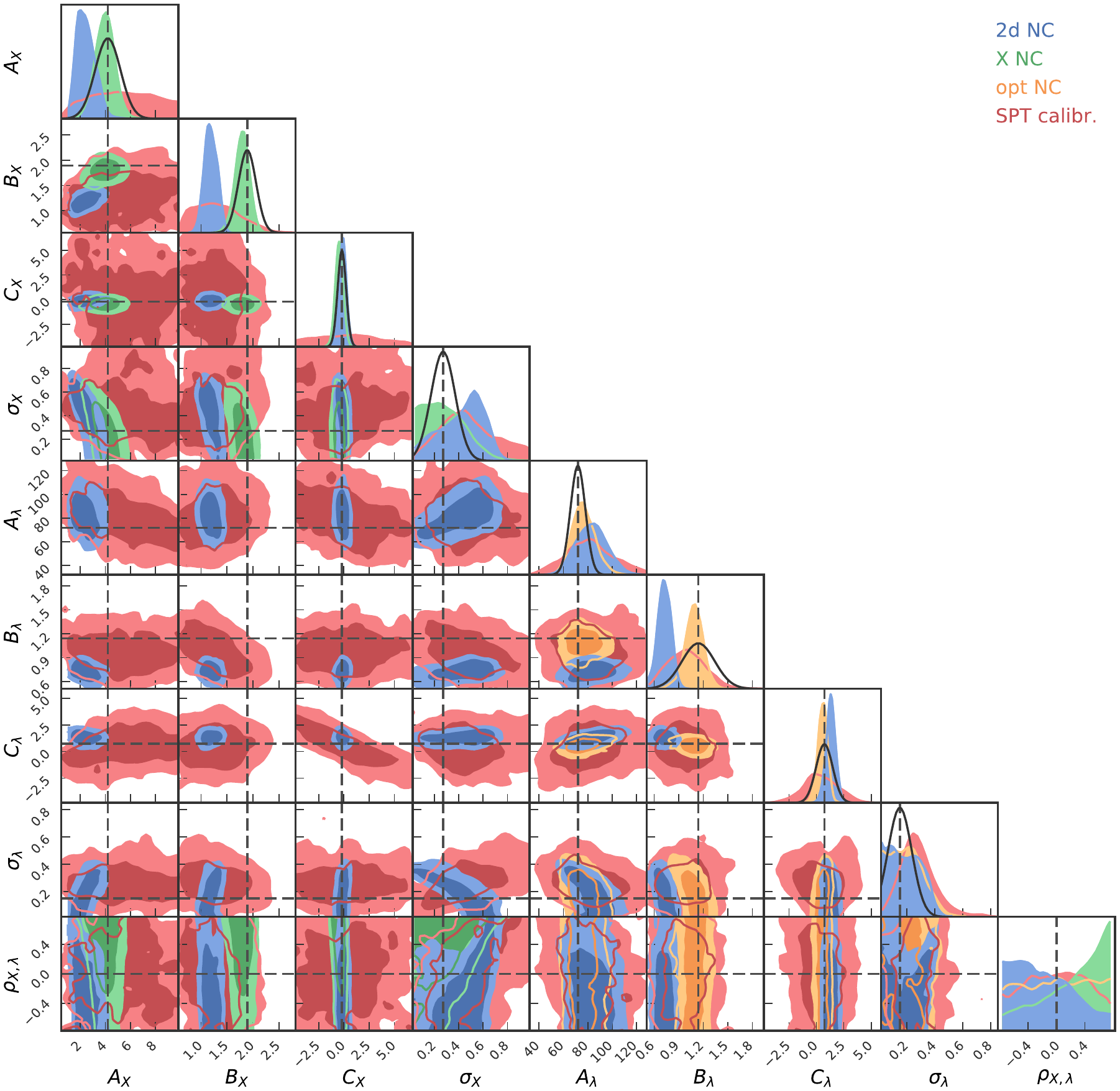}
	\vskip-0.10in
    \caption{Marginal posterior contours of the free parameters in the SPT-SZ cross-calibration (SPT calibr., red), the 
number counts in X-ray flux and redshift (X NC, green), the number counts in richness and redshift (opt NC, orange), 
and the number counts in X-ray flux, richness and redshift (2D NC, blue). In black the literature values from \citetalias{bulbul19} and 
\citetalias{saro15}. The SPT-SZ cross-calibration shows good agreement with the different number counts constraints and the literature. 
With the exception of the mass slope of the X-ray flux--mass relation inferred from 2D number counts, the 
constraints from the different number count experiments also show good agreement with the literature values. This provides strong evidence 
that our selection function modeling is adequate.}
\label{fig:SR_params_all}
    \label{fig:SRparams}
\end{figure*}

\begin{table*}
	\centering
	\caption{Mean and standard deviation estimated from the one dimensional marginal posterior plots for the parameters of the X-ray 
	scaling relation and the richness scaling relation.
	 Besides the constraints of the mass trend of the X-ray--mass relation and the corresponding intrinsic scatter, 
	 we find good agreement  among our different analysis methods and with the literature values. This provides strong evidence that 
	 our selection function modeling is adequate.}
	\label{tab:results}
	\begin{tabular}{lcccccccc} 
		\hline
		 & $A_\text{X}$ & $B_\text{X}$& $C_\text{X}$ & $\sigma_\text{X}$ & $A_\lambda$ & $B_\lambda$ & 
$C_\lambda$ & $\sigma_\lambda$\\
		\hline
		liter. & 4.20$\pm$0.91 & 1.89$\pm$0.18 & -0.20$\pm$0.50 & 0.27$\pm$0.10 & 71.9$\pm$6.1 & 1.14$
\pm$0.20 & 0.73$\pm$0.76 & 0.15$\pm$0.08 \\
		SPT calibr. & 5.42$\pm$2.48 & 1.31$\pm$0.43 & -- & 0.48$\pm$0.23 & 81.6$\pm$19.3 & 1.00$\pm$0.22 & 
0.39$\pm$1.55 & 0.28$\pm$0.13 \\
        X NC & 3.97$\pm$0.75 & 1.79$\pm$0.14 & -0.46$\pm$0.38 & 0.28$\pm$0.17 &  &  &  &  \\
        opt NC &  &  &  &  & 76.5$\pm$9.3 & 1.09$\pm$0.11 & 0.57$\pm$0.44 & 0.20$\pm$0.12 \\
        2D NC & 2.45$\pm$0.71 & 1.19$\pm$0.12 & -0.13$\pm$0.37 & 0.42$\pm$0.17 & 83.1$\pm$12.3 & 0.72$
\pm$0.08 & 1.31$\pm$0.43 & 0.19$\pm$0.11 \\
	\end{tabular}
\end{table*}

\begin{figure*}
	\includegraphics[width=2\columnwidth]{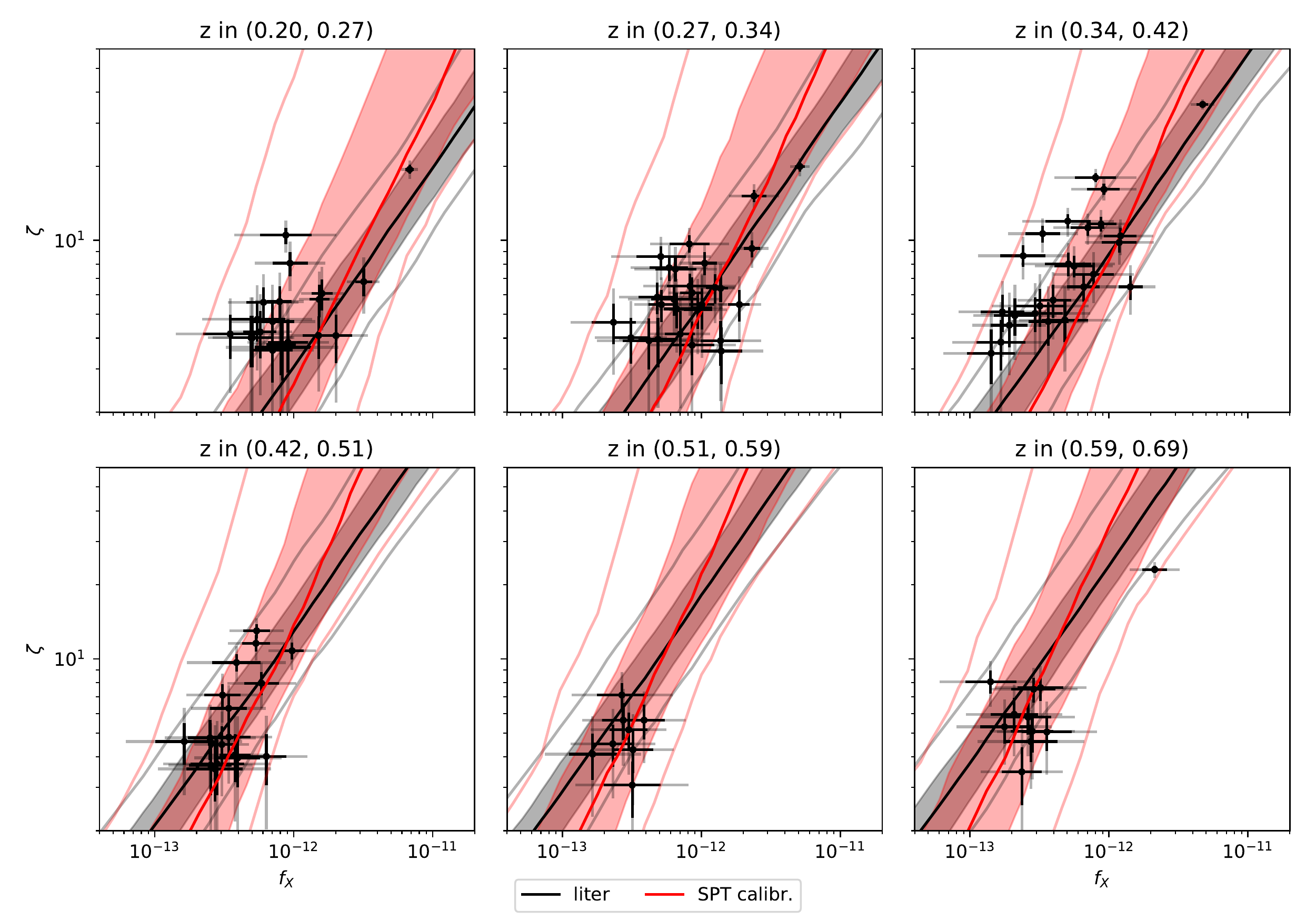}
	\vskip-0.10in
    \caption{Black points mark the intrinsic X-ray flux $f_\text{X}$ and SZE-signal to 
    noise $\zeta$ inferred from the 
    respective error models for our cross-matched sample.
     The X-ray flux--SZE-signal relation is shown either marginalized over the literature priors (black and grey) or over the posterior of 
     our cross-calibration to SPT-SZ (red).  The full lines are the median values, the filled region covers the range from the 16th to the 84th percentile and the transparent lines show the 2.5th and 97.5th percentile.
     While both sets of scaling relation parameters are consistent, we find a tendency for a weaker mass trend in the 
     X-ray observable than reported in the literature.
      }    
    \label{fig:SPT_flux_calibr}
\end{figure*}

\begin{figure*}
	\includegraphics[width=2\columnwidth]{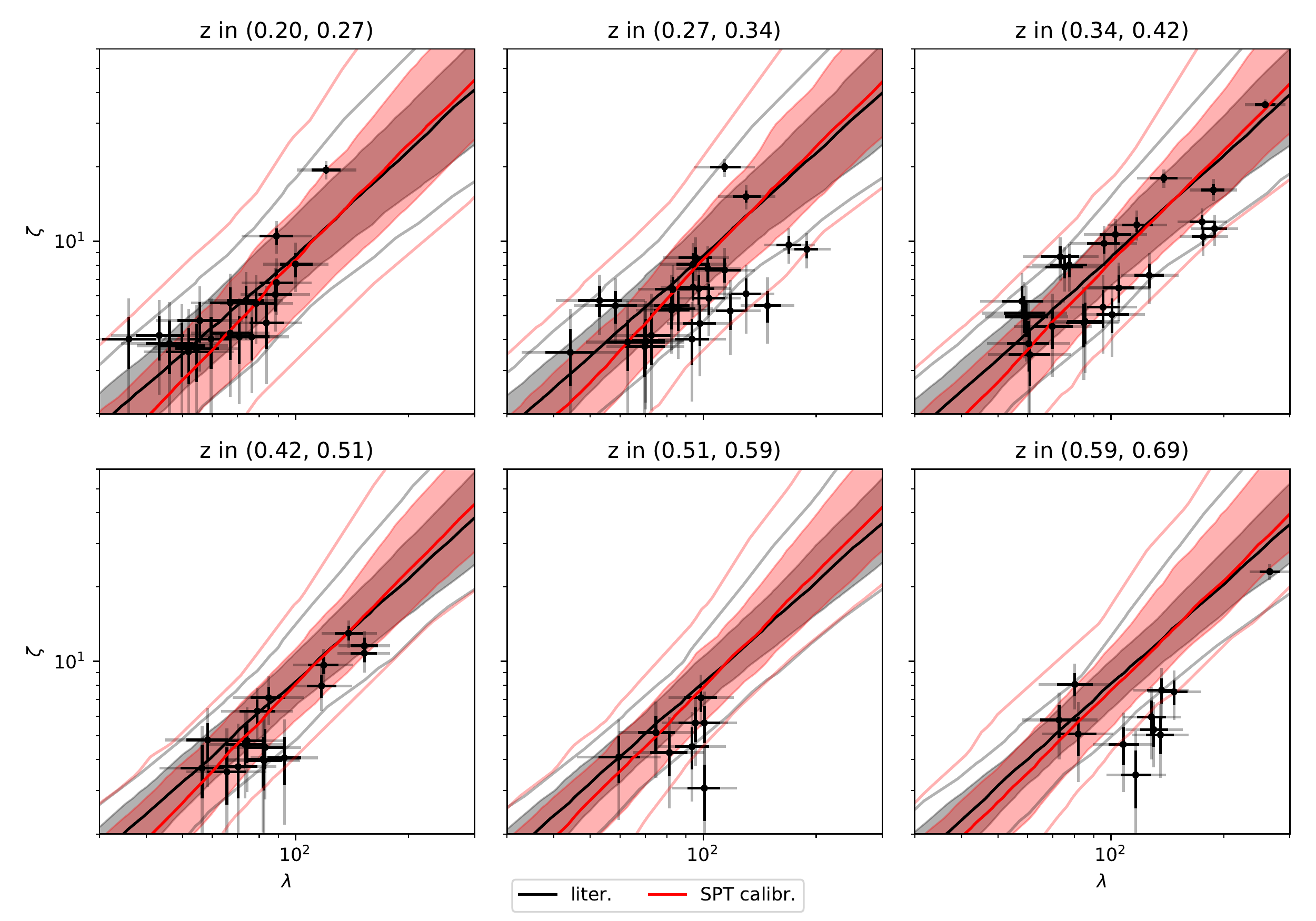}
	\vskip-0.10in
    \caption{Black points mark the the intrinsic richness $\lambda$ and SZE-signal-to-noise 
    $\zeta$ inferred from the respective error models for our cross-matched sample.
     The richness--SZE-signal relation is shown either marginalized over the literature priors (black) or over the posterior of our 
     cross-calibration to SPT-SZ (red). The full lines are the median values, the filled region covers the range from the 16th to the 84th percentile and the transparent lines show the 2.5th and 97.5th percentile.
     Both sets of scaling relation parameters are consistent, at 0.51 < z < 0.69, however, they fail to describe a part of the population with low SZE-signal and high richness. 
    }\label{fig:SPT_lambda_calibr}
    \label{fig:SPT_lam_calibr}
\end{figure*}

\section{Application to MARD-Y3 and SPT-SZ}
\label{sec:results}

In this section we present the results of validation tests on the MARD-Y3 sample by way of
examining the consistency of the X-ray--mass and the richness--mass scaling relations derived using different methods. 
First, we present the cross-calibration of the fluxes and richnesses using the externally calibrated SPT-SZ sample (Section~\ref{sec:SPTSZxCal}).  Then in Section~\ref{sec:numcounts}, 
we derive the parameters of the X-ray--mass scaling relation from the X-ray number 
counts, the parameters of the richness--mass scaling relation from the optical number counts, and 
then explore the constraints on both scaling relations from a joint 2-dimensional X-ray and optical number counts analysis.
We explore the implied cluster masses in Section~\ref{sec:mass_prediction}, 
and in Section~\ref{sec:xmatch} we validate our selection functions by computing the probabilities of each cluster in one
sample having a counterpart in the other and comparing these probabilities to the actual set of matched pairs and unmatched
single clusters in each sample. This last exercise allows us to study outliers in observables
beyond the measured scaling relation and observational scatter and has implications for 
the incompleteness in the SPT-SZ sample and the contamination in the
MARD-Y3 sample.

\subsection{Validation using SPT-SZ cross-calibration}\label{sec:SPTSZxCal}

As implied in the methods discussion in Section~\ref{sec:xcalmethod}, 
the results of the SPT-SZ cross-calibration of the MARD-Y3 mass indicators X-ray flux and richness are extracted by 
sampling the likelihood in equation~(\ref{eq:lnLsptcc}). The free parameters of this fit are the parameters of the 
X-ray scaling relation ($A_\text{X}$, $B_\text{X}$, $C_\text{X}$, $\sigma_\text{X}$), of the richness scaling relation 
($A_\lambda$, $B_\lambda$, $C_\lambda$, $\sigma_\lambda$) and the correlation coefficients between the intrinsic 
scatters ($\rho_{\text{X,} \lambda}$, $\rho_{\text{X,SZ}}$, $\rho_{\lambda\text{,SZ}}$). We put priors on the 
parameters of the SZE-signal--mass relation ($A_\text{SZ}$, $B_\text{SZ}$, $C_\text{SZ}$, $\sigma_\text{SZ}$) and 
on the cosmological parameters ($H_\text{0}$, $\Omega_\text{M}$, $\sigma_8$), as described in 
Section~\ref{sec:priors}.

The resulting marginal posterior contours on the parameters without priors are shown in red (SPT calibr.) in 
Fig.~\ref{fig:SR_params_all} and in Table~\ref{tab:results}. 
The same figure also shows as a black line the literature values for these parameters, where we use
\citetalias{bulbul19} for the X-ray parameters, and \citetalias{saro15} for the optical parameters. 
Our constraints are in agreement with these works, but 
display comparable or larger uncertainties despite the larger number of objects. 
This is due to different effects.  

The difference between the sizes of the uncertainties on the richness--mass relation in this work and in
 \citetalias{saro15} are mainly due to the tighter priors on the SZE-signal--mass relation parameters utilized by  \citetalias{saro15}. For instance, in \citetalias{saro15}  the prior on the amplitude of the SZE-signal--mass relation is four times smaller than the one used in this work. That being said, we here analyse a 4 times larger sample, which warrants at best an improvement of the constraints by a factor of 2. Our larger uncertainties on the richness--mass relation parameters are thus reflecting our more conservative treatment of systematic uncertainties on the SZE inferred masses.

This does not, however, explain why our constraints on the luminosity--mass relation are weaker than those reported by \citetalias{bulbul19}, as that work used priors on the SZE-signal--mass relation comparable to ours. Two different effects play a role in this case. (1) The measurement uncertainty on the luminosities extracted from pointed XMM observations is much smaller than on RASS based luminosities. (2)~Marginalizing over the systematic uncertainty on the matter density $\Omega_\text{M}$ and the Hubble parameter $H_\text{0}$ leads via the cosmological dependence of the luminosity distance and $E(z)$ to a systematic uncertainty $\delta C_\text{X}\sim0.37$. This source of uncertainty is not considered in \citetalias{bulbul19}. In summary, our data is considerably less constraining then the XMM measurements, which themselves were analysied ignoring an important systematic uncertainty.

In Fig.~\ref{fig:SPT_flux_calibr} are plotted in different redshift 
bins the scaling relation between the intrinsic X-ray flux inferred from the X-ray flux error model 
(equation~\ref{eq:P_X_meas_i}) and the 
intrinsic SZE signal-to-noise inferred from the SZE error model (equation~\ref{eq:P_SZ_meas}), 
as black points with 1 and 
2 sigma uncertainties. We also plot the predicted X-ray flux--SZE-signal relation obtained by combining the respective 
scaling relations. We show (black and grey) their marginalization over the \citetalias{bocquet19} cosmological parameter and 
SZE-scaling parameter priors, the \citetalias{bulbul19} X-ray-scaling parameter priors, 
and over the posterior of the SPT-SZ cross-calibration (red). 
As already noted from the contour plots of the marginal posteriors, our inferred scaling relation parameters 
are statistically consistent with the literature. Our calibration, however, prefers a steeper 
relation, which manifests also in the
lower inferred value on the X-ray mass trend $B_\text{X}$.

The results for the SPT-SZ cross-calibration of the richness--mass relation are shown in Fig.~\ref{fig:SPT_lam_calibr}. In 
different redshift bins we plot as black point the intrinsic richness $\lambda$ and 
the intrinsic SZE-signal $\zeta$ 
inferred from the respective error models (equations~\ref{eq:Pmeasopt} and \ref{eq:P_SZ_meas}). 
We also plot the 
richness-SZE scaling derived from combining the richness--mass and the SZE-signal--mass 
relation. The resulting 
relation is shown with the uncertainties derived from the literature priors and the cross-calibration 
posteriors. The two 
constraints are in very good agreement. Yet, at high redshift $z>0.5$, we note the presence of a high richness, 
low SZE-signal 
population, not well described by either the relation in the literature or our cross-calibrated relation.
These objects will be discussed in more detail in Section~\ref{sec:mardy_on_spt}.

\begin{figure*}
	\includegraphics[width=2\columnwidth]{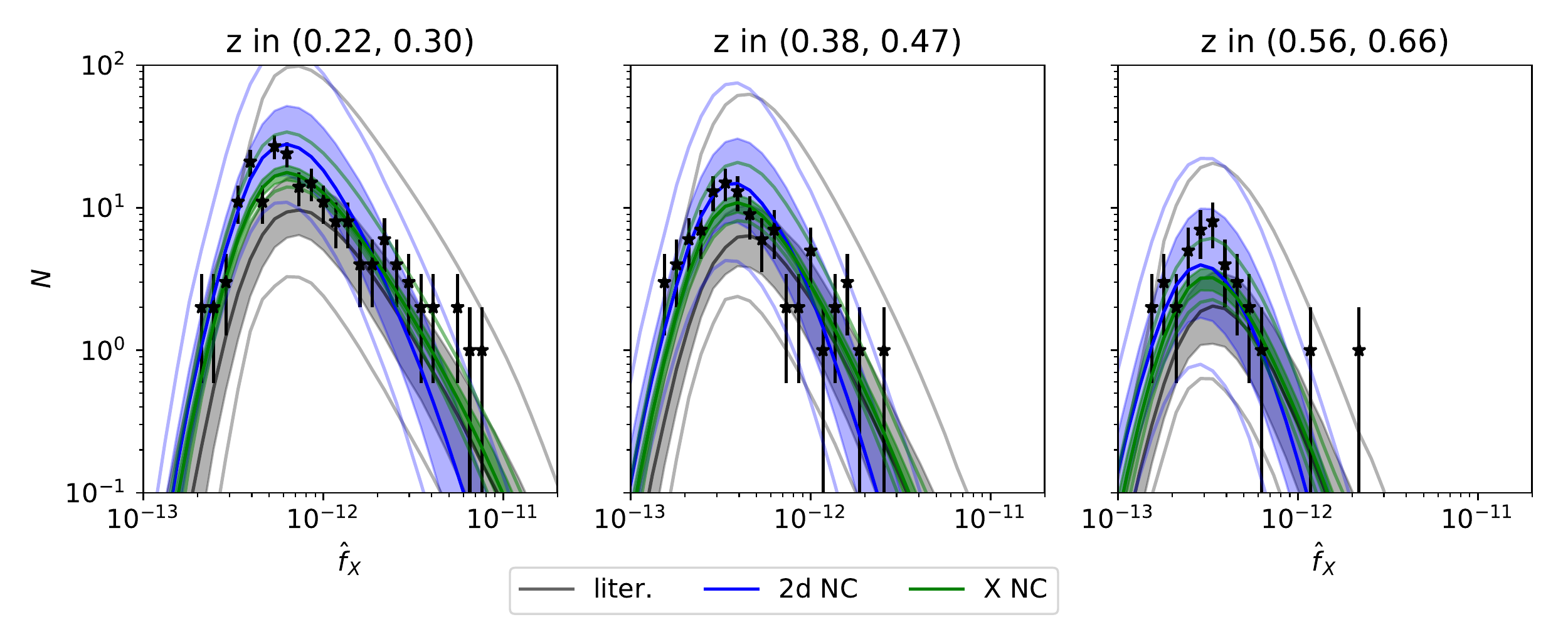}
	\vskip-0.10in
    \caption{Measured number of clusters in bins of measured X-ray flux for different redshift bins as black points with Poissonian error bars. 
    We over-plot the prediction for the number of objects with the 
    uncertainties derived from the literature values (black and gray), from our 1 d fit (green), and from our 2D fit (blue).  The full lines are the median values, the filled region covers the range from the 16th to the 84th percentile and the transparent lines show the 2.5th and 97.5th percentile.
    The latter captures adequately both the increasing rarity of high redshift objects, 
    as well as the effect of X-ray incompleteness at low flux.
    The measurement is also consistent with the literature values, although as discussed in Section~\ref{sec:XNC}, our assumption that
    cosmological and scaling relation parameters are 
    uncorrelated leads to an over-estimation of the uncertainty.}
    \label{fig:1d_NC_X}
\end{figure*}


\subsection{Validation using number counts}\label{sec:numcounts}

As described in the method discussion in Section~\ref{sec:numbercountsmethod}, 
we perform three different number counts experiments in this work: 
(1) we infer the X-ray flux--mass relation by fitting for the number counts of cluster as a function of measured flux and redshift; 
(2) we constrain the richness--mass relation by fitting for the number counts as a function of measured richness and redshift; and 
(3) we determine both relations by fitting the number of objects as a function measured flux, measured richness and redshift. 

\subsubsection{X-ray number counts}\label{sec:XNC}

While sampling the likelihood of the number counts in X-ray flux (equation~\ref{eq:lnL_NC_X}), we let the parameters of the 
X-ray flux--mass relation ($A_\text{X}$, $B_\text{X}$, $C_\text{X}$, $\sigma_\text{X}$) float within wide, flat priors. We 
adopt priors on the relevant cosmological parameters ($H_\text{0}$, $\Omega_\text{M}$, $
\sigma_8$) as described in Table~\ref{tab:priors}. 
We also put priors on the richness-mass relation parameter ($A_\lambda$, $B_\lambda$, $C_\lambda$, $
\sigma_\lambda$). Furthermore, we empirically constrain the relation between X-ray detection significance $
\xi_\text{X}$, measured flux $\hat f_\text{X}$ and exposure time $t_\text{exp}$ from the sample. As described in more 
detail in Sections~\ref{sec:X_selection} and \ref{sec:X_selection_disc}, 
this results in four tightly constrained nuisance parameters that impact the 
X-ray selection function. The resulting posteriors on the X-ray scaling relation parameters are shown in green in 
Fig.~\ref{fig:SRparams}. We find tight agreement with the literature values, at comparable accuracy on the marginal 
uncertainties.

\begin{figure*}
	\includegraphics[width=2\columnwidth]{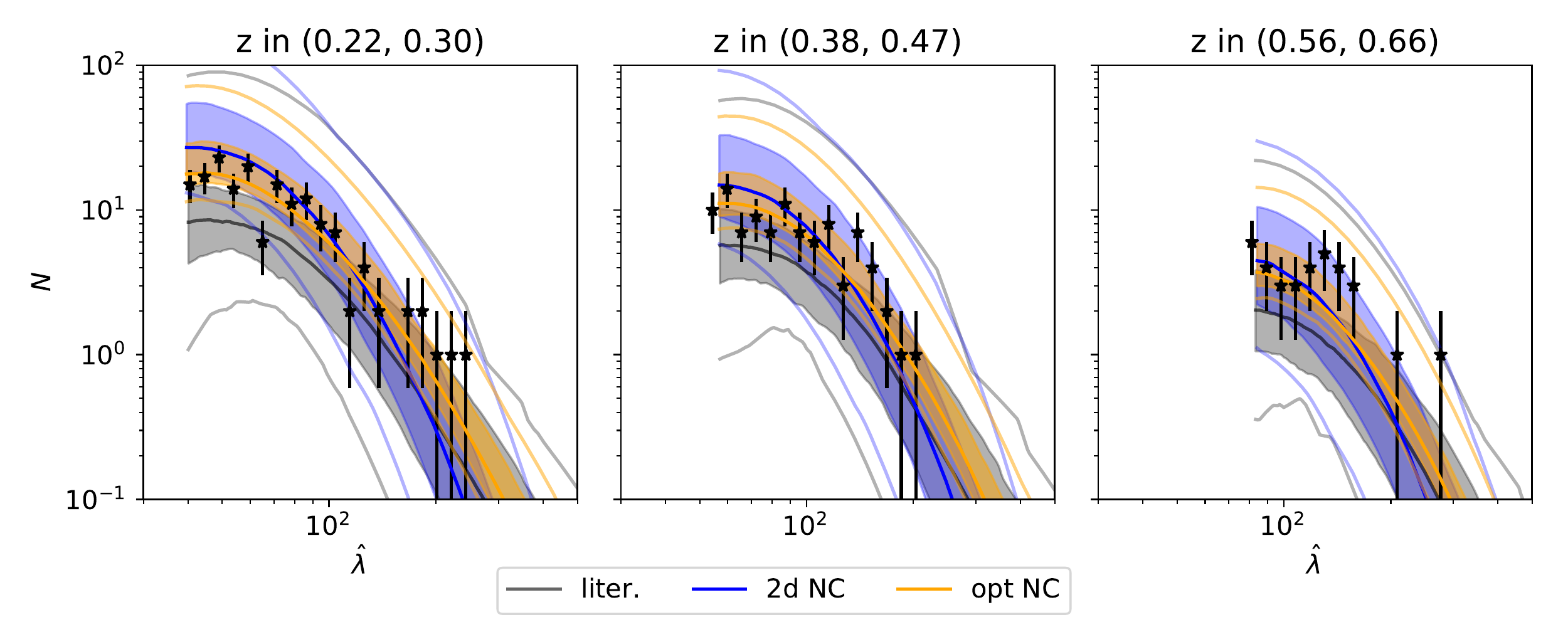}
	\vskip-0.10in
    \caption{Number of objects in bins of measured richness $\hat \lambda$ in different redshift bins as black points 
    with Poissonian error bars. 
    Over-plotted is the expected number of objects as a function of measured richness for 
    the same bins marginalized over the uncertainties from the literature (black and grey), from our 1D fit (orange), and from our 2D fit (blue). The full lines are the median values, the filled region covers the range from the 16th to the 84th percentile and the transparent lines show the 2.5th and 97.5th percentile.
    The shape of the abundance at low richness would in principle be closer to a power-law, 
    but at low richness the X-ray selection of the samples leads to a decrease in the number of objects, 
    which is well fit by our selection model.}
    \label{fig:1d_NC_opt}
\end{figure*}

In Fig.~\ref{fig:1d_NC_X} we plot the number counts in measured X-ray flux bins in three different redshift bins with 
the respective Poissonian errors. We also plot the prediction for the number of objects in the same bins, once 
marginalized over the literature values (black and grey), over our 1D fit (green), and our 2D number counts fit (blue, c.f. section~\ref{sec:2dnc}). The 1D fit provides an accurate fit to the data, with the exception regime where X-ray incompleteness sets in, where it tends to slightly underestimate the number of clusters.
The prediction from the literature provides a statistically consistent description of the data, albeit systematically more than 1, and less than 2 sigma low at low mass. These trends, while not statistically significant, are confirmed by inspecting the inferred masses from our posterior (see below section~\ref{sec:mass_prediction}).

\subsubsection{Optical number counts}\label{sec:opt_nc}

Just as the number counts as a function of measured flux can be used to infer the X-ray scaling relation parameters, 
the number counts in richness can be used to infer the richness--mass relation parameters. 
To this end, we sample the likelihood of number counts in richness bins (equation~\ref{eq:lnL_NC_opt}). 
We let the parameters of the richness--mass relation ($A_\lambda$, $B_\lambda$, $C_\lambda$, $\sigma_\lambda$) free, 
while we adopt priors from the literature on the cosmological parameters ($H_\text{0}$, $\Omega_\text{M}$, $\sigma_8$). 
Importantly, modeling the X-ray incompleteness in the space on measured richness requires a way to 
transform from measured richness to X-ray flux. 
Thus, while the transformation from richness to mass is fit, we need to assume a transformation from mass to X-ray flux. 
This is done by putting priors on the X-ray scaling relation parameters ($A_\text{X}$, $B_\text{X}$, $C_\text{X}$, $\sigma_\text{X}$).
As for the X-ray number counts, we empirically constrain the relation between X-ray detection significance $\xi_\text{X}$, 
measured flux $\hat f_\text{X}$ and exposure time $t_\text{exp}$ from the sample and predict the X-ray selection function on the fly.

The resulting marginal posterior contours are shown in Fig.~\ref{fig:SRparams} in orange. 
We find good agreement with the literature values and with the SPT-SZ cross-calibration. 
The marginal uncertainties are comparable to the literature values, despite being marginalized over cosmological parameters. 
We also find that the constraints from the number counts are more stringent than those derived from the SPT-SZ cross-calibration.

One can visually assess the quality of the resulting fit in Fig.~\ref{fig:1d_NC_opt}, 
where we plot the number of objects in measured richness 
for different redshift bins as black points with Poissonian error bars. 
We also plot the predicted number of objects with the uncertainties derived from the literature priors (black and grey) from our 1D  fit (orange), and our 2D number counts fit (blue, c.f. section~\ref{sec:2dnc}). 
Also in this case we note that the literature prediction is systematically between 1 and 2 sigma low, which manifests also in different mass estimates (see below in  section~\ref{sec:mass_prediction}).

\begin{figure*}
	\includegraphics[width=2\columnwidth]{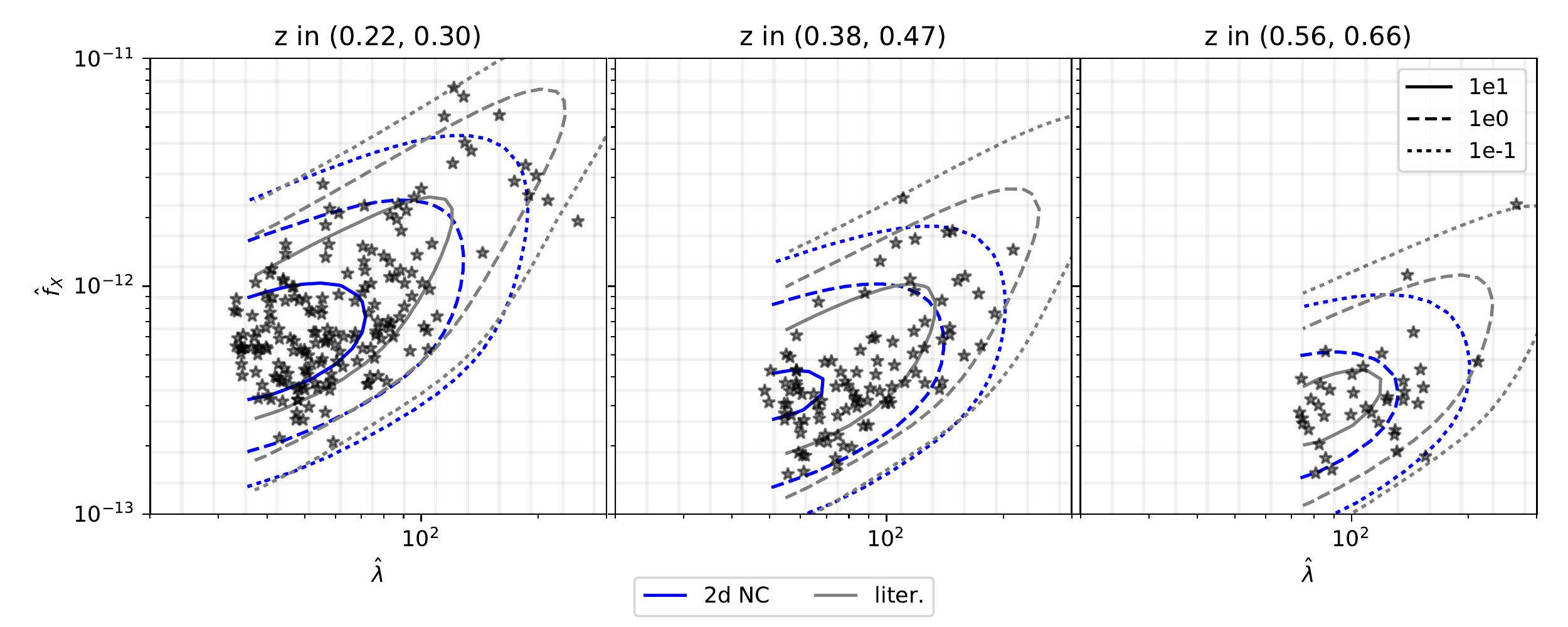}
	\vskip-0.10in
    \caption{Number counts in both measured richness $\hat \lambda$ and measured X-ray flux $\hat f_\text{X}$ visualized by presenting in different redshift bins the distribution of our sample (points), 
    the contours of the 2-dimensional predicted number of objects for the literature values (grey, liter.) 
    and for the maximum likelihood point of our fit to the data (blue, 2D NC). The number of objects shown in the contours (10, 1, 0.1 objects contours as full, dashed and dotted lines respectively) refers to the bins shown in the overlaid grid.
    Our 2D fit prefers larger scatter and provides a better description of the data than does the literature prediction. }
    \label{fig:2d_NC}
\end{figure*}

\subsubsection{Combined X-ray and optical number counts}\label{sec:2dnc}

We also fit for the abundance of clusters as a function of measured X-ray flux $\hat f_\text{X}$, 
measured richness $\hat \lambda$ and redshift, which we will refer to a `2D number counts', 
by sampling the likelihood in equation~(\ref{eq:lnL_NC_2d}). 
We allow the parameters of both the X-ray scaling relation ($A_\text{X}$, $B_\text{X}$, $C_\text{X}$, $\sigma_\text{X}$) 
and the richness scaling relation ($A_\lambda$, $B_\lambda$, $C_\lambda$, $\sigma_\lambda$) float within wide, flat priors. 
We adopt priors on the cosmological parameters from Table~\ref{tab:priors}. 
Furthermore, we empirically constrain the relation between X-ray detection significance $\xi_\text{X}$, 
measured flux $\hat f_\text{X}$ and exposure time $t_\text{exp}$ from the MARD-Y3 sample 
and predict the X-ray selection function on the fly (c.f. Sections~\ref{sec:X_selection} and \ref{sec:X_selection_disc}).

In Fig.~\ref{fig:SRparams} we show the marginal posterior contours on the scaling relation
parameters in blue. We find good agreement with the results from the SPT-SZ cross-calibration on 
all parameters. When comparing the constraints from 2D number counts (blue) on the X-ray 
scaling relation parameters to the constraints from the number counts in X-ray flux (green), we 
find good agreement on the values of the amplitude and redshift evolution. However, we find a 
shallower X-ray observable mass trend than from the X-ray number counts, and 
we see a similar shift in the optical mass trend parameter, although in this case the statistical significance is small.
Given the agreement of 
the X-ray number counts result is with \citetalias{bulbul19}, 
the results from the 2D number counts are in some tension with both. As show in Section~\ref{sec:mass_prediction} below, these constraints however do not results in statistically inconsistent mass estimates. Nevertheless, possible systematic effects impacting our validation tests are discussed in section~\ref{sec:X_selection_disc}~and~\ref{sec:outcome_val}.

Of interest is also the constraint the 2D number counts put on the two intrinsic scatters in X-ray flux and richness. Inspecting their joint marginal posterior in Fig.~\ref{fig:SRparams} reveals a distinct degeneracy in the form of an arc. This is the natural result of the fact that the 2D number counts can only constraint the total scatter between the two observables, but not the two individual scatters between each observable and mass. The total scatter between observables, being the squared sum of the individual scatter, sets the radius of the arc. Noticeably, this arc-like degeneracy excludes the possibility that both the X-ray and the richness scatter are small.

For visual inspection of the 2D number counts fit in Fig.~\ref{fig:2d_NC} we present the  
distribution in measured X-ray flux and measured richness of our sample in different redshift 
bins as black stars. We also plot the contours of the predicted number of objects in 
equally spaced logarithmic bins (shown by the overlaid grid): in blue the prediction for the best fit value of the 2D number counts, 
while in grey the prediction from the literature. The selection in richness due to the $f_\mathrm{cont}<0.05$ cut
is at every redshift a sharp cut in measured richness, as can be seen up to the intra bin scatter due to the large bins used for 
plotting. The effect of the X-ray selection function is harder to see, but can be appreciated 
in the shape of the contours at low flux: they show a bend, predicting very small numbers 
of objects at the lowest fluxes. Notably, the distribution of the data displays a large 
dispersion, which is better captured by our fit (blue) than by the prediction from the literature (grey). 
This confirms that the measurement of a larger X-ray scatter is indeed a feature of the 
data visible in the 2 dimensional cluster abundance. Despite the larger intrinsic scatter, 2D number counts posterior provide also a prediction of the X-ray and optical 1D number counts that is consistent with the data within the systematic and statistical uncertainties, as can be see by the blue predictions in Fig~\ref{fig:1d_NC_X}~and~\ref{fig:1d_NC_opt}.

\begin{figure*}
	\includegraphics[width=\columnwidth]{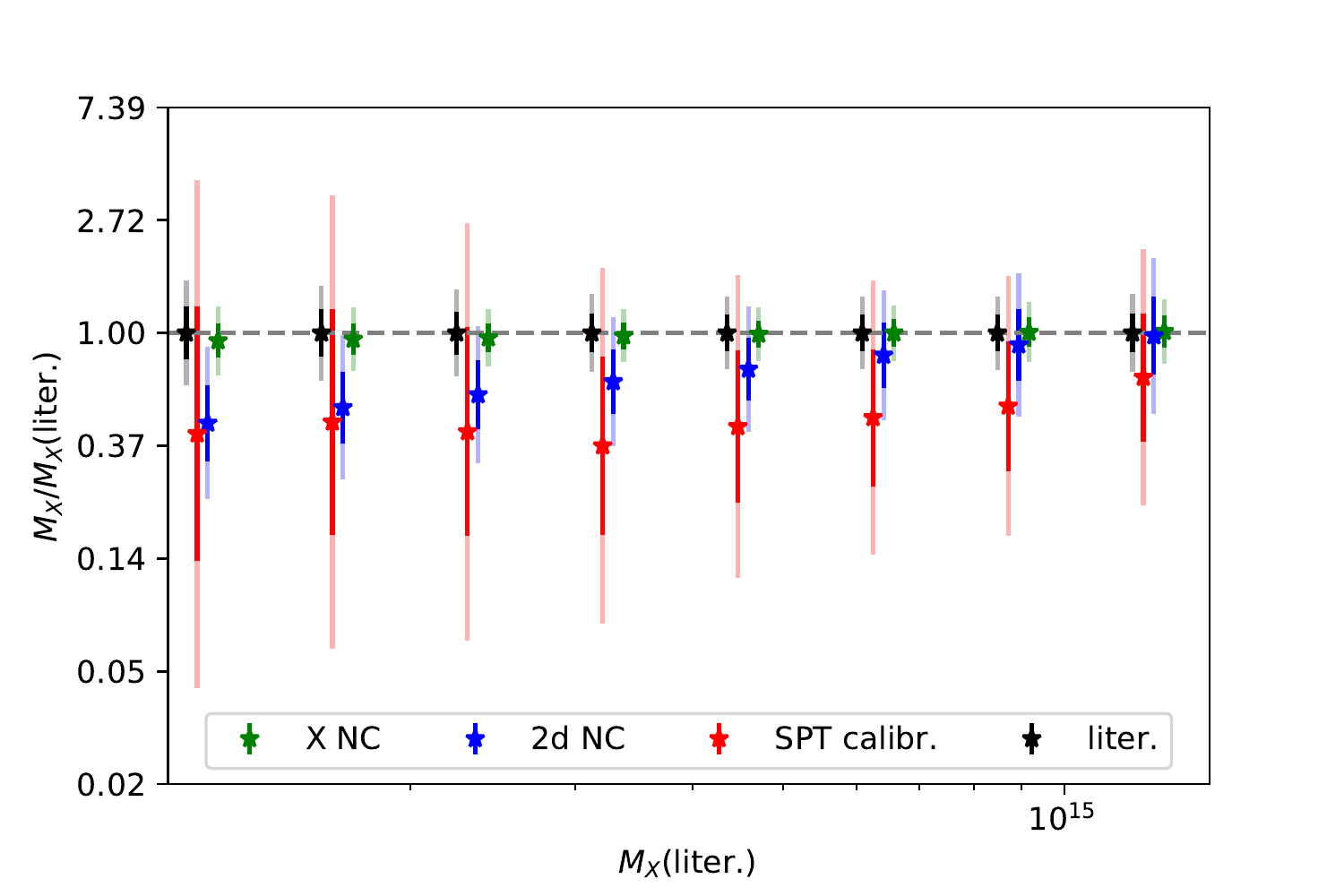}
	\includegraphics[width=\columnwidth]{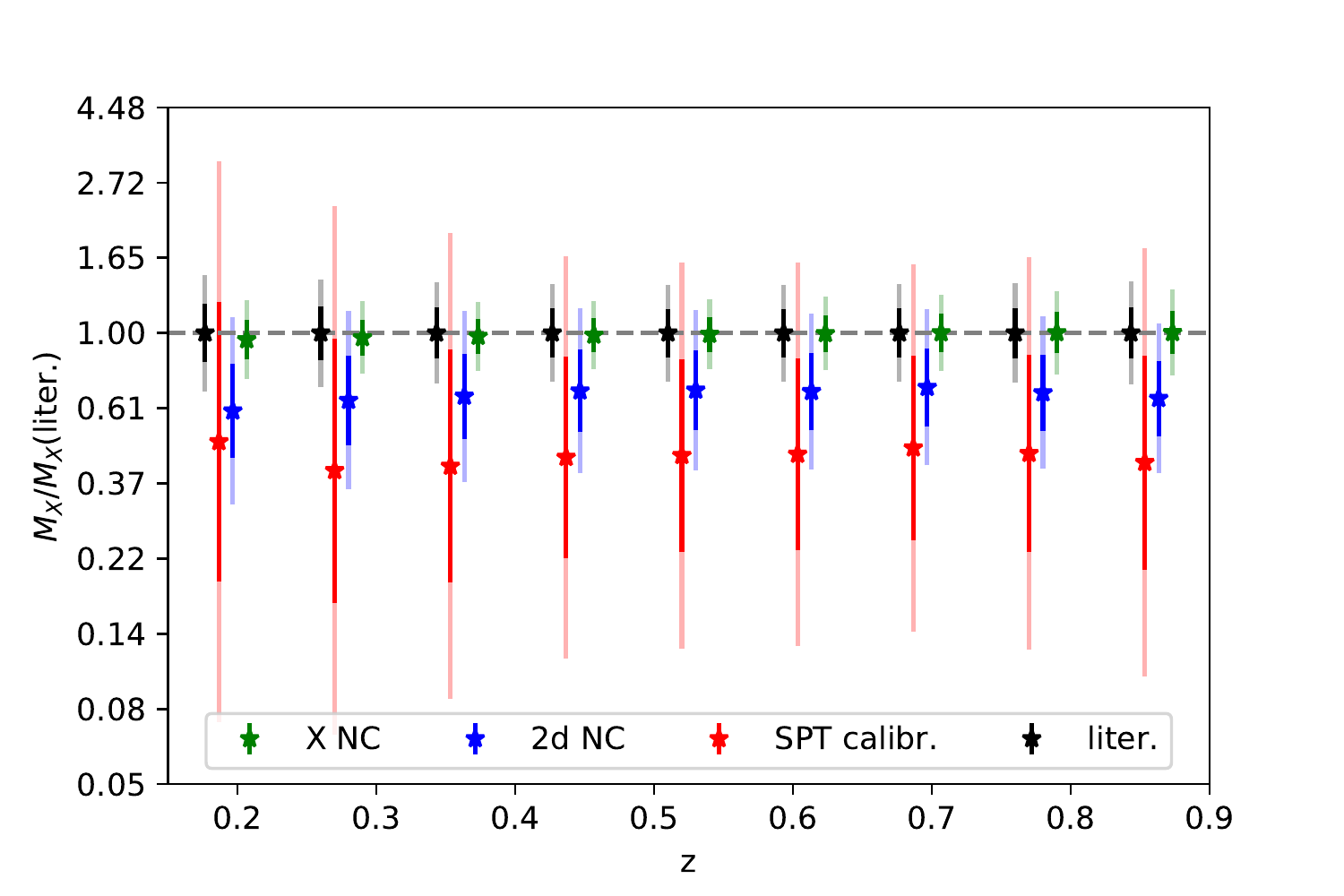}
	\includegraphics[width=\columnwidth]{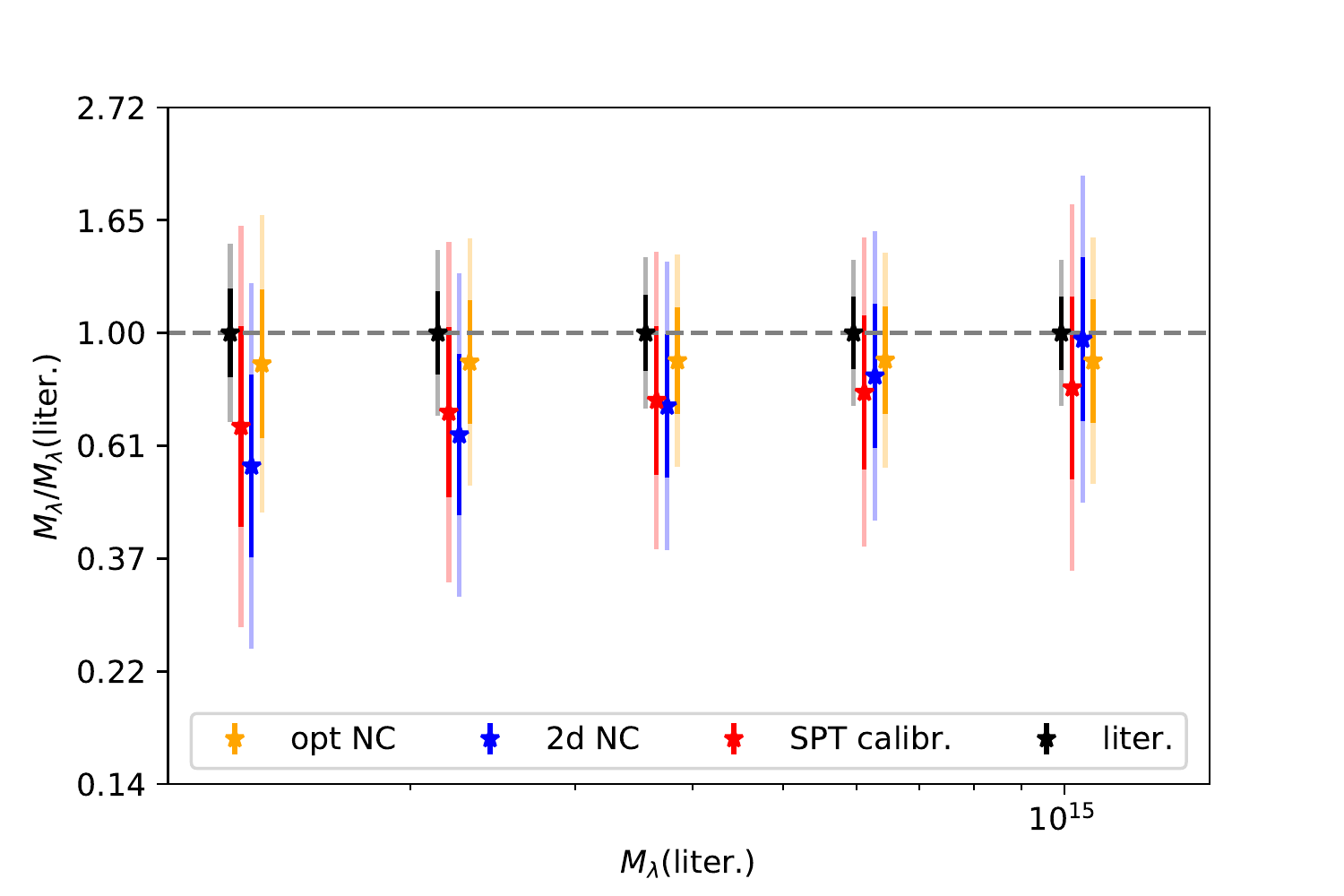}
	\includegraphics[width=\columnwidth]{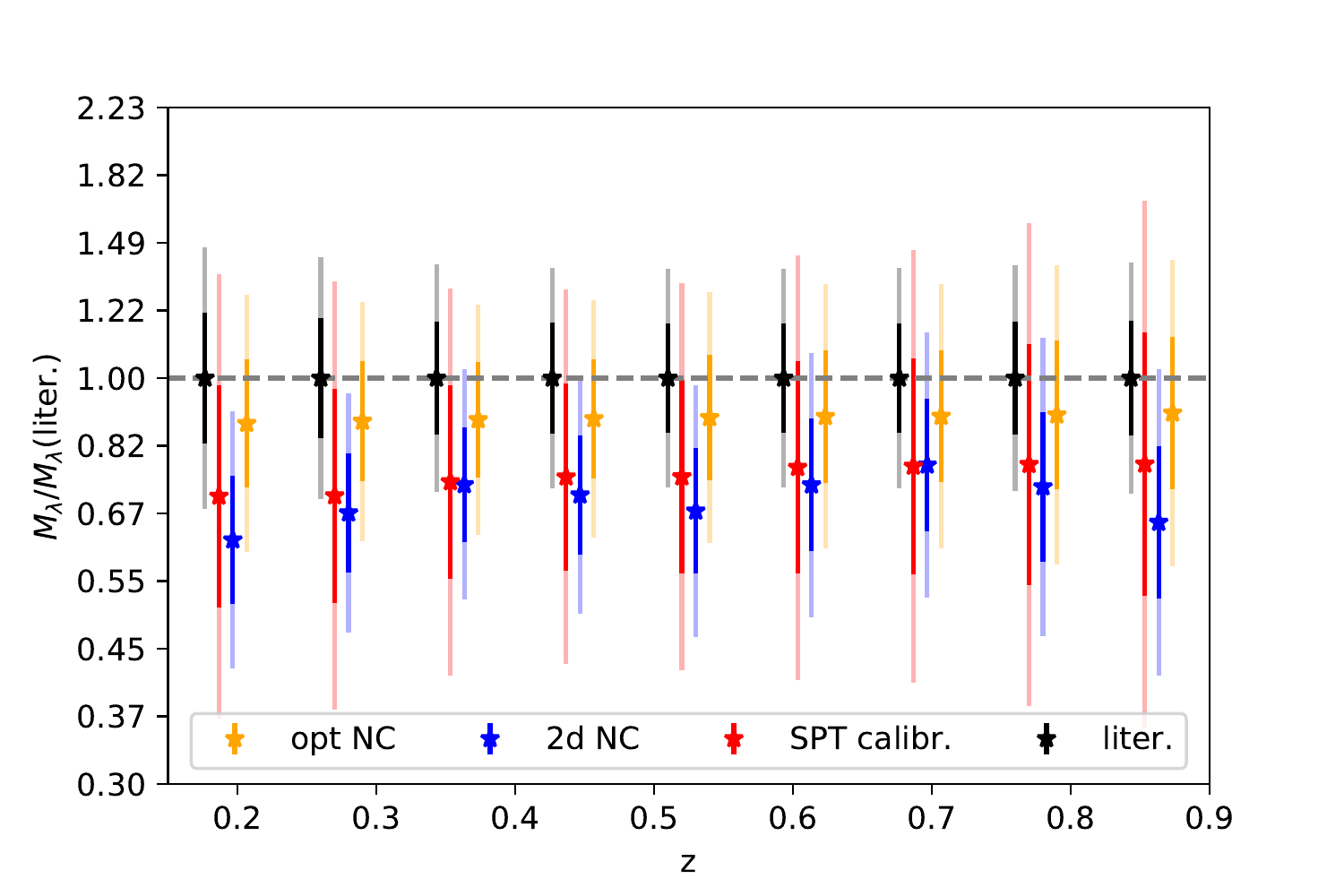}
	\vskip-0.10in
    \caption{Ratio of the masses derived by our analysis methods and the masses derived from 
    the literature values, for masses inferred from their measured X-ray flux ($M_\text{X}$, 
    upper row) and their measured richness ($M_\lambda$, lower row), as functions of mass (left column) and of redshift (right column), together with the 1 and 2-sigma 
    systematic uncertainties on the individual masses due to the incomplete knowledge of the scaling relation parameters. All masses refer to spherical over densities 500 times the critical density of the Universe. The masses we recover from SPT-SZ cross-calibration (SPT 
    calibr, red) and different flavours of number counts, while being in most cases 
    systematically low, are statistically consistent with the masses inferred by adopting the 
    literature values. Tension beyond 1 sigma, but still smaller than 2 sigma appears at the low mass end of the inferred X-ray masses.}
    \label{fig:masses}
\end{figure*}

\subsection{Validation using cluster masses}\label{sec:mass_prediction}

In this section we investigate the prediction of the individual halo masses derived from the different constraints on 
the scaling relation parameters extracted above. Given the that the number of objects as a function of mass is known, this section quantifies the relative goodness of fit of the number counts between the different fits we performed (X-ray, optical, and combined).

To estimate the masses for each cluster given its measured X-ray flux $\hat f_\text{X}^{(i)}$ (or analogously the measured richness $\hat \lambda^{(i)}$), we compute the distribution of probable masses

\begin{equation}\label{eq:P_M_given}
\begin{split}
P(M|\hat f_\text{X}^{(i)}, z^{(i)}, \vec{p}) \propto \int \text{d}f_\text{X} & P(\hat f_\text{X}^{(i)}|f_\text{X}) P(f_\text{X}| M, 
z^{(i)}, \vec{p}) \\
& \frac{\text{d}N}{\text{d}M} \Big|_{M, z^{(i)},\vec{p}},
\end{split}
\end{equation}
where $P(f_\text{X}| M, z^{(i)})$ is the mapping between intrinsic flux and mass obtained by 
only considering the first component of equation~(\ref{eq:P_introbs_m}). Note also that the above 
equation needs to be normalized in such a way that $\int \text{d} M \, P(M|\hat f_\text{X}^{(i)}, z^{(i)}, \vec{p}) = 1 $,
which sets the proportionality constant. 

The X-ray mass $M_\text{X}$ (and analogously the optical mass $M_\lambda$) can then be estimated to be 
\begin{equation}
\ln M_\text{X}^{(i)}|_{\vec{p}} = \int d\text{M} \,P(M|\hat f_\text{X}^{(i)}, z^{(i)}, \vec{p}) \ln M.
\end{equation}
Note that these masses naturally take account of the Eddington bias, which is fully described by equation~(\ref{eq:P_M_given}).

The X-ray and optical masses are affected by systematic uncertainties in the scaling relation and 
cosmological parameters. We capture this uncertainty in each case by marginalising the mass posterior over
the appropriate posterior distribution of the parameters that we determined above. We 
marginalize the mass over different scaling relation parameter posteriors, including those from
the literature (liter.), those from the SPT-SZ cross-calibration (SPT calibr.), and those from the combined X-ray and optical 
number counts (2D NC), the X-ray number counts (X NC) and the optical number counts (opt NC). The mass posteriors are 
derived for all clusters in the MARD-Y3 sample. 

In the upper row of Fig.~\ref{fig:masses} we present the ratio between the X-ray masses derived from our posteriors to the X-ray masses obtained from the literature \citepalias{bulbul19} as a function of inferred literature mass (left panel) and of redshift (right panel). We find that the mass inferred from the number counts in X-ray flux is consistent with 
the literature values, while the masses inferred from the 2D number counts and the SPT-SZ 
calibration are lower than the literature masses. In the case of the SPT masses the difference never exceeds one sigma at all redshifts and masses we considered. For the 2D number count masses, we find that they are 1 sigma low at all redshifts, and up to 2 sigma low at masses of 1-2 10$^{14}~M_\odot$. At masses of around 10$^{15}~M_\odot$ they are in perfect agreement with the other mass estimates. This is due to the different values of inferred mass trend. As a function of redshift, the masses inferred from 2D number counts and the SPT-SZ 
calibration are also lower, reflecting on one side the prevalence of low mass systems. On the other side, this shift is also be due to the larger intrinsic 
scatter recovered from the 2D number counts and the SPT-SZ calibration, that together with the 
shallower mass slope leads to a larger intrinsic mass scatter. This results in 
larger Eddington bias corrections and ultimately lower inferred masses. At the current level of 
statistical and systematic uncertainty we conclude that different methods predict mutually 
consistent individual masses from the X-ray flux at less than 2 sigma. Yet the magnitude of the intrinsic scatter of the X-ray 
luminosity at fixed mass and redshift, together with its mass trend, are indications of possible internal tensions and unresolved systematics. These trends where already noted when comparing our best fit number count models to the data (see above in section~\ref{sec:numcounts}) and will be discussed further in Section~\ref{sec:outcome_val}.

In the lower row of Fig.~\ref{fig:masses} we also show the ratio between the optical mass 
inferred from our fits to the value taken from the literature \citepalias{saro15}. Here we find that all our 
methods provide a lower, yet statistically consistent mass estimate. The difference is likely 
due to an analysis choice in the literature values. Namely, \citetalias{saro15} utilizes priors for the 
SZE-scaling relation parameters derived from fitting the SZE number counts at fixed cosmology. 
In that work, however, the CMB derived cosmology from \citet{planck13cosmo} was used, which 
results in $A_\text{SZ, \citetalias{saro15}} = 4.02 \pm 0.16$, and therefore is an overestimation of masses by 
$\sim18\%$ compared to our work. This shift accounts for most of the shifts seen in $M_\lambda$ here. Even without this correction, at the current level of systematical 
uncertainties, the individual optical masses inferred from our different analysis methods are mutually 
consistent. This is expected because our $A_\text{SZ}$ prior is consistent with the value used by \citetalias{saro15}. 
Furthermore, while 2D number counts predict a shallower mass trend than all other methods, in the mass range we consider this
 does not lead to significant tension with the other analysis methods.

This consistency check of mass estimates underscores the importance of weak lensing mass calibration as a component 
of the validation of cluster samples.  If the cosmology marginalized constraints on cluster masses from weak
lensing are not consistent with those from cluster counts, then that would be clear evidence of an inadequacy
in the selection model or an unaccounted for bias in the weak lensing calibration analysis.  As noted 
previously, we will examine the validation with the weak lensing constraints in a forthcoming analysis.

\subsection{Validation using independent cluster samples}\label{sec:xmatch}

\begin{figure}
    \includegraphics[width=\columnwidth]{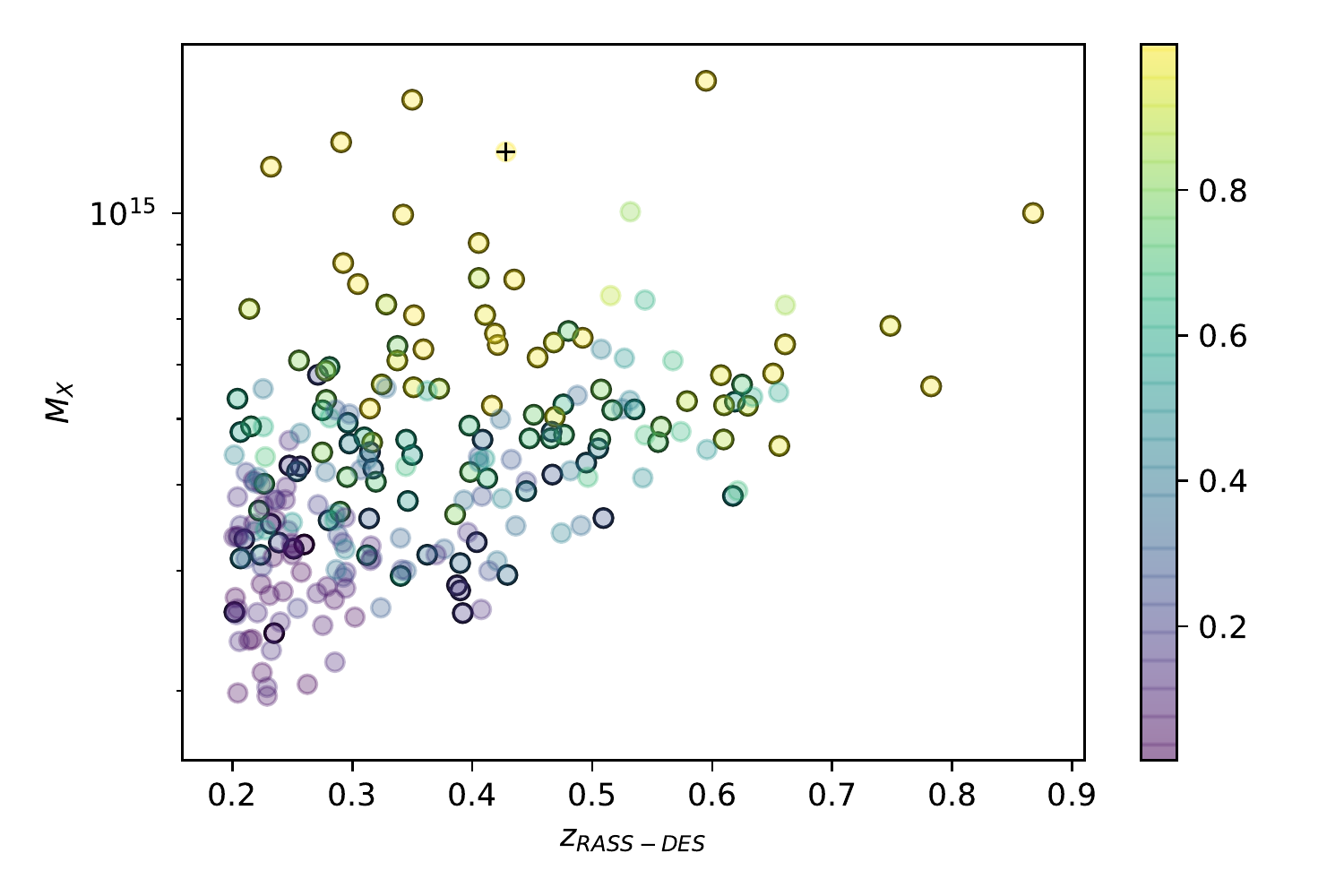}
	\vskip-0.10in
    \caption{MARD-Y3 sample in the joint SPT-DES Y3 footprint at redshift $z>0.2$. Color encodes the probability 
of an SPT-SZ detection for each object, showing the characteristic mass selection of the SPT-SZ catalog. Black circles indicate 
matched clusters, while the cross marks the missing SPT-SZ confirmations of MARD-Y3 objects. The only missed SPT confirmation is due to a catastrophic redshift error in the application of MCMF to the SPT-SZ sample.}.
    \label{fig:rass_pdet_full}
\end{figure}

Having established in the section above that our selection function modeling allows us to infer the masses of the MARD-Y3 clusters
consistently within the 
systematic uncertainties, we now move on to a further test of the selection functions of the two samples. 

As described in the methods Section~\ref{sec:consistencymethod}, we 
investigate the SPT-SZ and MARD-Y3 selection functions by comparing the 
probability of each MARD-Y3 object being detected by SPT-SZ to 
the actual occurrence of such a detection.  As established in section~\ref{sec:matched_sample}, there are 123 clusters in the cross-matched sample,
but the validation we do here also uses information from unmatched clusters.
We first consider the MARD-Y3 sample and compute the SPT-SZ detection probability for each of these objects. Comparing these probabilities to the actual occurrence of matches
provides an estimate of SPT-SZ incompleteness as well as MARD-Y3 
contamination. We then consider the SPT-SZ sample and compute the probability that an SPT-SZ cluster is detected in 
MARD-Y3.  In this case, we also constrain the outlier fraction beyond the log-normal scatter, more precisely the fraction 
of objects with an abnormally high X-ray flux and optical richness, or a surprisingly low SZE-signal.

\subsubsection{SPT-SZ detection of MARD-Y3 clusters}\label{sec:SPTdetMARDY3}

In Fig.~\ref{fig:rass_pdet_full} we show the MARD-Y3 cluster sample in the joint SPT-DES Y1 
footprint, plotted as a function of the X-ray derived mass and the redshift presented by 
\citetalias{klein19}. Note that the mass used in this plot is used solely for presentation 
purposes, and does not go into any further calculation. We color-code the MARD-Y3 clusters 
based on their SPT-SZ detection probability $p^{(i)}_\text{S|M}$, computed following 
equation~(\ref{eq:psptmardy3}). This prediction reflects the mass information contained in each 
cluster's measured flux $\hat f_\text{X}^{(i)}$ and measured richness $\hat \lambda^{(i)}$. It 
also nicely visualizes the approximate mass selection at $M\gtrapprox3\times 10^{14} M_\odot$ of 
the SPT-SZ sample.

We place black circles around the matched clusters. When determining the detection 
probabilities using the literature values for the scaling relation parameters, we identify six 
clusters that have high detection probability, but are not matched, so-called missed SPT confirmations of MARD-Y3 objects. However, when determining the detection probabilities either from the posterior of 
our SPT-SZ cross-calibration or the 2D number counts, only one of these systems is identified as 
a missed confirmation: 2RXS J033045.2-522845. 

This object coincides in the sky with the NE part of A3128. It has been found to be a $z\sim0.43$ cluster by  \citet{werner07} using XMM observations, by ACT observations \citep{hincks10} and by the SPT-SZ survey \citep{bleem15},  despite the large number of $z\sim0.05$ galaxies in the foreground (also visible the DES image in the upper left panel of Fig.~\ref{fig:gal2}). The redshift $z\sim0.43$ is confirmed by MARD-Y3. However, application of MCMF on the SPT-SZ sample found $z_\text{SPT}=0.056$, sourced by the foreground galaxies. Consequently, this object is erroneously not included in our SPT-SZ sample, which has the redshift selection $z_\text{SPT}\in (0.2, 1.)$. Noticeably, MCMF run on SPT-SZ finds also a structure with $f_\text{cont, SPT}=0$ at $z\sim0.43$. It is however discarded by the automated highest peak selection as also the z$\sim0.05$ structure has $f_\text{cont, SPT}=0$. In summary, this object is missing in our SPT-SZ sample due to a catastrophic MCMF failure when run on SPT-SZ. To keep the pipeline automated and avoid human decision making, we do not apply any special treatment to this object.

We also aim to constrain the occurrence of contamination in the MARD-Y3 sample by introducing 
the probability $\pi_\text{c}$ that a MARD-Y3 object is not a cluster, and should not therefore  be detected by SPT. Simultaneously, we also introduce the SPT-SZ incompleteness 
$\pi_\text{i}$, the probability that any MARD-Y3 cluster that should have been detected in the SPT-SZ survey was not (c.f. Fig.~\ref{fig:sptdetmardy}). This allows us to use the actual list of 
detections and non-detections together with the raw probabilities of detection to constrain 
these extra probabilities, as discussed in equation~(\ref{eq:lnL_cont_inc}). We find that  
$\pi_\text{c}$ and $\pi_\text{i}$ are degenerate parameters,with only the difference between the two values constrained by our data, rather than the two values separately. Under the assumption of a MARD-Y3 
contamination of $\pi_\text{c}=0.025$ , as derived by \citetalias{klein19} for the 
$f_\text{cont}<0.05$ sample used here, we find $\pi_\text{i}= 0.284 \pm 0.043 \text{(stat.)} ^{+0.108}_{-0.186} \text{(sys.)}$, 
when marginalising over the literature priors. When marginalizing over the SPT-SZ calibration 
posterior we find $\pi_\text{i}< 0.030 \text{ (stat.) }$ and $\pi_\text{i} <0.030 \text{ (sys.)}$ at 68$\%$ 
confidence, while we find $\pi_\text{i}< 0.047 \text{ (stat.) }$ and $\pi_\text{inc} <0.231 \text{ (sys.)}$ at 
68$\%$ confidence when marginalising over the 2D number counts constraints together with the 
priors from \citetalias{bocquet19} on the SZE-signal scaling relation. 

The difference in inferred central value for the SPT-SZ incompleteness is due to the different 
mass predictions when using the literature priors as compared to our fits. As discussed in 
Section~\ref{sec:mass_prediction}, our SPT-SZ cross-calibration and our 2D number counts analysis 
imply lower X-ray and optically derived masses than the literature priors. This 
systematically lowers the SPT-SZ detection probability of MARD-Y3 clusters, resulting in different 
incompleteness probabilities when comparing to the actual number of matched objects. We 
interpret this as another piece of evidence that the SPT-SZ cross-calibration and the 2D number 
counts provide a more accurate picture of the observable--mass relation than the literature 
priors. In fact, they reveal that the scatter around our luminosity--mass relation is larger 
than the scatter found by \citetalias{bulbul19}. Yet, within the statistical and systematic uncertainties the 
results are still in agreement.

Another interesting aspect is the magnitude of the statistical and systematic uncertainty on 
the SPT-SZ contamination. Note that the statistical uncertainties when marginalizing over the 
different posteriors are comparable. This reflects the fact that they are derived from a sample of a 
given size. The minor differences can be appreciated by noting that in equation~(\ref{eq:lnL_cont_inc}) the 
individual clusters likelihood of $\pi_\text{inc}$ are weighted by the detection probabilities 
$p^{(i)}_\text{S|M}$, which are different depending on which posterior is used to compute them. 
On the other hand, the magnitude of the systematic uncertainty introduced by the marginalization 
over the different posteriors is quite different. Marginalizing over the SPT-SZ cross-calibration 
posterior provides the smallest systematic uncertainty. This is expected when considering that 
the SPT-SZ cross-calibration constrains 
$P(\zeta | \hat f_\text{X}^{(i)}, \hat \lambda^{(i)}, z^{(i)}) $ (c.f. 
equations~\ref{eq:P_zeta_xopt}-\ref{eq:lnLsptcc}), which is the major source of systematic 
uncertainty when computing the SPT-SZ detection probabilities of MARD-Y3 cluster (c.f. 
equation~\ref{eq:psptmardy3}). These distributions are predicted less accurately by the literature priors and the 2D number counts.

\begin{figure}
	\includegraphics[width=\columnwidth]{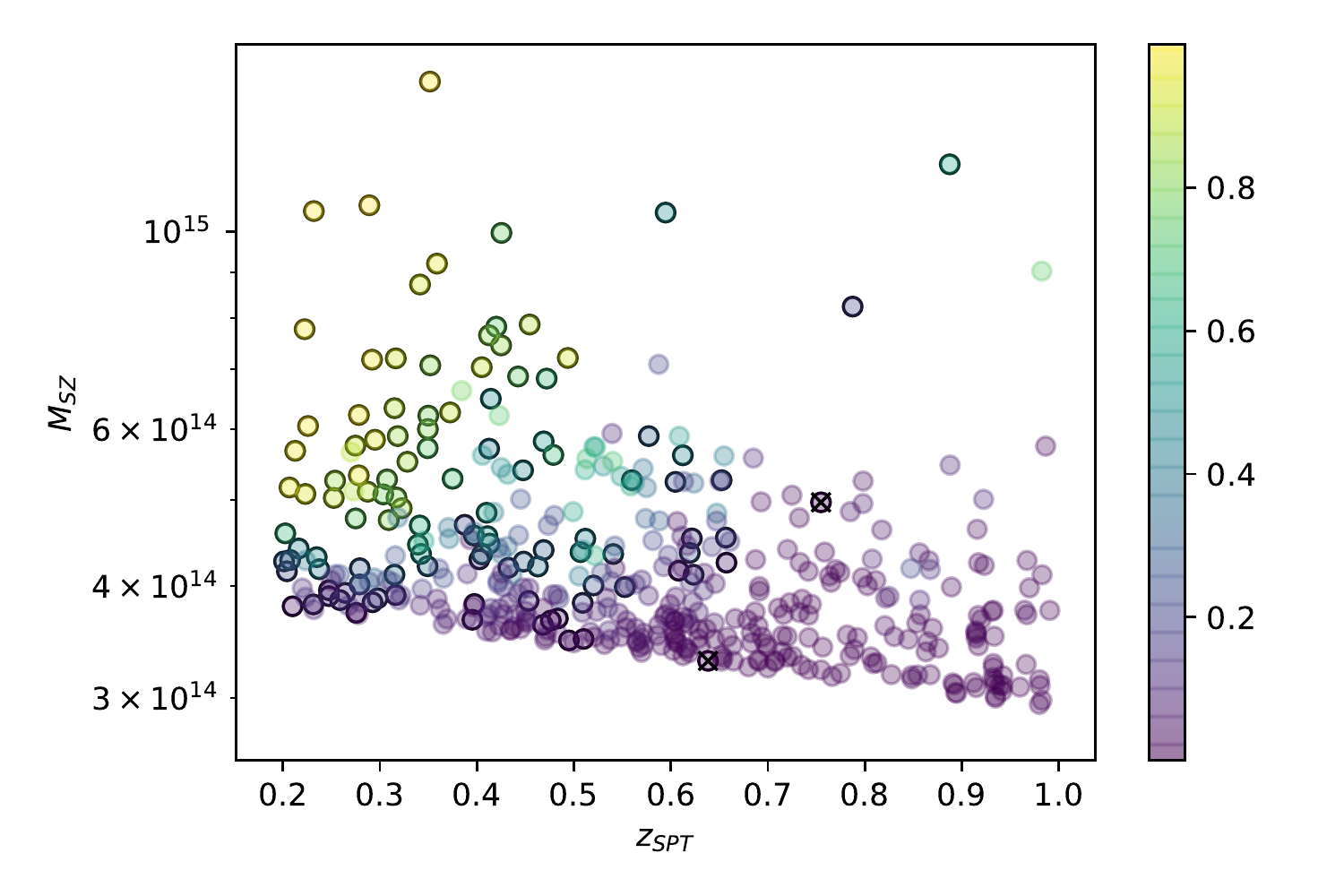}
	\vskip-0.10in
    \caption{SPT-SZ selected sample in the joint SPT-DES Y3 footprint. Color encodes the probability of MARD-Y3 
detection, showing the characteristic flux selection of an X-ray survey. Black circles indicate matched clusters, while 
crosses mark unexpected MARD-Y3 confirmations of SPT-SZ objects (having a MARD-Y3 detection probability $<0.025$ but matched nonetheless).}
    \label{fig:spt_pdet}
\end{figure}

\subsubsection{MARD-Y3 detection of SPT-SZ clusters}\label{sec:mardy_on_spt}

We also test the MARD-Y3 selection function by computing the probability of detecting each of the SPT-SZ clusters in 
the DES-Y3 footprint. In Fig.~\ref{fig:spt_pdet} we show the SPT-SZ sample as a function of 
redshift and SZE derived mass. Note that the SZE derived mass shown in this figure is only used 
for presentation purposes. Color encodes the MARD-Y3 detection probability, computed 
via equation~(\ref{eq:p_mardy3_spt}). The color coding reflects the approximate flux selection 
of the MARD-Y3 sample. We highlight the matched clusters with black circles.

Out of the 123 clusters in the cross-matched sample, we identify 5 unexpected MARD-Y3 confirmations, which are SPT-SZ 
clusters that show up in MARD-Y3 even though they have a MARD-Y3 detection probability $<0.025$, 
as calculated by marginalising over the 
literature values. This list is expanded by 3 more unexpected confirmations when marginalising over the 2D number counts posterior.
When marginalising 
over the SPT-SZ cross-calibration posterior, we find 2 unexpected confirmations, all of which are in common 
with the aforementioned (`SPT-CLJ0324-6236', visual inspection in Fig.~\ref{fig:gal3}, and `SPT-CLJ0218-4233', visual inspection in Fig.~\ref{fig:gal4}).  These clusters are marked in Fig.~\ref{fig:spt_pdet} with crosses.  
Visual inspection of this object reveals that in both cases a clear X-ray structure coincides with a significant SZE signal and a red sequence galaxy over-density at the cluster redshift. Thus, these objects are likely genuine clusters with multi-wavelength properties that are not captured by our scaling relation model and the scatter around it. These objects also do not display any exceptional behaviour in w.r.t. the mean distribution in X-ray flux, richness.

Given these indications, we investigate the probability that an 
SPT-SZ object that should not be matched by MARD-Y3 is matched anyway.
It cannot {\it a priori} be excluded that the distribution of X-ray luminosities or SZE-signals at fixed mass  has in actuality tails extending beyond the log-normal scatter model we assumed in Section \ref{sec:cluster_pop_model}. Such tails would lead to unexpected detections. The probability of a cluster living in such a tail, i.e. being an outlier, is given by the parameter $\pi_\text{t}$ (see section~\ref{sec:mardy3detonspt}). Taking 
account of the detection probabilities and the actual occurrence of detections, we use the 
likelihood presented in equation~(\ref{eq:lnL_ptail}). We find 
$\pi_\text{t}= 0.059 \pm0.017 \text{(stat.)} \pm 0.031 \text{(sys.)}$, when 
marginalising over the literature 
priors. When marginalizing over the SPT-SZ cross-calibration posterior we find 
$\pi_\text{t}= 0.061 \pm 0.018 \text{(stat.)} \pm 0.040 \text{(sys.)}$, while we 
find $\pi_\text{t}= 0.095 \pm 0.018 \text{(stat.)} \pm 0.019 \text{(sys.)}$ when 
marginalising over the 2D number counts constraints together with the priors from 
\citetalias{bocquet19} on the 
SZE-signal scaling relation. 

These constraints are mutually consistent in a statistical sense. Yet, the significance of the detection of the tail beyond log-normality ranges from 1.4 sigma for the literature priors, over 1.6 sigma for the SPT-SZ cross-calibration to 3.6 sigma for the 2D number counts constraints. Internal inconsistencies in the number counts (discussed in sections~\ref{sec:numcounts},~\ref{sec:mass_prediction},~\ref{sec:X_selection_disc}~and~\ref{sec:outcome_val}) might affect the latter result. Better mass information is required to distinguish whether our findings are a statistical fluke, the result of an unresolved systematic or stem from a genuine signal. If the presence of a log normal tail would be confirmed, more detailed observations are needed to understand the source of the outliers we selected.  For example, high-angular resolution X-ray or mm-wave observations, in combination with the spectroscopic optical data, would help to rule out any astrophysical confusion in either the X-ray or SZ measurements and identify any lower mass structures or objects along the line of sight, which could be affecting any of the observables.

\section{Discussion}
\label{sec:discussion}

Here we first summarize the findings from the previous section and then
discuss implications. We focus on 
different aspects, including: (1) internal indications for unresolved systematics in the selection function modelling, (2) the outcome of our validation, (3) the impact of optical incompleteness and 
the resulting benefits from its modeling, and finally (4) the implications of this work for cosmological studies.

\subsection{X-ray selection function systematics}\label{sec:X_selection_disc}

In section~\ref{sec:residual_trends} we discussed potential unresolved redshift trends of the selection function fit. Given that the X-ray selection spans a mass range of factor of 3 (see for instance Fig.~\ref{fig:sample}) from low redshift to high redshift, residual redshift trends in the X-ray selection functions are likely to impact the inferred mass trend as much as they are likely to impact the redshift trend of the X-ray flux--mass relation. This systematic manifests itself in different places, as discussed in the following. 

When sampling the X-ray number counts (c.f. Section~\ref{sec:XNC}) we 
sample the parameters of the richness--mass scaling relation with priors from the literature to estimate the 
effect of optical incompleteness of the sample. While the prior on the redshift evolution is 
$C_\lambda=0.73\pm0.76$, the posterior is $C_\lambda=0.34\pm0.53$, indicating that the X-ray number counts likelihoods slightly prefer a weaker redshift trend of the richness, effectively making the optical incompleteness larger at low redshift than at high redshift. This preference may be compensation for the fact that our model seems to predict too large an X-ray 
selection function at low redshift and too small an X-ray selection function at high redshift. 

Similarly, when 
sampling the optical number counts, we rely on priors on the X-ray flux--mass scaling relation to propagate the X-ray 
selection function to the space of measured richness. Also in this case the prior $C_\text{X}=-0.20\pm0.50 $ is altered 
to a posterior $C_\text{X}=-0.50\pm0.38$. Consequently, a weaker redshift trend is preferred by the number counts, 
possibly as in an attempt to compensate the same residual systematic effect. Lastly, we find that the X-ray, as 
well as the optical number counts, pull the prior we placed on $\Omega_\text{M}= 0.276\pm0.047$ to a posterior $
\Omega_\text{M}= 0.296\pm0.038$ from X-ray number count, and $\Omega_\text{M}= 0.302\pm0.037$ from optical 
number counts, respectively. If these shifts result in biases of the cosmological results once direct mass information 
from weak lensing is available, they should be further investigated.

\begin{figure}
	\includegraphics[width=\columnwidth]{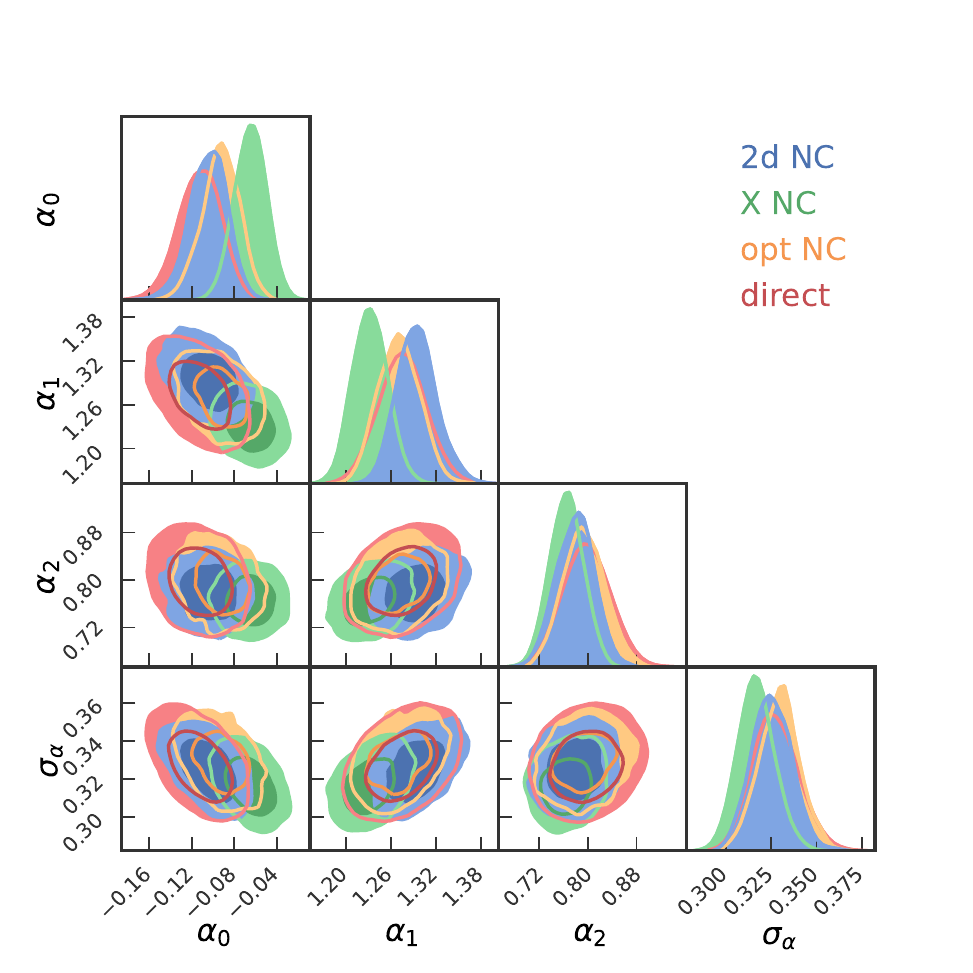}
	\vskip-0.10in
    \caption{Marginal posterior contours on the extra nuisance parameters controlling the mapping between X-ray flux and detection significance, and hence the X-ray selection function from the direct fit to the data (red), the sampling of that fit with the 2D number counts (blue), with the X-ray number counts (green), and the optical number counts (orange). Shifts of the contours with respect to the constraints from the data alone are indicative of residual systematics.}
    \label{fig:sel_params}
\end{figure}

As described in the case of a putative redshift residual, the empirical calibration of the 
selection function provides an opportunity  
to uncover unresolved systematics. From this perspective it offers advantages in comparison to selection functions 
determined from image simulations. 
For instance,  consider in Fig.~\ref{fig:sel_params} the posterior constraints on the significance--flux scaling parameters resulting from fitting either directly to the relevant catalog data by sampling equation~\ref{eq:anc_like} (red) or adding different number counts likelihoods (2D in blue, X-ray in green, and 
optical in orange). In principle, we expect no extra information from the number counts on the scaling governing the 
X-ray selection function. Yet the posterior of the X-ray number counts in particular display shifts compared to the direct fit. This might hint at unresolved systematic effects in the X-ray number counts. Indeed, we find 
that the X-ray number counts predict a smaller intrinsic scatter $\sigma_\text{X}$ and a steeper mass slope than both the SPT-SZ cross-calibration 
and the 2D number counts. While at the current stage these putative systematics are smaller than the statistical 
uncertainties, the empirical methods here already prove to be potent tools for validating the number counts. We plan to 
include such tests as unblinding conditions for the forthcoming cosmological analysis of this catalog.

\begin{figure}
	\includegraphics[width=\columnwidth]{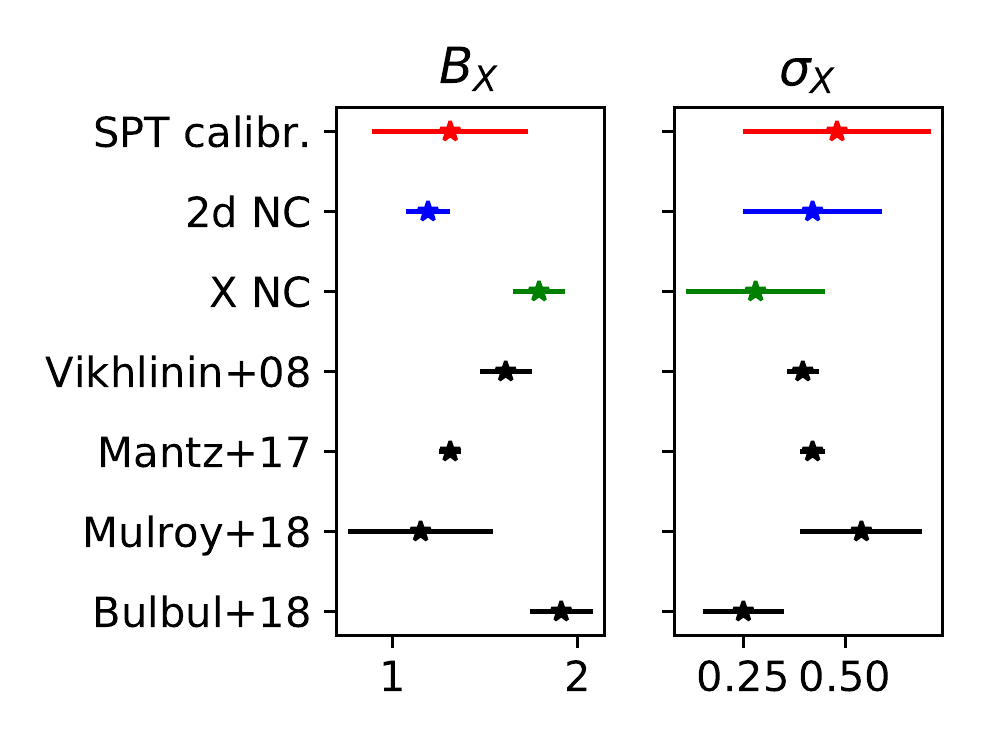}
	\vskip-0.10in
    \caption{Compilation of results on the mass trend $B_\text{X}$ and the intrinsic scatter 
    $\sigma_\text{X}$ of the luminosity--mass relation, compared to our results. While all our 
    results lay within the dispersion of the literature results, this dispersion among the 
    results is larger than the reported error bars, indicating that no consensus has yet been 
    reached.}
    \label{fig:LX-M_comparison}
\end{figure}

\subsection{Outcome of the validation}\label{sec:outcome_val}

As outlined in Section~\ref{sec:mass_prediction}, different methods with different sensitivities to the selection function provide statistically consistent masses. This provides strong evidence for the adequacy of the selection functions we constructed in this work. Interestingly, however, non-significant tensions appear on different parameters, mainly in the scaling relation parameters derived from 1D X-ray number counts and 2D number counts. These tensions are also visible in the comparison of the predicted and the measured number count (section~\ref{sec:numcounts}), as well as in the comparison of the inferred masses (section~\ref{sec:mass_prediction}), We identify two main scenarios: low intrinsic scatter and steep luminosity--mass trend, preferred by X-ray number counts, and large intrinsic
scatter and shallow slope, preferred by 2D number counts. In the following, we will discuss evidence for these two scenarios.

Comparison to the literature does not provide clear guidance on which scenario is more plausible, as can be seen in Fig.~\ref{fig:LX-M_comparison}. The low scatter scenario is in very good agreement with the results from \citetalias{bulbul19} on XMM luminosities of SPT-SZ selected clusters. On the other hand, weak lensing calibrated measurements of the luminosity--mass relation on RASS selected clusters by \citet{Mantz15} and \citet{mulroy18} find shallower mass trends and larger intrinsic scatter in good agreement with our large scatter scenario. In analysing number counts of RASS selected clusters with X-ray mass information, \citet{vikhlinin09b} found a mass trend and scatter value consistent with both scenarios. 

Further evidence for the amount of intrinsic scatter can be obtained by comparing different measurements of 
the luminosities. \citetalias{klein19} show that there is 
significant scatter among the luminosities measured by \citet{boller16} and those 
reported by \citet{piffaretti11}. Namely, a log-normal scatter of $0.48\pm0.05$ for $0.15<z<0.3$ and $0.40\pm0.10$ for $0.3<z$. This in unsettling, 
considering that the luminosities reported by 
\citet{piffaretti11} are measured on the same ROSAT data as the ones by \citet{boller16}. Given that this effect might be partially sourced by the fixed aperture measurements by \citet{boller16}, we can not exclude that the X-ray flux measurement introduces mass dependent trends.
Further investigation of the systematics in flux measurement methods is clearly required.

The hypothesis of larger scatter in the X-ray mass scaling is further supported by the constraints on the SPT-SZ 
incompleteness derived from the different posteriors (see Section~\ref{sec:SPTdetMARDY3}). Compared to the literature priors, which prefer small scatter but 
predict high incompleteness, both the SPT-SZ cross-calibration and the 2D number counts predict incompletenesses 
consistent with zero, mainly due to the larger X-ray intrinsic scatter. On the other hand, the mass calibrations of the 
SZE-mass scaling determined using different, independent methods \citep{capasso19a, stern19, dietrich19, chiu18} 
match with the masses emerging from a fully self-consistent cosmological analysis of the SPT-SZ cluster sample
\citep{bocquet15, dehaan16, bocquet19}. In the presence of high incompleteness, this agreement would be 
coincidental. Larger X-ray scatter is thus made even more plausible, because it predicts low SPT-SZ incompleteness.

In summary, the large scatter/shallower mass trend scenario is supported by the comparison of different luminosity measures, different literature results and the implications of these scenarios on the inferred SPT-SZ incompleteness. Furthermore, we find that the 2D number count fits introduce less internal tension on the parameters of the significance-flux scaling governing the X-ray selection function.

\begin{figure*}
	\includegraphics[width=2\columnwidth]{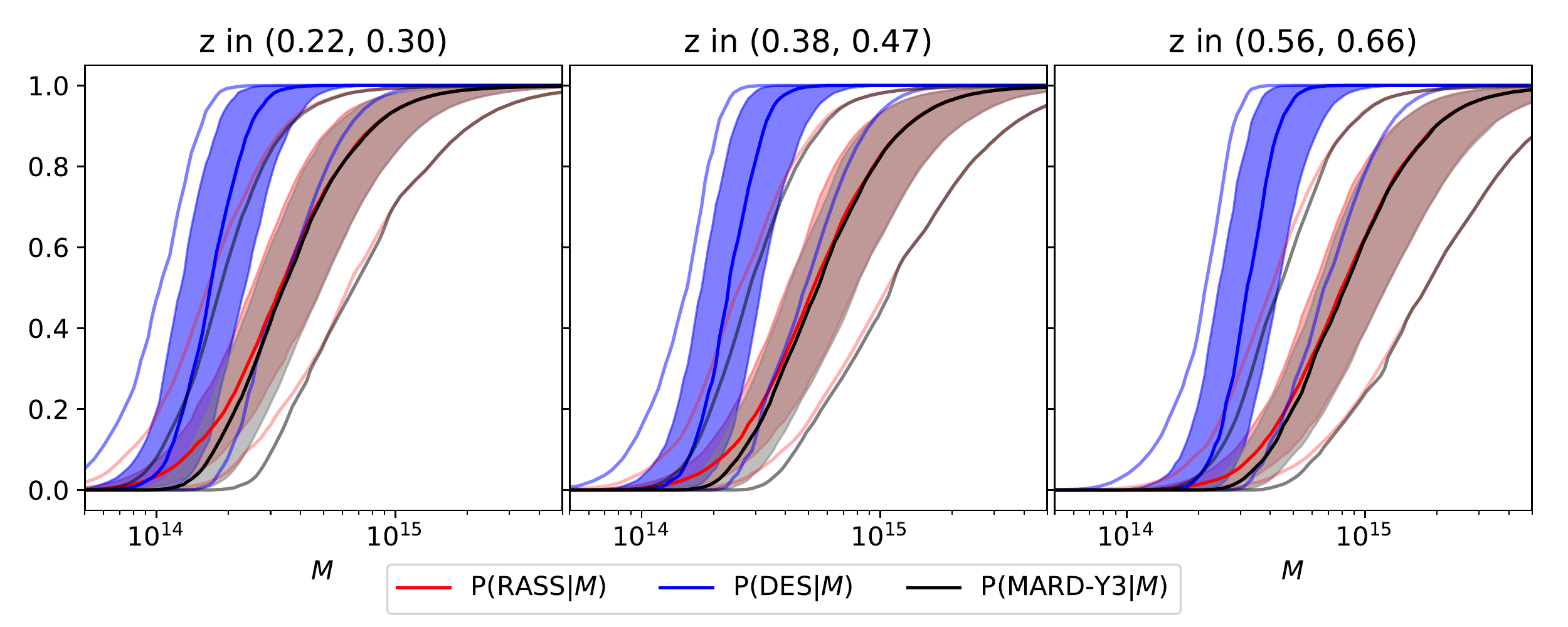}
	\vskip-0.10in
	\caption{X-ray (red), optical (blue) and combined (black and grey) selection functions as 
	functions of mass for different redshift bins, plotted with the systematic uncertainties 
	derived from the 2D number counts posterior on the scaling relation parameters. 
	The full lines are the median values, the filled region covers the range from the 16th to the 84th percentile and the transparent lines show the 2.5th and 97.5th percentile. All masses refer to spherical over densities 500 times the critical density of the Universe. While the 
	combined selection of the sample is clearly dominated by the X-ray selection function at 
	most masses, the optical cleaning introduces some extra incompleteness at low masses, 
	especially at low redshift.}
    \label{fig:selection_functions}
\end{figure*}

\subsection{Impact of the optical incompleteness}

As shown throughout this work, we model the selection of the MARD-Y3 sample in a two staged approach, which 
mirrors the operational creation of the catalog: (1) we determine an X-ray selection function based on the fact that the 
candidate catalog is selected with a X-ray detection significance threshold, and (2) we model the optical 
cleaning, which is operationally equivalent to a redshift dependent minimum value for the measured richness. The two result in 
selection functions in the space of X-ray flux and richness, respectively (c.f. Section~\ref{sec:selection_function1}).

For ease of representation, we utilise the observable--mass scaling relation to transform these observable selection functions into 
mass selection functions. This introduces systematic uncertainty through the widths of the posteriors on 
the scaling relation parameters. The mass 
selection functions in three redshift bins are shown in Fig.~\ref{fig:selection_functions}.  As stated above, 
the X-ray selection is dominant at most masses. 
Yet, the optical cleaning introduces an excess incompleteness at the lowest 
masses, leading to a suppression of the selection probability at those masses. 

The fact that the optical selection can not be completely ignored can be appreciated also from Fig.~\ref{fig:1d_NC_opt} and 
Fig.~\ref{fig:2d_NC}. Given that in these plots we show the number of clusters also as a function of 
measured richness, we can appreciate that the MARD-Y3 sample displays a sharp, redshift dependent cut in measured 
richness. This is the result of the optical cleaning process, which takes effect before the X-ray selection probability nears zero.

The fact that we can consistently infer the masses when marginalizing over a fiducial 
cosmology indicates that the two stage selection function modeling is adequately describing the sample. This in 
turn means that optical cleaning with MCMF can provide clean cluster samples even from highly contaminated 
candidate samples. At the cost of tracking an extra scaling relation, the richness--mass relation, this has the potential to 
significantly lowering the limiting mass of ongoing and future surveys with SPT, eROSITA or similar ICM observable based surveys while maintaining a similar contamination level. 
Given that all selected clusters in such samples would have a richness in addition to  an X-ray or SZE observable,
the richness--mass relation would be calibrated along side the X-ray
or SZE observable in the context of a direct mass calibration as we have demonstrated with our SPT-SZ cross calibration. Furthermore, the possibility to perform number counts not only in the 
X-ray or SZE observable, but in richness alone, or even in the combination of multiple observables, 
provides additional consistency checks that could be used to reveal unappreciated systematics.

\subsection{Implications for cosmological studies}

In this work we explore several techniques that allow us to validate the selection function of a cluster survey. 
However, we would like to caution that in this work we never directly determined the masses of our clusters. This would require 
either the measurement of the weak lensing signal around our clusters, or the study of the projected 
phase space distribution of spectroscopically observed cluster members. From a formal perspective, such studies can be 
treated analogously to our SPT-SZ cross-calibration. They will allow us to determine the parameters of the scaling 
relations to high accuracy, enabling the use of the number counts to study cosmology.

In contrast, our current work assumes the cosmology derived by \citetalias{bocquet19} in order to determine the scaling relation 
parameters from the number counts of the MARD-Y3 sample. Also the indirect mass information we use in form of the priors on the SZE--mass relation were derived by \citetalias{bocquet19} in the same analysis.  
So they, too, are contingent upon that analysis. The consistency of their result with our modeling is supported by the fact that we do not find a significant level of SPT-SZ incompleteness.

Our work then demonstrates several techniques
that we anticipate will be important for controlling systematics in future 
X-ray selected cluster samples, especially the sample detected by eROSITA \citep{Predehl10, merloni12}. First, we 
have shown that the X-ray selection function can be determined empirically from the selected sample. As such, the 
simplistic assumptions made in forecast works \citep[e.g.][]{grandis19} can easily be replaced by a more accurate 
description without introducing much numerical complexity. The empirical determination of the selection function also 
allows one to check for unresolved systematic effects, 
as demonstrated in Section~\ref{sec:residual_trends}. As an addition to 
the set of systematics tests, such techniques are likely to improve the systematics control within eROSITA cluster 
cosmological studies.

Our work also highlights the use of secondary mass proxies to inform the number counts experiment. We demonstrate 
that performing the number counts in optical richness despite the X-ray selection provides a valuable source of mass 
information. In the presence of a direct mass calibration, that mass information would be provided externally, and 
optical number counts would provide independent cosmological constraints. This in turn allows one to set up another 
important consistency check, ensuring a higher level of systematics control. On the same note, we also clearly 
demonstrate the value of additional mass proxies to put direct constraints on the scatter. Indeed, the analysis of the 
number counts in X-ray flux and richness space was central to revealing the larger scatter in X-ray observable. 
Given the planned application of MCMF to eROSITA such multi-observable number counts experiments 
can be undertaken also in that context.

Furthermore, we present here an expansion of earlier work by \citetalias{saro15} on detection probabilities of clusters 
selected by one survey in another survey. Our formalism tests
the selection functions of different surveys against each other and thereby gains precious
empirical constraints on those selection functions. This method depends on the shape of the mass function for the Eddington bias correction, and on the redshift--distance relation for the X-ray scaling relation. Importantly, however, it is independent of the distribution of clusters in observable and redshift. In turn, these are the major sources of 
cosmological information in the number counts experiment. Consequently, in the presence of direct mass information to constrain the scaling relation parameters, this technique provides a selection function test that is insensitive 
to the predicted number of clusters and its redshift evolution. As such this test is ideally suited to validated cluster number count experiments.

Our approach would not only benefit the systematics control in future X-ray and SZE surveys, but also future optical surveys.
The selection function in optical surveys remains a source of systematic uncertainty that has been mainly studied through simulations \citep{costanzi19}. Applying techniques like ours to empirically validate an 
optical survey cluster selection function offers important advantages and will become more
relevant with the upcoming next generation surveys from Euclid and LSST.

\section{Conclusions}\label{sec:conclusions}

We perform a multi-wavelength analysis of the MARD-Y3 sample \citepalias{klein19}. 
This sample was selected by performing an optical follow-up of the 
X-ray selected 2nd ROSAT faint source catalog \citep{boller16} using DES-Y3 data.  
The optical followup was carried out using
MCMF \citep{klein18}, which is a tool that includes spatial and colour filters designed to identify 
optical counterparts of ICM selected cluster candidates and to exclude random superpositions of 
X-ray and optical systems. The multi-wavelength dataset allows for an extensive set of cross-checks and
systematics probes of the MARD-Y3 sample, its selection function and the associated observable--mass scaling relations.

We model the selection function (see Section~\ref{sec:selection_function1}) 
of the MARD-Y3 sample as the combination of the X-ray selection function 
of the candidate sample together with a model of the incompleteness introduced by the optical cleaning of that sample.  We then 
proceed to calibrate the X-ray luminosity--mass and optical richness--mass relation using different sources of mass 
information to test whether
there is tension in the dataset or a flaw in the selection function.

First, we cross-match the MARD-Y3 and the SZE selected SPT-SZ cluster samples, and calibrate the MARD-Y3 
scaling relations using the published calibration of the SZE signal-to-noise--mass relation 
(see Section~\ref{sec:SPTSZxCal}).  Second, 
assuming priors on the cosmological parameters from the most recent SPT-SZ cluster cosmology analysis \citep{bocquet19}, we 
calibrate the observable mass scaling relations 
from the number counts of MARD-Y3 clusters (see Section~\ref{sec:numcounts}). 
In addition to the traditional number 
counts as a function of X-ray flux and redshift, we also use the number counts as a function of richness and redshift and the 
number counts as a function of X-ray flux, richness and redshift. 

We find that the different flavours of number counts provide scaling relation constraints that are statistically consistent 
with the constraints from the SPT-SZ calibration performed on the cross-matched sample. This validates the MARD-Y3
selection function, because the SPT-SZ calibration is independent of the 
MARD-Y3 selection function, while the number count experiments are highly sensitive to it. This leads us to the main conclusion of 
this work: {\it optical cleaning with MCMF allows one to create a clean cluster sample with a controllable selection function}. Once direct mass 
information is available, we will be able to study cosmology using the MARD-Y3 number counts. The fact that the incompleteness (primarily at low masses) introduced by optical cleaning can be modeled using the richness-mass relation implies
that much larger, reliable cluster samples extending to higher redshift  and lower masses can be constructed from ICM based surveys if 
appropriately deep optical and NIR data are available.

In these tests we identify some moderate tension between constraints on the luminosity--mass relation from 
X-ray number counts and 2D (optical+X-ray) number counts: while the former prefers small intrinsic X-ray scatter 
and a steep mass trend, the latter prefers a shallower mass trend and larger intrinsic scatter. This hints at some unresolved 
systematic on the X-ray side. As discussed in Section~\ref{sec:outcome_val}, the high scatter scenario is 
supported by the scatter among different measurements of luminosity on the same 
X-ray raw data highlighted in \citetalias{klein19}, a further indicator of systematics in the flux measurement. Nevertheless, the individual masses derived from the different 
scenarios are consistent within the uncertainties. Because there is no consensus in the literature, this question merits further 
investigation once direct mass information is available.

In Section~\ref{sec:mass_prediction} we present the implications for MARD-Y3 masses from different scaling relations that emerge from the tests described above. There is a tendency for these masses to lie below those calculated using externally calibrated relations from the literature \citep{saro15,bulbul19}, and the largest tensions occur at low masses.

We also study the MARD-Y3 selection function by comparing the matched and unmatched MARD-Y3 clusters in the SPT-SZ sample and vice-versa.  If the selection functions for MARD-Y3 and SPT-SZ are well understood then the number of matched and unmatched clusters should be fully consistent with the statistical expectations.  Simply stated, this test allows us to constrain MARD-Y3 contamination or SPT-SZ incompleteness (the two effects are degenerate in this test).  As discussed in Section~\ref{sec:SPTdetMARDY3}, in the large 
scatter luminosity--mass scenario, we find no evidence for either effect, while in the low scatter scenario we find evidence at the 2 sigma level for either contamination or incompleteness.  Given that the MARD-Y3 sample contamination is estimated to be 2.5\% \citepalias{klein19} and given that the SPT-SZ sample has been used to produce cosmological constraints in good agreement with independent probes \citep{dehaan16,bocquet19}, we take this as further evidence supporting the large scatter scenario.

Looking at the probability of a MARD-Y3 confirmation of an SPT-SZ selected cluster we find a subsample of clusters whose 
SZE properties suggest they should not have been detected in MARD-Y3, but they are. The size of this sample is susceptible to the scaling relation constraints assumed. As discussed in 
Section~\ref{sec:mardy_on_spt}, if we model this as an outlier fraction in the distribution of scatter about
the mass--observable relations (either abnormally high X-ray flux and richness, or low SZE signature), 
we find a preference for an outlier fraction of $\sim5\%$--$10\%$ with a detection significance ranging from 1.4 -- 3.6 sigma, depending on the scaling relation constraints assumed. More accurate and independent mass information is needed to further elucidate this aspect of the cluster population.

From a methodological perspective we demonstrate several new techniques:
\begin{enumerate}
    \item Optical follow-up allows for three different flavours of number counts. While we demonstrate the potential of 
    multi-observable number counts, the real novelty is that one can perform number counts as a function of optical richness 
    for a predominantly X-ray selected sample in a consistent manner. In a blinded WL-calibrated cosmological analysis we 
    would demand that the blinded cosmology from these three likelihoods be consistent.
    \item We improve the technique of studying matched and unmatched clusters in two independent samples 
    by including binomial statistics and marginalizing over the systematic 
    uncertainties associated with lack of knowledge of the observable--mass relation parameters. With the use of probability trees, extra 
    probabilities, such as those quantifying contamination, incompleteness or outlier fractions, can all be constrained in a statistically sound way. This 
    technique does not depend on the amplitude and redshift evolution of the number of objects, reducing its cosmological sensitivity.
    \item We present a flexible empirical method to determine the X-ray selection function from the data itself. It does not require 
    any assumptions about cluster morphology. The empirical nature of the constraint also marginalizes 
    over the inherent uncertainty of the selection function by sampling extra nuisance parameters. Shifts in these nuisance 
    parameters when, for example, calibrating the observable--mass relation using different sources of information
    can serve as a further test of systematic.
\end{enumerate}

The techniques highlighted here have the potential to enable better control of systematic effects in cosmological studies of current 
and upcoming cluster surveys. They also demonstrate the potential of multi-wavelength analysis of cluster samples not only to inform 
the selection function modeling of individual surveys, but also to identify interesting cluster populations. This will 
help exploit the wealth of information provided by deep and wide surveys in X-ray, optical, 
NIR and millimeter wavelengths.

\section*{Acknowledgements}

We acknowledge financial support from the MPG Faculty Fellowship program, the DFG Cluster of Excellence 
``Origin and Structure of the Universe'', the new DFG cluster "Origins" and the Ludwig-Maximilians-Universit\"at Munich.
Numerical computations in this work relied on the \texttt{python} packages \texttt{numpy} \citep{numpy} and 
\texttt{scipy} \citep{scipy}. The plots were produced using the package \texttt{matplotlib} \citep{matplotlib}. The 
marginal contour plots were created using \texttt{pyGTC} \citep{pygtc}. Posterior samples have been drawn from the likelihood functions and the priors using \texttt{emcee} \citep{emcee}. 

This work was performed in the context of the South-Pole Telescope scientific program.  SPT is supported by the  National  Science  Foundation  through  grant  PLR-1248097.  Partial support is also provided by the NSF Physics Frontier Center grant PHY-0114422 to the Kavli Institute  of  Cosmological  Physics  at  the  University  of Chicago,  the  Kavli  Foundation  and  the  Gordon  and Betty Moore Foundation grant GBMF 947 to the University of Chicago.  This work is also supported by the U.S.  Department  of  Energy.  AAS acknowledges support by US NSF AST-1814719.

This paper has gone through internal review by the DES collaboration. Funding for the DES Projects has been provided by the U.S. Department of Energy, the U.S. National Science Foundation, the Ministry of Science and Education of Spain, 
the Science and Technology Facilities Council of the United Kingdom, the Higher Education Funding Council for England, the National Center for Supercomputing 
Applications at the University of Illinois at Urbana-Champaign, the Kavli Institute of Cosmological Physics at the University of Chicago, 
the Center for Cosmology and Astro-Particle Physics at the Ohio State University,
the Mitchell Institute for Fundamental Physics and Astronomy at Texas A\&M University, Financiadora de Estudos e Projetos, 
Funda{\c c}{\~a}o Carlos Chagas Filho de Amparo {\`a} Pesquisa do Estado do Rio de Janeiro, Conselho Nacional de Desenvolvimento Cient{\'i}fico e Tecnol{\'o}gico and 
the Minist{\'e}rio da Ci{\^e}ncia, Tecnologia e Inova{\c c}{\~a}o, the Deutsche Forschungsgemeinschaft and the Collaborating Institutions in the Dark Energy Survey. 

The Collaborating Institutions are Argonne National Laboratory, the University of California at Santa Cruz, the University of Cambridge, Centro de Investigaciones Energ{\'e}ticas, 
Medioambientales y Tecnol{\'o}gicas-Madrid, the University of Chicago, University College London, the DES-Brazil Consortium, the University of Edinburgh, 
the Eidgen{\"o}ssische Technische Hochschule (ETH) Z{\"u}rich, 
Fermi National Accelerator Laboratory, the University of Illinois at Urbana-Champaign, the Institut de Ci{\`e}ncies de l'Espai (IEEC/CSIC), 
the Institut de F{\'i}sica d'Altes Energies, Lawrence Berkeley National Laboratory, the Ludwig-Maximilians Universit{\"a}t M{\"u}nchen and the associated Excellence Cluster Universe, 
the University of Michigan, the National Optical Astronomy Observatory, the University of Nottingham, The Ohio State University, the University of Pennsylvania, the University of Portsmouth, 
SLAC National Accelerator Laboratory, Stanford University, the University of Sussex, Texas A\&M University, and the OzDES Membership Consortium.

Based in part on observations at Cerro Tololo Inter-American Observatory, National Optical Astronomy Observatory, which is operated by the Association of 
Universities for Research in Astronomy (AURA) under a cooperative agreement with the National Science Foundation.

The DES data management system is supported by the National Science Foundation under Grant Numbers AST-1138766 and AST-1536171.
The DES participants from Spanish institutions are partially supported by MINECO under grants AYA2015-71825, ESP2015-66861, FPA2015-68048, SEV-2016-0588, SEV-2016-0597, and MDM-2015-0509, 
some of which include ERDF funds from the European Union. IFAE is partially funded by the CERCA program of the Generalitat de Catalunya.
Research leading to these results has received funding from the European Research
Council under the European Union's Seventh Framework Program (FP7/2007-2013) including ERC grant agreements 240672, 291329, and 306478.
We  acknowledge support from the Brazilian Instituto Nacional de Ci\^encia
e Tecnologia (INCT) e-Universe (CNPq grant 465376/2014-2).

This manuscript has been authored by Fermi Research Alliance, LLC under Contract No. DE-AC02-07CH11359 with the U.S. Department of Energy, Office of Science, Office of High Energy Physics.

\section*{Affiliations}
$^{1}$Faculty of Physics, Ludwig-Maximilians-Universit\"at, Scheinerstr. 1, 81679, Munich,  Germany\\
$^{2}$Excellence Cluster Origins, Boltzmannstr. 2, 85748, Garching, Germany\\
$^{3}$Max Planck Institute for Extraterrestrial Physics, Giessenbachstr. 85748, Garching, Germany \\
$^{4}$Cerro Tololo Inter-American Observatory, National Optical Astronomy Observatory, Casilla 603, La Serena, Chile \\
$^{5}$Departamento de F\'{i}sica Matem\'{a}tica, Instituto de F\'{i}sica, Universidade de S\~{a}o Paulo, CP 66318, S\~{a}o Paulo, SP, 05314-970, Brazil \\
$^{6}$Laborat\'{o}rio Interinstitucional de e-Astronomia - LIneA, Rua Gal. Jos\'{e} Cristino 77, Rio de Janeiro, RJ - 20921-400, Brazil\\
$^{7}$Fermi National Accelerator Laboratory, P. O. Box 500, Batavia, IL 60510, USA\\
$^{8}$Kavli Institute for Cosmological Physics, University of Chicago, 5640 South Ellis Avenue, Chicago, IL, USA 60637\\
$^{9}$Department of Astronomy and Astrophysics, University of Chicago, 5640 South Ellis Avenue, Chicago, IL, USA 60637\\
$^{10}$CNRS, UMR 7095, Institut d'Astrophysique de Paris, F-75014, Paris, France\\
$^{11}$Sorbonne Universit\'{e}s, UPMC Univ Paris 06, UMR 7095, Institut d'Astrophysique de Paris, F-75014, Paris, France\\
$^{12}$Department of Physics and Astronomy, Pevensey Building, University of Sussex, Brighton, BN1 9QH, UK\\
$^{13}$Department of Physics \& Astronomy, University College London, Gower Street, London, WC1E 6BT, UK\\
$^{14}$Kavli Institute for Particle Astrophysics \& Cosmology, P. O. Box 2450, Stanford University, Stanford, CA 94305, USA\\
$^{15}$SLAC National Accelerator Laboratory, Menlo Park, CA 94025, USA\\
$^{16}$Centro de Investigaciones Energ\'{e}ticas, Medioambientales y Tecnol\'{o}gicas (CIEMAT), Madrid, Spain\\
$^{17}$Department of Astronomy, University of Illinois at Urbana-Champaign, 1002 W. Green Street, Urbana, IL 61801, USA\\
$^{18}$National Center for Supercomputing Applications, 1205 West Clark St., Urbana, IL 61801, USA\\
$^{19}$Institut de F\'{i}sica d'Altes Energies (IFAE), The Barcelona Institute of Science and Technology, Campus UAB, 08193 Bellaterra (Barcelona) Spain\\
$^{20}$Oskar Klein Centre, Department of Physics, Stockholm University, AlbaNova University Centre, SE 106 91 Stockholm, Sweden\\
$^{21}$INAF-Osservatorio Astronomico di Trieste, via G. B. Tiepolo 11, I-34143 Trieste, Italy\\
$^{22}$Institute for Fundamental Physics of the Universe, Via Beirut 2, 34014 Trieste, Italy\\
$^{23}$Laborat\'{o}rio Interinstitucional de e-Astronomia - LIneA, Rua Gal. Jos\'e Cristino 77, Rio de Janeiro, RJ - 20921-400, Brazil\\
$^{24}$Observat\'{o}rio Nacional, Rua Gal. Jos\'{e} Cristino 77, Rio de Janeiro, RJ - 20921-400, Brazil\\
$^{25}$Department of Physics, IIT Hyderabad, Kandi, Telangana 502285, India\\
$^{26}$Department of Astronomy/Steward Observatory, University of Arizona, 933 North Cherry Avenue, Tucson, AZ 85721-0065, USA\\
$^{27}$Jet Propulsion Laboratory, California Institute of Technology, 4800 Oak Grove Dr., Pasadena, CA 91109, USA\\
$^{28}$Department of Astronomy, University of Michigan, Ann Arbor, MI 48109, USA\\
$^{29}$Department of Physics, University of Michigan, Ann Arbor, MI 48109, USA\\
$^{30}$Institut d'Estudis Espacials de Catalunya (IEEC), 08034 Barcelona, Spain\\
$^{31}$Institute of Space Sciences (ICE, CSIC),  Campus UAB, Carrer de Can Magrans, s/n,  08193 Barcelona, Spain\\
$^{32}$Kavli Institute for Cosmological Physics, University of Chicago, Chicago, IL 60637, USA\\
$^{33}$Instituto de Fisica Teorica UAM/CSIC, Universidad Autonoma de Madrid, 28049 Madrid, Spain\\
$^{34}$Department of Physics, Stanford University, 382 Via Pueblo Mall, Stanford, CA 94305, USA\\
$^{35}$Department of Physics, ETH Zurich, Wolfgang-Pauli-Strasse 16, CH-8093 Zurich, Switzerland\\
$^{36}$School of Mathematics and Physics, University of Queensland,  Brisbane, QLD 4072, Australia\\
$^{37}$Santa Cruz Institute for Particle Physics, Santa Cruz, CA 95064, USa\\
$^{38}$Center for Cosmology and Astro-Particle Physics, The Ohio State University, Columbus, OH 43210, USA\\
$^{39}$Department of Physics, The Ohio State University, Columbus, OH 43210, USA\\
$^{40}$Center for Astrophysics $\vert$ Harvard \& Smithsonian, 60 Garden Street, Cambridge, MA 02138, USA\\
$^{41}$Australian Astronomical Optics, Macquarie University, North Ryde, NSW 2113, Australia\\
$^{42}$Lowell Observatory, 1400 Mars Hill Rd, Flagstaff, AZ 86001, USA\\
$^{43}$George P. and Cynthia Woods Mitchell Institute for Fundamental Physics and Astronomy, and Department of Physics and Astronomy, Texas A\&M University, College Station, TX 77843,  USA\\
$^{44}$Department of Astrophysical Sciences, Princeton University, Peyton Hall, Princeton, NJ 08544, USA\\
$^{45}$Instituci\'{o} Catalana de Recerca i Estudis Avan\c{c}ats, E-08010 Barcelona, Spain\\
$^{46}$Astronomy Unit, Department of Physics, University of Trieste, via Tiepolo 11, I-34131 Trieste, Italy\\
$^{47}$Brookhaven National Laboratory, Bldg 510, Upton, NY 11973, USA\\
$^{48}$School of Physics and Astronomy, University of Southampton,  Southampton, SO17 1BJ, UK\\
$^{49}$Computer Science and Mathematics Division, Oak Ridge National Laboratory, Oak Ridge, TN 37831\\
$^{50}$Institute of Cosmology and Gravitation, University of Portsmouth, Portsmouth, PO1 3FX, UK\\
$^{51}$Universit\"ats-Sternwarte, Fakult\"at f\"ur Physik, Ludwig-Maximilians Universit\"at M\"unchen, Scheinerstr. 1, 81679 M\"unchen, Germany\\


\bibliographystyle{mnras}
\bibliography{manuscript} 


\appendix

\section{X-ray flux error model}\label{app:X_meas_uncert}

\begin{figure}
    \includegraphics[width=\columnwidth]{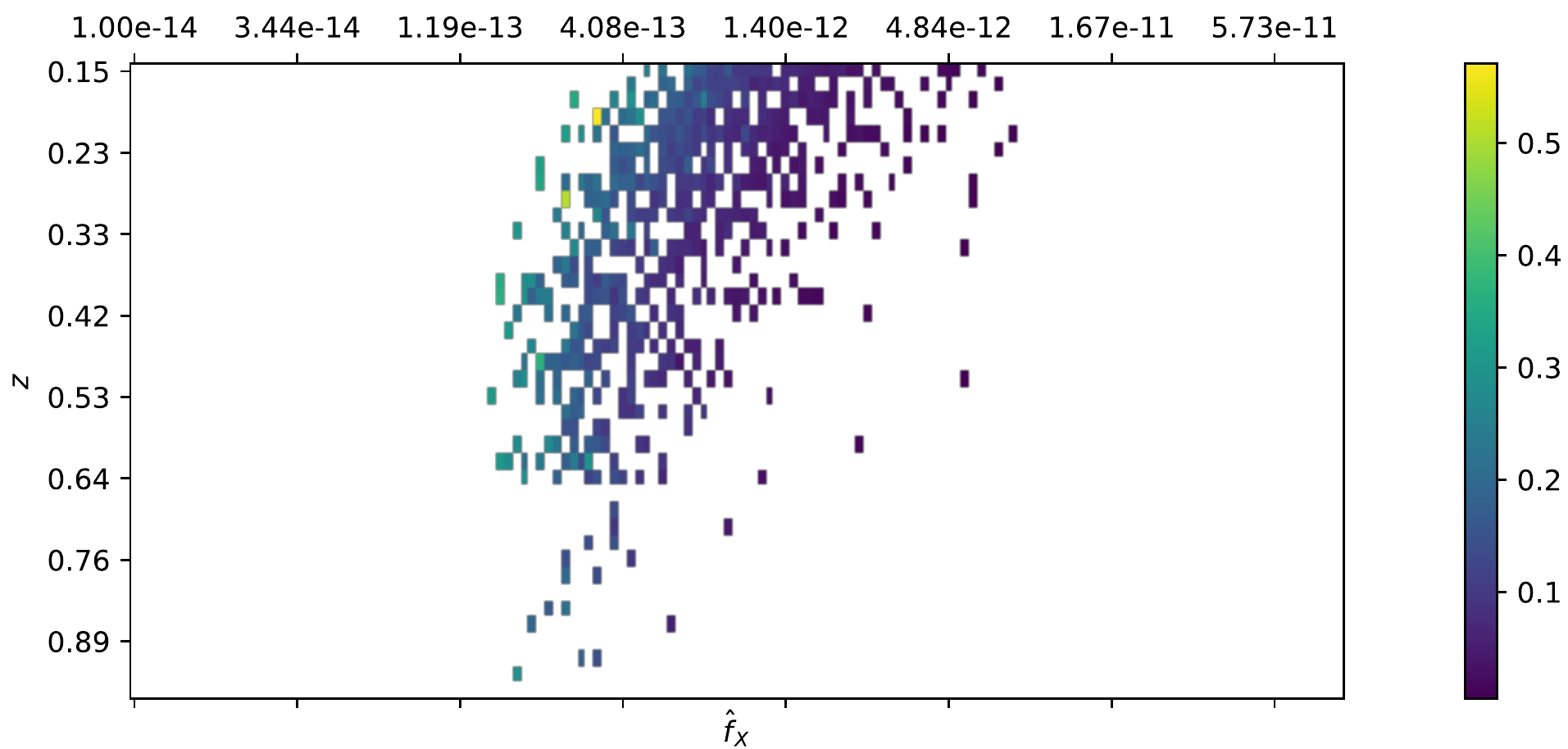}
	\includegraphics[width=\columnwidth]{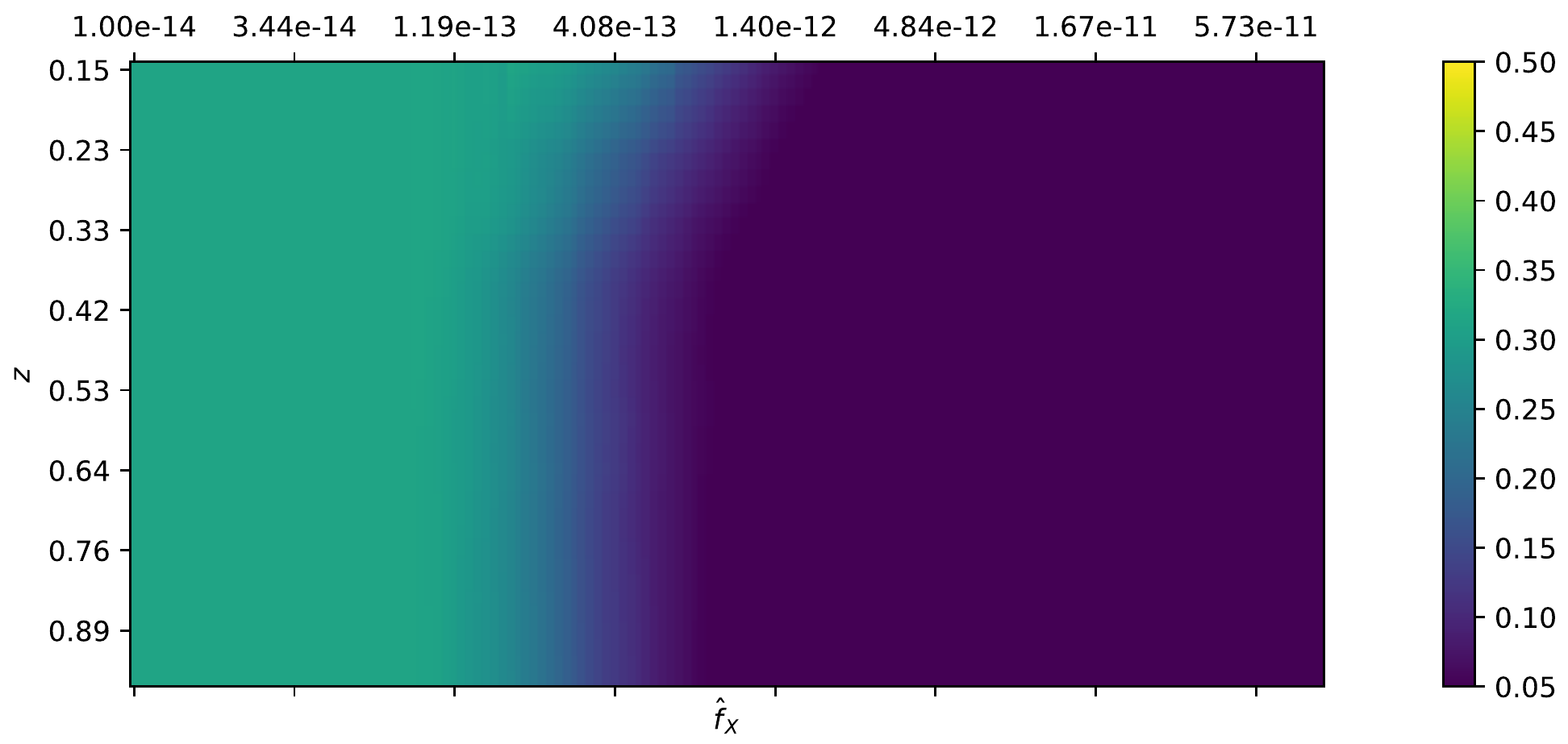}
	\vskip-0.10in
    \caption{Construction of the expected measurement uncertainty at the median redshift (lower panel) as a function of redshift (y axis) and measured flux (x axis), from the normalized measurement uncertainties reported in the catalog (upper panel). In the range where we have data, the predicted observational measurement uncertainty nicely extrapolates the trends in the catalog.}
    \label{fig:constr_meas_error}
\end{figure}

As outlined in Section~\ref{sec:treat_meas_err} in some application it is not sufficient to know the measurement 
uncertainty only for the objects in the catalog, but the measurement uncertainty is also needed for arbitrary values of measured flux 
$\hat f_\text{X}$ and redshift $z$. We therefore seek to predict $\hat \sigma_\text{X}^2 (\hat f_\text{X}, z, t_\text{exp})$ from the 
measured entries $\hat \sigma_\text{X}^{(i)}$. First we note that the measurement uncertainties in the catalog scale with the 
exposure time approximately like $\hat \sigma_\text{X}^{(i)} \sim t_\text{exp}^{-0.5}$. We thus bin the quantity $(\hat 
\sigma_\text{X}^{(i)})^2 \,t_\text{exp}/ 400\text{s}$ in fine redshift and measured flux bins, as shown in the upper panel 
of Fig.~\ref{fig:constr_meas_error}. This is then extrapolated and smoothed to provide a prediction of the 
measurement uncertainty $\sigma_\text{pred}^2 (\hat f_\text{X}, z)$ at each measured flux $\hat f_\text{X}$ and 
redshift $z$, if the exposure time was $t_\text{exp}=400\text{s}$, shown in the lower panel of 
Fig.~\ref{fig:constr_meas_error}. This prediction can than be scaled to the desired exposure time assuming the scaling 
above, i.e.

\begin{equation}
\hat \sigma_\text{X}^2 (\hat f_\text{X}, z, t_\text{exp}) = \sigma_\text{pred}^2 (\hat f_\text{X}, z) \frac{400 \text{s}}
{t_\text{exp}}.
\end{equation}

Applying this prediction the cluster in our catalog and comparing the resulting uncertainties to the actual measurement 
uncertainty leads to a mean relative error of $5.6\%$. Furthermore, these residuals display no strong trends with 
background brightness, neutral hydrogen column density or measured extent. Given the small magnitude, we choose 
to ignore this source of systematic uncertainty, which could be included at the cost of sampling extra nuisance 
parameters.

\section{Gallery of Multi-wavelength Cluster Images} \label{app:gallery}

\newpage
\onecolumn
    
\begin{figure*}
	\includegraphics[width=0.32\columnwidth]{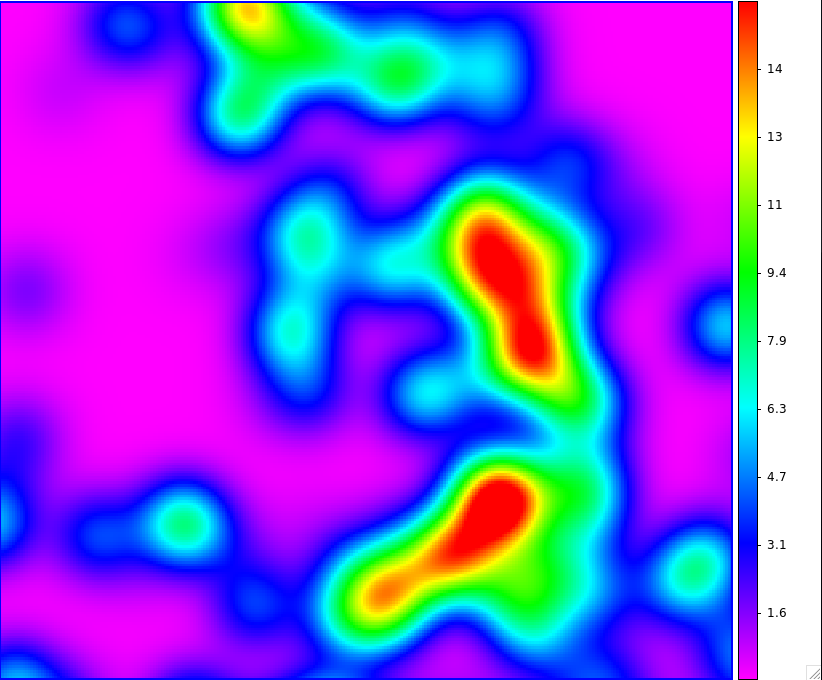}
	\includegraphics[width=0.32\columnwidth]{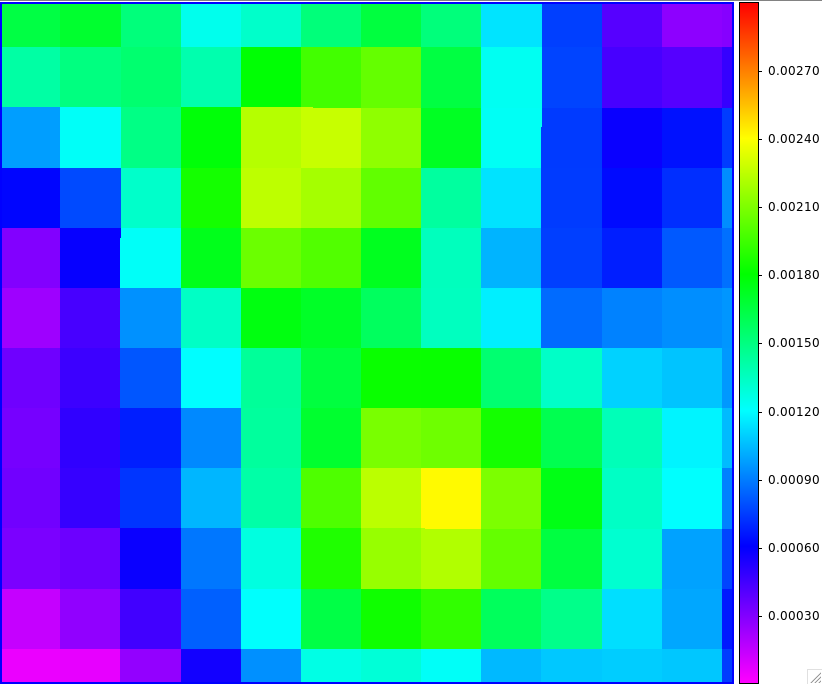}
	\includegraphics[width=0.32\columnwidth]{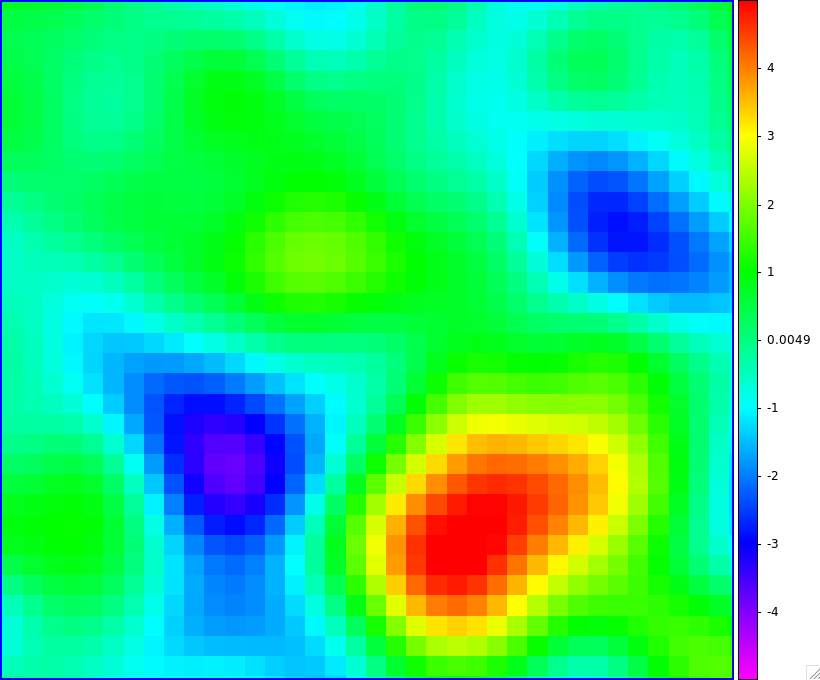}
	    \vskip-0.1in
	    \caption{SPT-CL J2358-6129. Red sequence galaxy density map in DES at $z=0.403$ (left panel), exposure and background corrected count rate from RASS (center panel), SPT signal-to-noise map in the smaller filter scale (right panel). One pixel in the RASS count rate maps has diameter 45''. The two MARD-Y3 detected clusters (X-ray peaks in the center panel) are matched to the same SPT-SZ detection (right panel) (c.f. section~\ref{sec:matched_sample}).}
        \label{fig:gal1}
\end{figure*}   
 
\begin{figure*}
	\includegraphics[width=0.32\columnwidth]{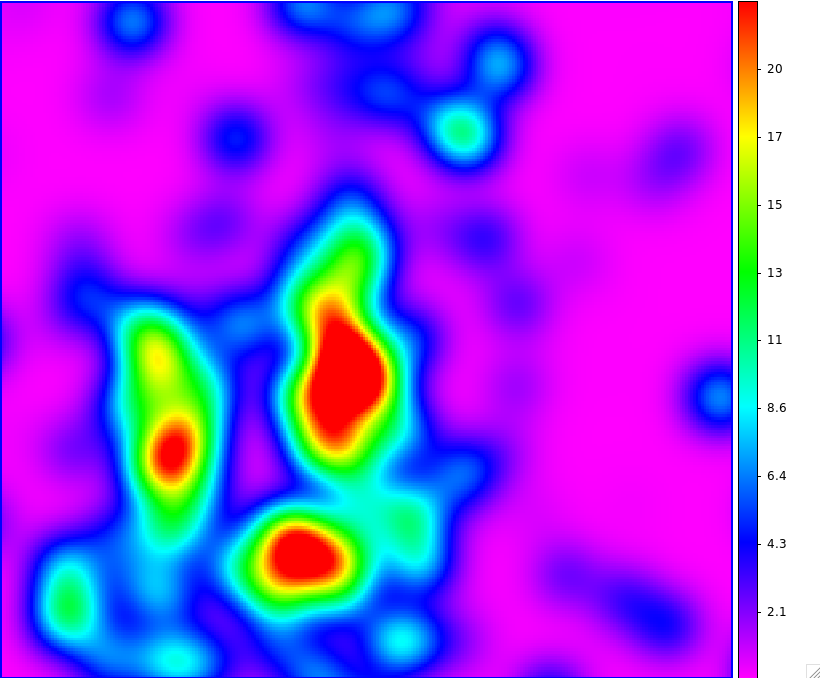}
	\includegraphics[width=0.32\columnwidth]{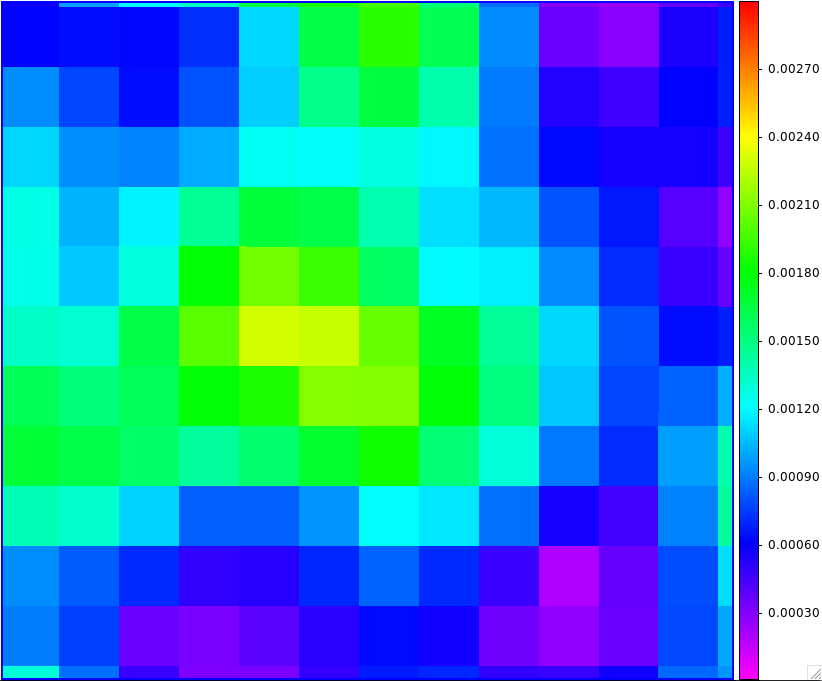}
	\includegraphics[width=0.32\columnwidth]{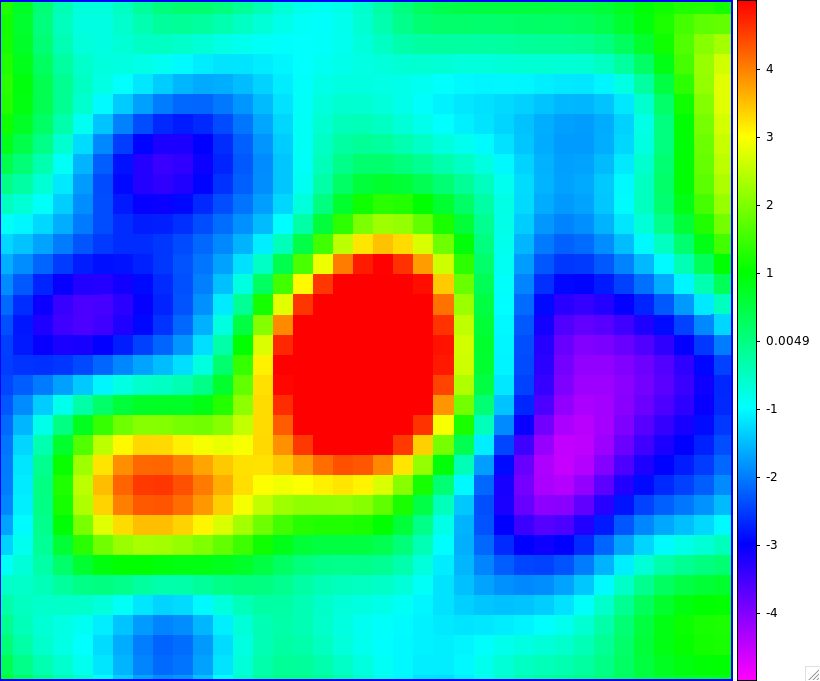}
	    \vskip-0.1in
	    \caption{SPT-CL J2331-5051 and SPT-CL J2331-5053. Red sequence galaxy density map in DES at $z=0.577$ (left panel), exposure and background corrected count rate from RASS (center panel), SPT signal-to-noise map in the smaller filter scale (right panel). One pixel in the RASS count rate maps has diameter 45''. The two SPT-SZ detected clusters (right panel) are matched to the same MARD-Y3 detection (right panel) (c.f. section~\ref{sec:matched_sample}).}
        \label{fig:gal_doublespt}
\end{figure*}    

\begin{figure*}
	\includegraphics[width=0.32\columnwidth]{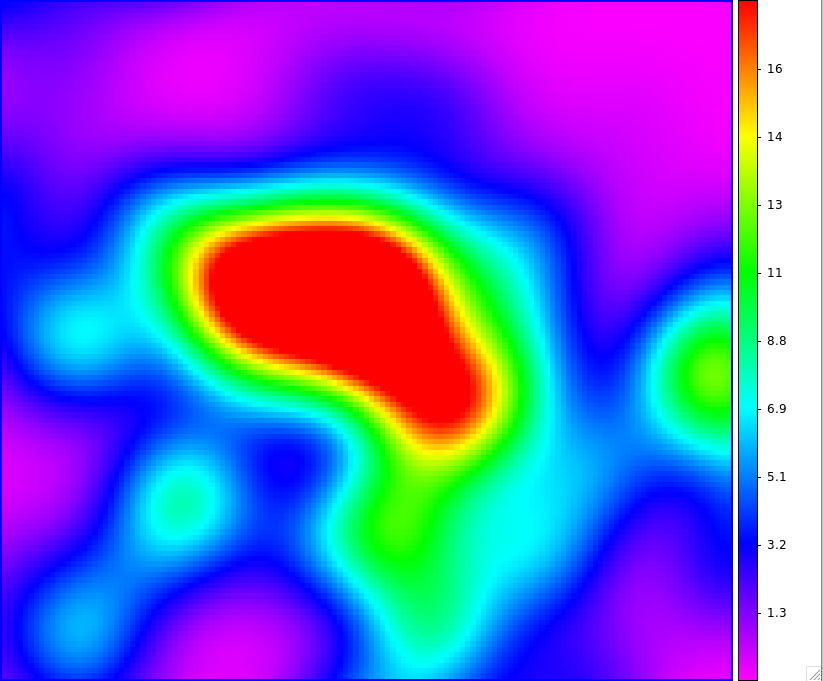}
	\includegraphics[width=0.32\columnwidth]{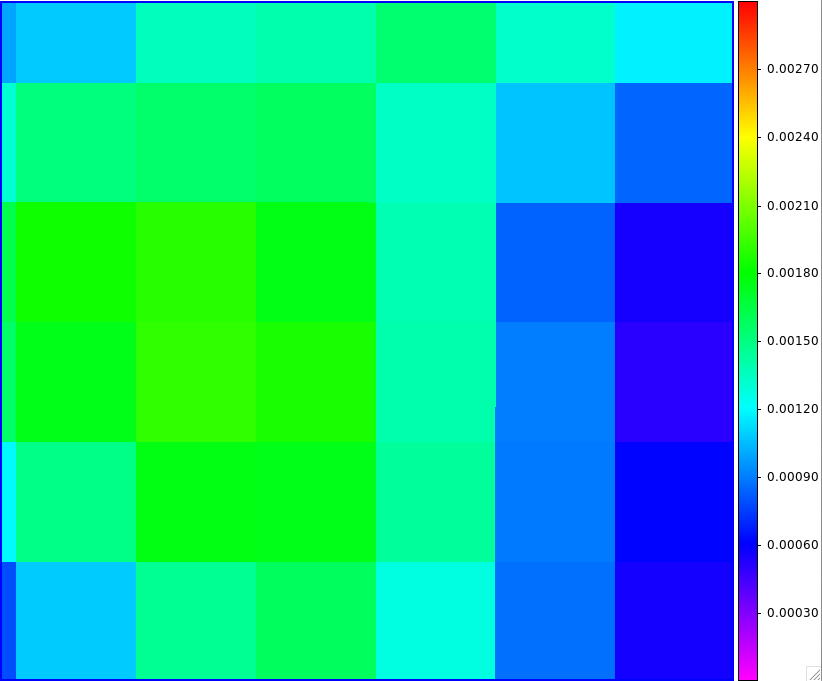}
	\includegraphics[width=0.32\columnwidth]{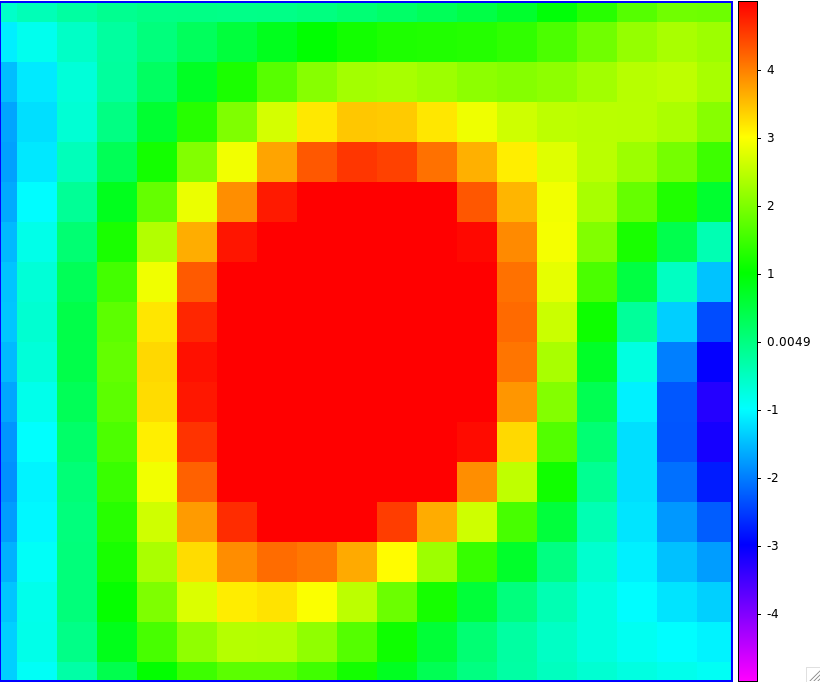}
	    \vskip-0.1in
	    \caption{SPT-CL J0218-4233. Red sequence galaxy density map in DES at $z=0.755$ (left panel), exposure and background corrected count rate from RASS (center panel), SPT signal-to-noise map in the smaller filter scale (right panel). One pixel in the RASS count rate maps has diameter 45''. This SPT-SZ detection was unexpectedly confirmed by MARD-Y3, c.f. section~\ref{sec:mardy_on_spt}.}
        \label{fig:gal4}
\end{figure*}   

\begin{figure*}
	\includegraphics[width=0.32\columnwidth]{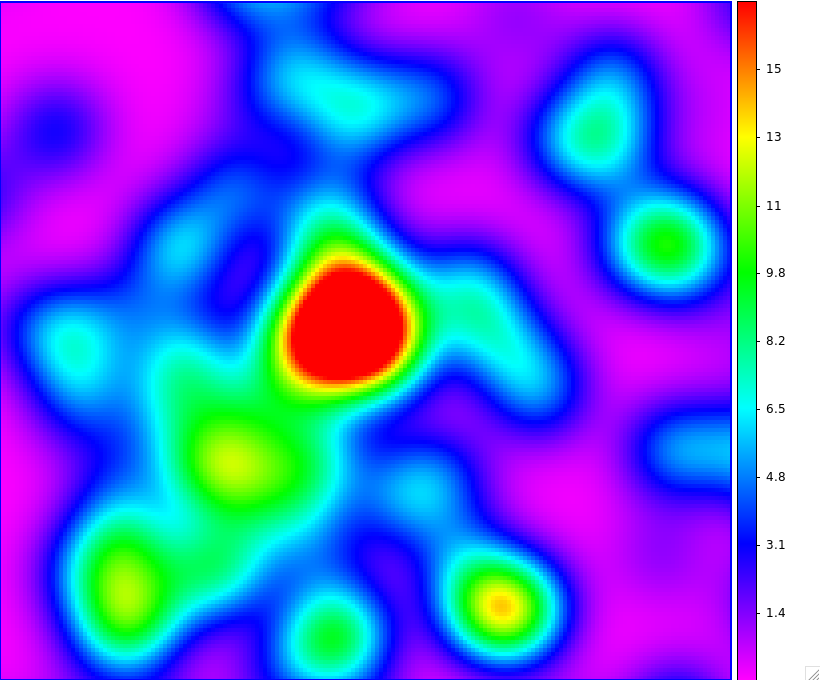}
	\includegraphics[width=0.32\columnwidth]{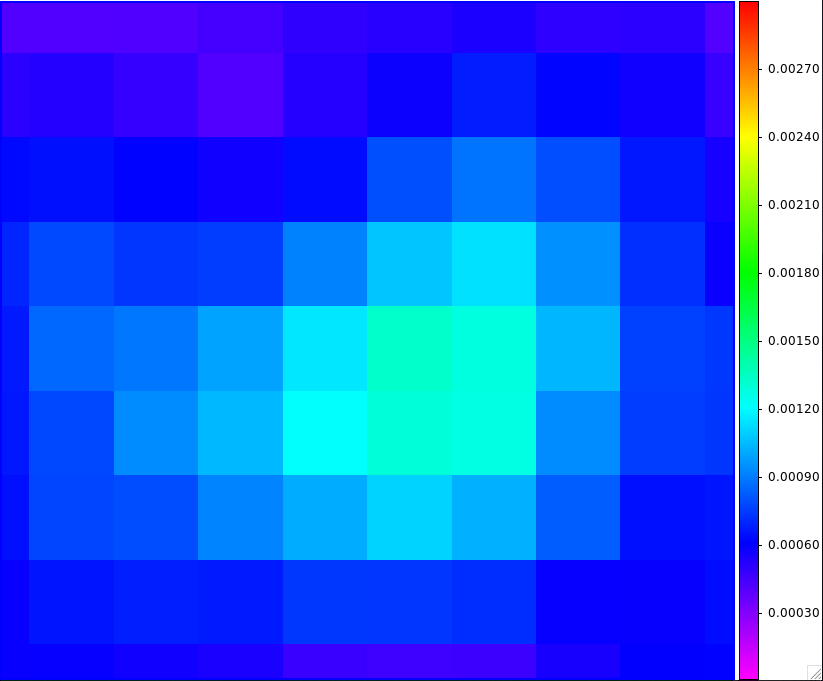}
	\includegraphics[width=0.32\columnwidth]{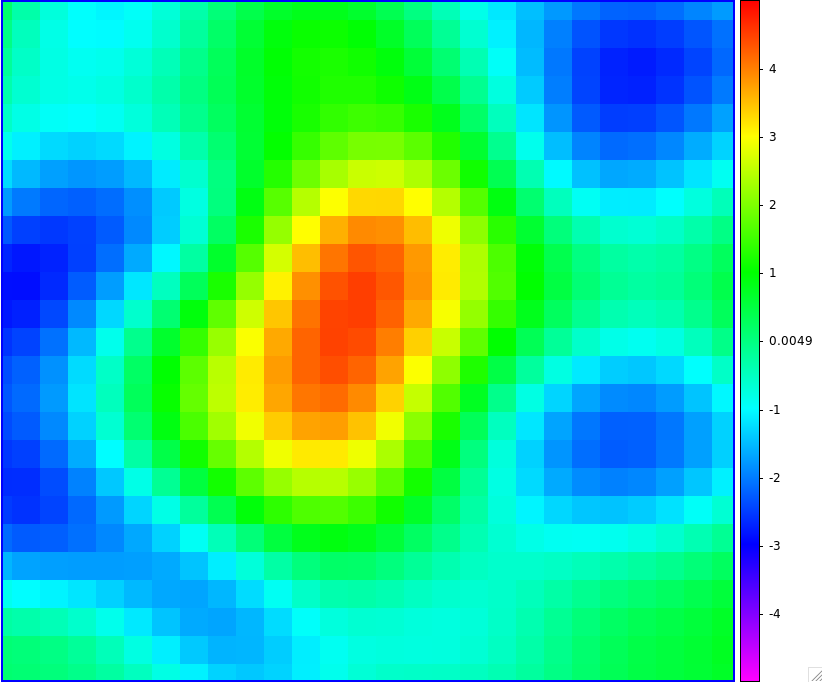}
	    \vskip-0.1in
	    \caption{SPT-CL J0324-6236. Red sequence galaxy density map in DES at $z=0.638$ (left panel), exposure and background corrected count rate from RASS (center panel), SPT signal-to-noise map in the smaller filter scale (right panel). One pixel in the RASS count rate maps has diameter 45''. This SPT-SZ detection was unexpectedly confirmed by MARD-Y3, c.f. section~\ref{sec:mardy_on_spt}.}
        \label{fig:gal3}
\end{figure*}    

\begin{figure*}
	\includegraphics[width=0.285\columnwidth]{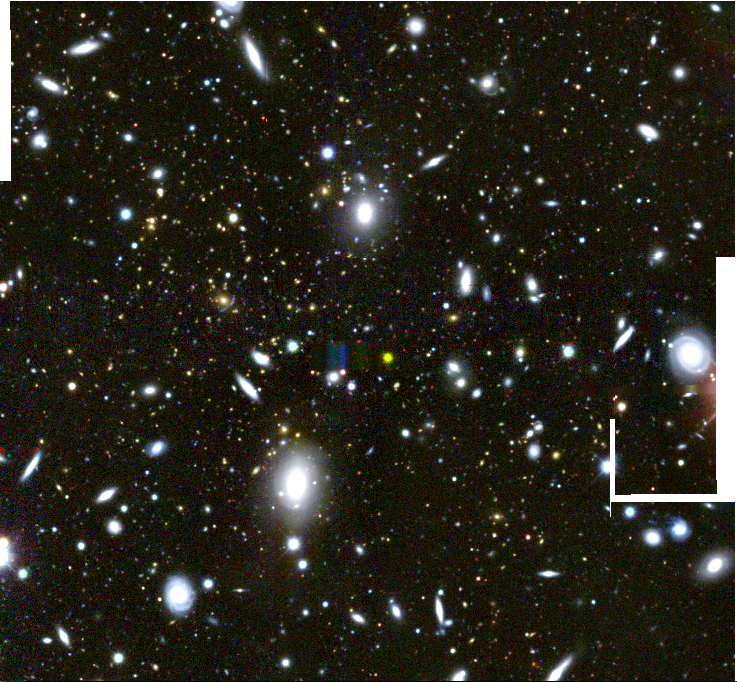}
	\includegraphics[width=0.32\columnwidth]{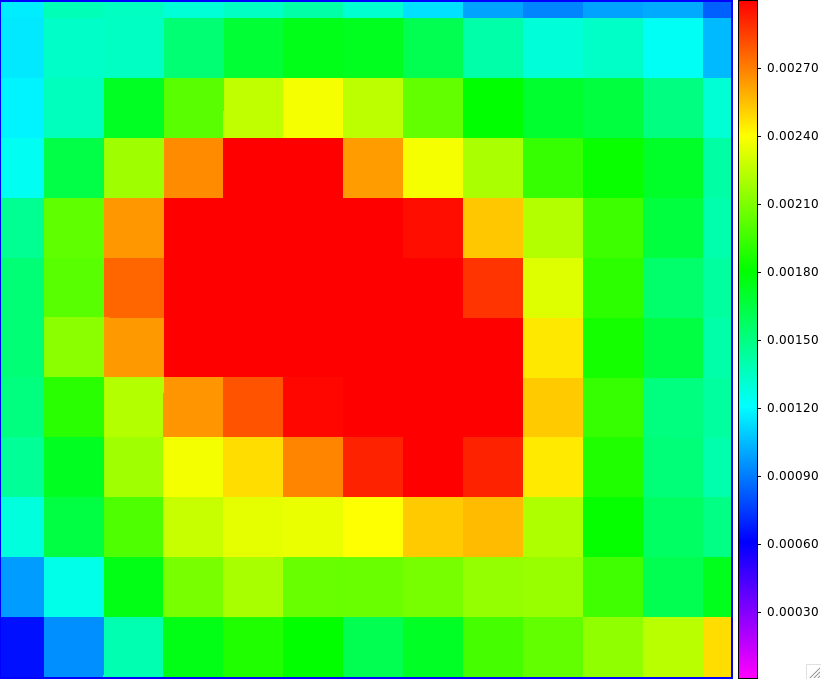}
	\includegraphics[width=0.32\columnwidth]{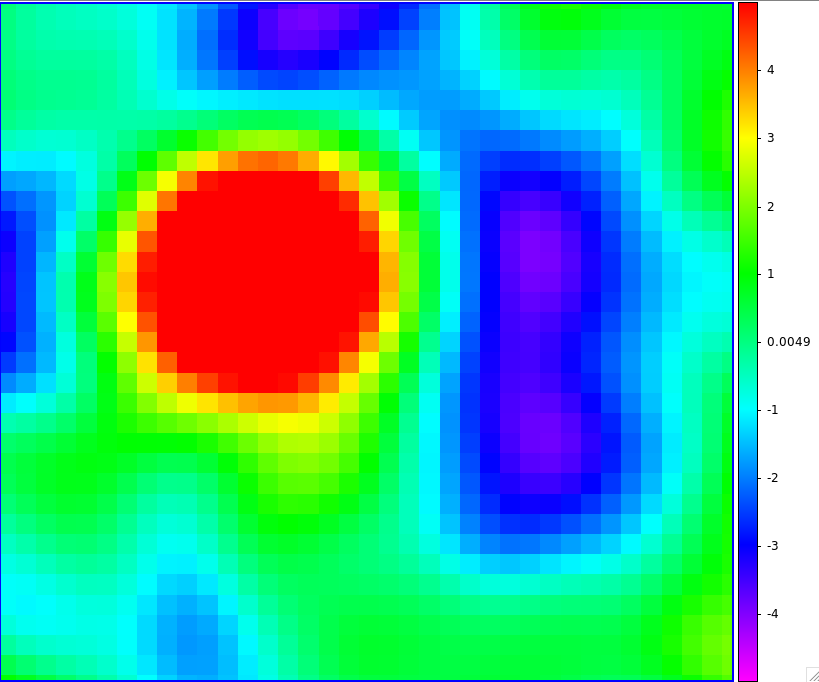}
	\includegraphics[width=0.32\columnwidth]{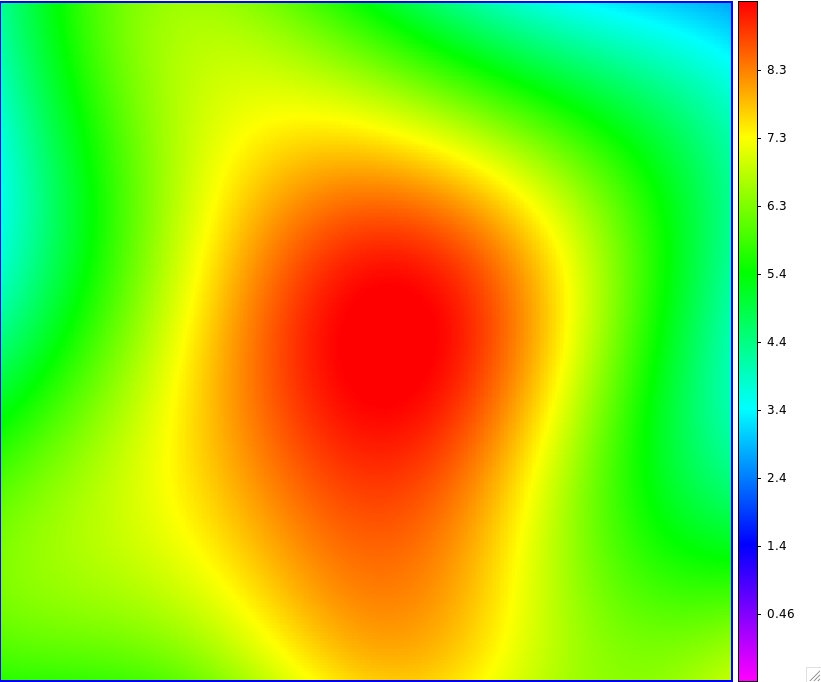}
	\includegraphics[width=0.32\columnwidth]{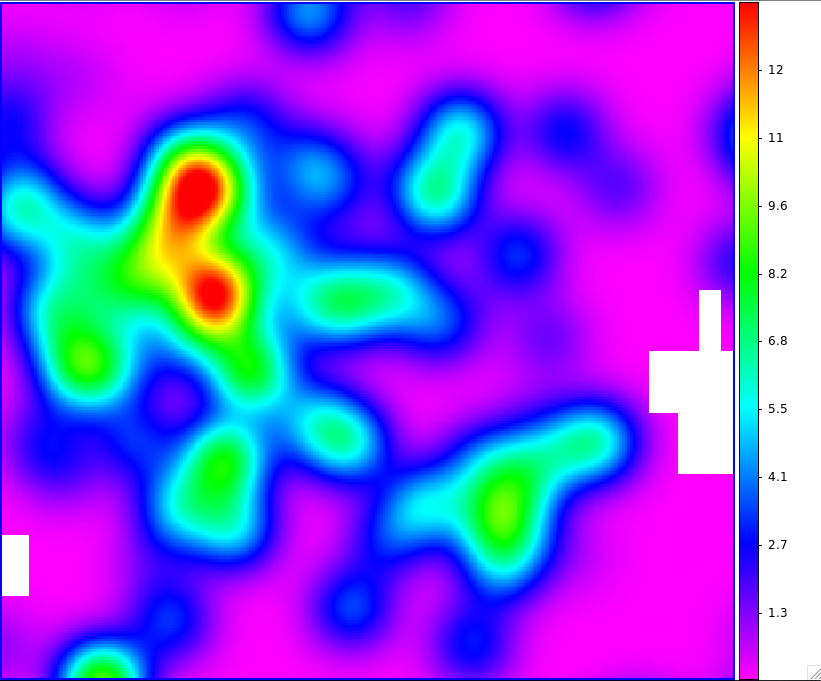}
	    \vskip-0.1in
	    \caption{2RXS J033045.2-522845. DES image (upper left panel), exposure and background corrected count rate from RASS (upper center panel ), SPT signal-to-noise map in the smaller filter scale (upper right panel), Red sequence galaxy density map in DES at $z=0.056$ (lower left panel) and at $z=0.428$ (lower right panel). One pixel in the RASS count rate maps has diameter 45''. This MARD-Y3 detection was unexpectedly missed by SPT-SZ, c.f. section~\ref{sec:SPTdetMARDY3}. This is due to a catastrophic redshift failure of MCMF when run on SPT-SZ detections: it selected the low redshift group (white galaxies in the upper left panel, red sequence galaxy density in the lower left panel), while the actual structure is at intermediate redshift (galaxy density in the lower right corner). When zooming into DES image, a blue Einstein arc around the brightest central galaxy can be seen. }
        \label{fig:gal2}
\end{figure*}


\label{lastpage}
\end{document}